\documentclass[aps,twocolumn,prd,superscriptaddress,preprintnumbers,floatfix,showpacs,showkeys,letter]{revtex4-1}

\usepackage{newtxtext}
\usepackage[latin1]{inputenc}
\usepackage[T1]{fontenc}
\usepackage{lmodern}

\usepackage{empheq}
\usepackage{mathtools}
\usepackage{amsmath}

\usepackage{amscd}
\usepackage{amsmath}
\usepackage{amssymb}
\usepackage{graphicx}%
\usepackage{scalerel}

\usepackage{dsfont}

\usepackage[ruled]{algorithm2e}
\usepackage{needspace}
\usepackage{newtxtext}
\usepackage{booktabs}
\usepackage{bm,comment}

\newcommand{\indep}{\perp\!\!\!\perp}

\usepackage[colorlinks=true, pdfstartview=FitV, linkcolor=blue,
            citecolor=blue, urlcolor=blue]{hyperref} 
\usepackage{subcaption}
\usepackage{sidecap}
\usepackage{grffile}
\usepackage{xcolor}

\providecommand{\U}[1]{\protect\rule{.1in}{.1in}}
\newenvironment{proof}[1][Proof]{\noindent\textbf{#1.} }{\ \rule{0.5em}{0.5em}}

\definecolor{rred}{rgb}{0.7,0,0.1}
\definecolor{ccyan}{rgb}{0,.5,1}
\definecolor{greenrb}{rgb}{0.2,0.6,0.2}
\definecolor{faintpink}{rgb}{1.0, 0.9, 0.95}

\graphicspath{{./Fig_LRT_jumps/}}

\def\W{{\boldsymbol{W}_t}}

\def\L{{\mathcal{L}}}
\def\X{{\mathcal{X}}}

\newcommand{\norm}[1]{\left\lVert#1\right\rVert}

\newcommand{\x}{\bm{x}}
\newcommand{\z}{\bm{z}}
\def\bea{\begin{equation} \begin{aligned}}
\def\eea{\end{aligned} \end{equation}}
\def\beas{\begin{equation*} \begin{aligned}}
\def\eeas{\end{aligned} \end{equation*}}
\def\bes{\begin{equation*}}
\def\ees{\end{equation*}}

\def\d{\, \mathrm{d}}

\def\be{\begin{equation}}
\def\ee{\end{equation}}
\def\adots{
  \mathinner{\mkern1mu\raise1pt\hbox{.}\mkern2mu\raise4pt\hbox{.}
  \mkern2mu\raise7pt\vbox{\kern7pt\hbox{.}}\mkern1mu}}

\def\x{\bm{x}}
\def\y{\boldsymbol{y}}

\def\G{\mathcal{G}}

\newtheorem{thm}{Theorem}[section]
\newtheorem{lem}{Lemma}[section]
\newtheorem{defi}{Definition}[section]

\newtheorem{rem}{Remark}[section]
\newtheorem{cor}{Corollary}[section]
\def\bt{\begin{thm}}
\def\et{\end{thm}}
\def\bl{\begin{lem}}
\def\el{\end{lem}}
\def\bd{\begin{defi}}
\def\ed{\end{defi}}
\def\bc{\begin{cor}}
\def\ec{\end{cor}}
\def\bp{\begin{proof}}
\def\ep{\end{proof}}
\def\br{\begin{rem}}
\def\er{\end{rem}}
\def\bi{\begin{itemize}}
\def\ei{\end{itemize}}

\begin{document}

\title[Kolmogorov Modes and Linear Response of L\'evy-driven Nonlinear Dynamics]{Kolmogorov Modes and Linear Response of Jump-Diffusion Models}

\author{Micka{\"e}l D. Chekroun}
\email{mchekroun@atmos.ucla.edu} 
\affiliation{Department of Atmospheric and Oceanic Sciences, University of California, Los Angeles, CA 90095-1565, USA}
\affiliation{Department of Earth and Planetary Sciences, Weizmann Institute, Rehovot 76100, Israel} 
\author{Niccol{\`o} Zagli}
\affiliation{Nordic Institute for Theoretical Physics, Stockholm University, 10691 Stockholm, Sweden}
\affiliation{School of Computing and Mathematical Sciences, University of Leicester, Leicester, LE17RH, UK}
\author{Valerio Lucarini}
\affiliation{School of Computing and Mathematical Sciences, University of Leicester, Leicester, LE17RH, UK}

\date{October 22, 2025}

\begin{abstract}
We present a generalization of linear response theory for mixed jump-diffusion models---which combine both Gaussian and L\'evy noise forcings that interact with the nonlinear dynamics---by deriving a comprehensive set of response formulas that accounts for perturbations to both the drift term and the jumps law.  This class of models is particularly relevant for parameterizing the effects of unresolved scales in complex systems. 
Our formulas are thus particularly relevant to quantify uncertainties in either what needs to be parameterized (e.g.~the jumps law), or to measure dynamical changes  due to perturbations of the drift term (e.g.~parameter variations).  By generalizing the concepts of Kolmogorov operators and Green's functions, we obtain new forms of fluctuation-dissipation relations. 
The resulting response is decomposed into contributions from the eigenmodes of the Kolmogorov operator, providing a fresh look into the intimate relationship between a system's natural and forced variability.  We demonstrate the theory's predictive power with two distinct climate-centric applications. First, we apply our framework to a paradigmatic El Ni\~no-Southern Oscillation (ENSO) model subject to state-dependent jumps and additive white noise, showing how the theory accurately predicts the system's response to perturbations and how Kolmogorov modes can be used to diagnose its complex time variability. In a second, more challenging application, we use our linear response theory to perform accurate climate change projections in the Ghil-Sellers (GS) energy balance climate model, which is a spatially-extended model forced here by a spatio-temporal $\alpha$-stable process. This work provides a comprehensive approach to climate modeling and prediction that enriches Hasselmann's program, with implications for understanding climate sensitivity, detection and attribution of climate change, and assessing the risk of climate tipping points. Our results may find applications beyond the realm of climate, and seem of relevance for epidemiology, biology, finance, and quantitative social sciences, among others.
\end{abstract}



\maketitle

\tableofcontents

\section{Introduction}\label{Intro}
Two key tasks in understanding complex systems consist of analyzing their modes of spatial-temporal variability and assessing their sensitivity to perturbations. Unperturbed variability reveals the underlying (nonlinear) processes driving the system's dynamics across various scales. Sensitivity to perturbations, on the other hand, is linked to the strength of the system's internal feedbacks. Critical transitions occur when positive feedbacks overwhelm negative ones, making the system vulnerable to external influences.

The fluctuation-dissipation theorem (FDT) is a fundamental principle in statistical physics that connects the response of a system to external perturbations with the fluctuations of its internal variables, for systems near thermodynamic equilibrium. 
The FDT roughly states that for such systems, the average response to small external perturbations can be calculated through the knowledge of suitable time-lagged correlation functions of the unperturbed statistical system \cite{Kubo1966,marconi2008fluctuation,kubo2012statistical}.

Beyond isolated conservative systems, linear response theory (LRT) has been (rigorously) extended to non-equilibrium systems that include  chaotic systems exhibiting a strange attractor  \cite{ruellegeneral1998,butterley2007smooth,ruelle2009}. However, the situation is much more complex here. Indeed, the attractors of chaotic systems have typically very intricate fractal structure, 
making the direct link between free fluctuation statistics and response challenging as the asymptotic statistics collected on the strange attractor do not contain information about the dynamics outside it. Constructing response operators from unperturbed flow statistics becomes then computationally demanding \cite{Wang2013,Ni2017,Chandramoorthy2022,Ni2023}.   Alternatively, carefully designed perturbed experiments (exploiting covariant Lyapunov vectors) offer a practical approach to circumvent  these computational challenges  \cite{LucariniSarno2011,gritsun2017}.

For complex systems though, it is a common practice to add a small amount of Gaussian noise to chaotic dissipative systems. This regularization technique can simplify the analysis by smoothing the attractor. Such noise can be physically justified as a representation of unresolved physical processes (stochastic parameterization \cite{palmer2005representing,MSM2015,berner2017stochastic,santos2021reduced,LucariniChekroun2023}) or as resulting  from eliminating fast variables \cite{majda2001mathematical,chekroun2021stochastic}.

The increased regularity due to noise significantly simplifies the analysis, overcoming challenges encountered in deterministic systems  \cite{baladi2014,Baladi2017a}. This regularization enables for the rigorous derivation of linear response theory for a wide class of forced-dissipative stochastic systems relevant to climate dynamics  \cite{Hairer_Majda},  and the derivation of explicit and usable formulas for the response operators \cite{Santos2022}.
However, these results are primarily concerned with systems driven by Gaussian white noise.

In many complex systems, non-Gaussian noise, particularly noise with jumps, is more appropriate for capturing non-smooth dynamics. Jump-diffusion models, which accounts for Gaussian diffusion and stochastic processes with jumps such as L\'evy processes \cite{Applebaum2009}, have found significant applications in various fields, including finance \cite{tankov2003financial}, geophysics, and climate science \cite{penland2008modelling,Lovejoy2013}. These models provide valuable insights into multistable behaviors  \cite{Zheng2016,serdukova2017metastability,zheng2020maximum,Lucarini2022NPG} and have been used to study paleoclimate  \cite{ditlevsen1999observation,rypdal2016late}, chaotic transport  \cite{solomon1993observation}, particle detection \cite{stubenrauch2024furutsu}, atmospheric dynamics \cite{khouider2003coarse,stechmann2011stochastic,penland2012alternative,thual2016simple,chen2017simple,chen2024stochastic}, cloud physics \cite{horenko2011nonstationarity,Chekroun_al22SciAdv}, and other complex systems \cite{Wu2017,Singla2020}.

Jump-diffusion models have recently received significant attention, particularly in the context of {\it epidemiological modeling}---a need that was greatly amplified during the evolution of the COVID-19 pandemic \cite{Thanh2020,Kharrazi2021,Tesfay2021,Sabbar2024,Albani2024}. Beyond these compartmental models, spatially-extended systems driven by jump processes are also becoming increasingly relevant across various applications \cite{debussche2013dynamics,Lucarini2022NPG}. For instance, jump processes offer a flexible modeling tool that can replace non-smooth, discontinuous dynamics often represented by "if-then" conditions or thresholds in numerical models, leading to more mathematically tractable and computationally efficient simulations \cite{chen2024stochastic}.

While response theory for systems driven by Gaussian noise is well-established, the theory for jump-diffusion models remains relatively underdeveloped; see though the recent work \cite{stubenrauch2024furutsu}. This work extends key concepts and tools of response theory to systems with non-Gaussian noise, particularly those with jump processes. Section  \ref{Sec_thispaper} elaborates on this generalization.   To better appreciate the latter, we provide below a brief overview of the FDT Green's function formalism for the leading-order response of stochastic systems driven by Gaussian white noise.

\subsection{FDT for stochastic systems: Gaussian noise}\label{Sec_FDT_Gaussian}

We consider the following class of It\^o stochastic differential equation (SDEs):
\be\label{Eq_Gauss}
\d X_t = {\bm F} (X_t) \d t +{\bm \Sigma}(X_t) \d \W,
\ee  
where $X_t$ is a $d$-dimensional state vector, ${\bm F}$ is a vector field on $\mathbb{R}^d$, ${\bm \Sigma}$ is a $d\times p$ matrix-valued function on $\mathbb{R}^d$, and $\W$ is a $p$-dimensional Brownian motion. We assume that ${\bm F}$  and ${\bm \Sigma}$ are sufficiently smooth to ensure the existence of a unique ergodic invariant measure $\mu$, representing the system's statistical equilibrium. Conditions for this to hold can be found in \cite{Hairer_Majda,Chekroun_al_RP2} and references therein. The noise term in Eq.~\eqref{Eq_Gauss} often represents unresolved small-scale variables or processes. Such stochastic systems can be rigorously derived from chaotic systems under appropriate timescale separation assumptions  \cite{majda2001mathematical}, and even under less restrictive assumptions \cite{chekroun2021stochastic}. Hasselmann proposed in \cite{hasselmann1976} to use such stochastic systems to study the dynamics of slow chaotic climatic variables influenced by fast weather variables, modeled as the stochastic component in Eq.~\eqref{Eq_Gauss}. A recent review on this topic is provided in \cite{LucariniChekroun2023}.


Assume that ${\bm F}$ is perturbed to ${\bm F}+ \epsilon g(t) {\bm G}$, where $\epsilon$ is sufficiently small, $g$ is a bounded time-dependent function, and ${\bm G}$  is a smooth vector field on $\mathbb{R}^d$. Using standard perturbative arguments at the leading order in $\epsilon$ for the solution to the Fokker-Planck equation associated with Eq.~\eqref{Eq_Gauss}, we can derive a useful formulation of the FDT, often referred to as linear response theory (LRT) \cite{majda2005information,Santos2022}. 

The goal of response theory is to provide formulas that rely solely on the structural characteristics and statistics of the unperturbed system, enabling the prediction of the time evolution of the system's statistical quantities when a perturbation is applied. These statistical quantities are typically ensemble averages $\langle \Psi \rangle_{\rho_\epsilon^t}$ (of arbitrary observable $\Psi$) with respect to the system's probability distribution $\rho_\epsilon^t$ at time $t$, that satisfies the following (perturbed) Fokker-Planck equation:
\bea\label{Eq_FKE}
\partial_t \rho_\epsilon= -\mbox{div} ({\bm F} \rho_\epsilon )-\epsilon g(t)&\mbox{div}({\bm G} \rho_\epsilon ) \\
&+ \frac{1}{2}\sum_{i,j=1}^d \partial_{ij} \big(a_{ij}(\x) \rho_\epsilon \big),
\eea
where the $a_{ij}(\x)$ are the coefficients of the covariance matrix  $ {\bm \Sigma}(\x){\bm \Sigma}(\x)^T$. 
 In other words, one wishes to quantify the impact of the term $\epsilon g(t)\mbox{div}({\bm G} \cdot)$ on the ensemble average $\langle \Psi \rangle_{\mu}$, i.e.~when statistics are evaluated with respect to the unperturbed system's statistical equilibrium $\mu$, the stationary solution of Eq.~\eqref{Eq_FKE} when $\epsilon=0$. The LRT provides an answer to this problem. It predicts that  
\be\label{Eq_LRT}
\langle \Psi \rangle_{\rho_\varepsilon^t} - \langle \Psi \rangle_{\mu}=\delta^{(1)}[\Psi] (t) +O(\epsilon^2),
\ee
where the first-order response operator $\delta^{(1)}[\Psi] (t)$  is given, explicitly, by:
\be\label{Eq_LRF}
\delta^{(1)}[\Psi] (t) = \epsilon \int_{-\infty}^{\infty}  \G_{\Psi,G} (t-s) g(s)\d s.
\ee
Here, $\G_{\Psi,G}$ is the system's Green function associated with the observable $\Psi$.  It is given as
\be\label{Eq_Green_intro}
\G_{\Psi,G}(t)=\Theta(t) \hspace{-1ex}\int  \hspace{-1ex} \bigg(e^{t  \mathcal{L}_K}
 \Psi(\x)\big[L_{\bm G}\log(\mu)\big] (\x)\bigg)  \mu(\d \x),
\ee
where $\Theta(t)$ is the Heaviside function ensuring causality  \cite{ruelle2009,Lucarini2017,Lucarini2018JSP}, and $L_G= -\mbox{div} ({\bm G} \cdot )$ and $\mathcal{L}_K$ denotes the Kolmogorov operator associated with Eq.~\eqref{Eq_Gauss}: 
\begin{equation}\label{Eq_Kop}
\mathcal{L}_K \psi (\x) ={\bm F} (\x) \cdot \nabla \psi+ \sum_{i,j=1}^d a_{ij}(\x) \partial_{ij} \psi.
\end{equation} 
 Note that the linear character lies here in the linear dependence of the response operator $\delta^{(1)}[\Psi] (t)$ on the term $\epsilon g(t)$, controlling the magnitude of the perturbation of the drift term in Eq.~\eqref{Eq_Gauss}. We refer to  \cite{Santos2022} for formulas of response operators  when the (Gaussian) diffusion term is perturbed.   
 At higher-order in $\epsilon$, these response operators involve nonlinear dependences on the perturbation terms; see \cite{ruelle_nonequilibrium_1998,lucarini2008,LucariniColangeli2012,Lucarini_Chekroun_PRL24}.

Equations \eqref{Eq_LRT}-\eqref{Eq_Kop} offer significant practical and theoretical insights. Firstly, they provide a generalized version of the FDT \cite{Kubo1966,pavliotisbook2014}, as Eq.~\eqref{Eq_Green_intro} enables us to interpret  the Green's function in terms of time-lagged correlations \cite{majda2005information,Baiesi2013,Santos2022} as we recall below.

In that respect, recall that the operator $e^{t  \mathcal{L}_K}$  corresponds to the Markov semigroup associated with Eq.~\eqref{Eq_Gauss} and thus $e^{t  \mathcal{L}_K} \Psi(x)=\mathbb{E} \Psi(X_t)$, with $X_t$ solving Eq.~\eqref{Eq_Gauss} given $X_0 = x$ and with $\mathbb{E}$ denoting the expectation  over noise paths \cite{Chekroun_al_RP2}.

If the system given by Eq.~\eqref{Eq_Gauss} is ergodic and $\mu$ has a probability density $\rho_{eq}$ with respect to the Lebesgue measure, the integral in Eq.~\eqref{Eq_Green_intro} can be computed through the following time average along a typical solution:
\bes
\G_{\Psi,G}(t)=-\lim_{T\rightarrow \infty} \frac{1}{T} \mathbb{E} \int_0^T \Psi (X_{s+t})\frac{\nabla \cdot\left( {\bm G}(X_s) \rho_{eq}(X_s) \right)}{\rho_{eq}(X_s)} \d s,
\ees
for $t > 0.$

The Green's function thus corresponds to time-lagged correlations between the observables $\Phi=L_{\bm G}\log(\mu)$  and $\Psi$, involving only the statistics of the unperturbed system. This formula can be practically applied by replacing the unperturbed density $\rho_{eq}$ with a suitable analytical ansatz. This is the quasi-Gaussian approximation approach introduced by Leith in his seminal paper \cite{Leith75}, which has been successfully applied to various systems \cite{majda2005information}, from prototype geophysical fluid models \cite{gritsun1999barotropic,abramov2007} to more advanced atmospheric general circulation models \cite{gritsun2007}.

Another important aspect of Equations \eqref{Eq_LRT}-\eqref{Eq_Kop}, as demonstrated in \cite{Santos2022}  by relying on  the framework of  \cite{Chekroun_al_RP2}, is that they allow for decomposing the Green's function into contributions from the system's modes of (unforced) variability. These modes, which encode decay of correlations and power spectra  \cite{Chekroun_al_RP2}, are given by the eigenelements of the Kolmogorov operator.
They  reduce to the Koopman modes when ${\bm \Sigma}=0$ \cite{budivsic2012applied}, opening thus new perspectives on linking forced variability to such modes, otherwise learnable from observational data \cite{williams2015data}.

\subsection{This Paper}\label{Sec_thispaper}

In this paper, we present a comprehensive generalization of linear response theory for mixed jump-diffusion models. We derive novel and comprehensive response formulas that account for perturbations to both the drift term and, crucially, to the jumps' law. This class of models is particularly relevant for parameterizing the effects of unresolved scales in complex systems, such as in stochastic climate dynamics. Our formulas are therefore particularly relevant for quantifying uncertainties in what needs to be parameterized (e.g., the jumps' law), or for measuring dynamical changes due to perturbations of the drift term (e.g., parameter variations). By generalizing the concepts of Kolmogorov operators and Green's functions, we obtain a comprehensive set of fluctuation-dissipation relationships. A key feature of our approach is the decomposition of the resulting response into contributions from the eigenmodes of the Kolmogorov operator, providing a fresh and interpretable look into the intimate relationship between a system's natural and forced variability. Furthermore, we present new results about the diffusion limit for such systems.

We demonstrate the theory's predictive power with two distinct climate-centric applications. First, we apply our theoretical framework to the paradigmatic El Ni\~no-Southern Oscillation (ENSO) model by Jin \cite{jin1997equatorial} that is subject to state-dependent jumps and additive white noise. Leveraging recent advances in stochastic modeling \cite{Chekroun_al22SciAdv}, we show that the model exhibits shear-induced chaos \cite{Young2016} arising from the subtle interplay between noise and nonlinear dynamics. Our approach enables us to accurately construct and verify the validity of our response operators for this complicated dynamics.

In a second, more challenging application, we use our linear response theory to perform accurate climate change projections in the Ghil-Sellers (GS) energy balance climate model \cite{Sellers,Ghil1976,Bodai2015}. This spatially-extended model is forced by a spatio-temporal $\alpha$-stable process, making it a realistic and rigorous test case for dealing with climate change in presence of extreme events \cite{Lucarini2022NPG}.

Two key points emerge from our findings. First, in a climate context, our results allow for extending Hasselmann's program \cite{hasselmann1976,imkeller2001stochastic} by providing a more general framework for unresolved processes, capable of incorporating both Gaussian and jump processes. Second, our work aligns with the response theory framework of \cite{Lucarini_Chekroun_PRL24}, enabling a dynamical foundation for the optimal fingerprinting method---previously formulated heuristically and statistically---for detecting and attributing climate change \cite{Hasselmann1997, Allen2003, Hannart2014}. As such, the extension of response theory and the investigation of its breakdown presented here further supports the link between climate change signals and forcing factors, even when accounting for "climate shocks" modeled as jump processes \cite{Lucarini2022NPG}.

Our paper is structured as follows. In Section \ref{Sec_FKPE_Levy} we carefully discuss how to derive a generalized Fokker-Planck equation able to describe the evolution of probability distributions characterizing the statistics of jump-diffusion models. Section \ref{Sec_RPs} is devoted, taking inspiration from \cite{Chekroun_al_RP2}, to developing the theory of Ruelle-Pollicott resonances \cite{pollicott1986meromorphic,ruelle1986locating} for this class of systems, and of the corresponding Kolmogorov modes. In Sections  \ref{Sec_Green_Kolmo} and  \ref{Sec_general_formula}, we present the derivation of the comprehensive response formulas, generalizing previous results \cite{Santos2022} and including the case of perturbations applied to the jumps' law. The approach retained enables us furthermore to compute these modes and resonances efficiently by leveraging on Ulam's method \cite{ulam1960collection} from observational data issued from such systems. In Section \ref{ENSO} we succesfully apply the theory to a jump-diffusion version of the Jin model for ENSO. In Section \ref{Ghil-Sellers} we investigate the response of the modified Ghil-Sellers model and show how response theory can be used to perform climate change projections also in this challenging case where jump processes are considered. Finally, in Section \ref{conclu}, we present our conclusions and perspectives for future works.

\section{Fokker-Planck Equation of L\'evy-driven Dynamics}\label{Sec_FKPE_Levy}

\subsection{Kolmogorov Operator of L\'evy-driven Dynamics}\label{Sec_Kolmo_Levy}
There is a vast literature on L\'evy processes which roughly speaking are processes given as the sum of a Brownian motion and a jump process \cite{tankov2003financial,peszat2007stochastic,Applebaum2009,kuhn2017levy,oksendal2019stochastic}. Unlike the case of Gaussian diffusion processes (SDEs driven by Brownian motions), the representation of Kolmogorov operators is non-unique in the case of L\'evy-driven SDEs, and may take the form of operators involving fractional Laplacians or singular integrals among other representations \cite{Kwasnicki:2017aa}. We adopt here the definition commonly used in probability theory \cite[Theorem 3.5.14]{Applebaum_2019} which presents the interest of being particularly intuitive as recalled below. 

We consider SDEs in $\mathbb{R}^d$ of the form:
\be\label{Eq_Levy_additive}
\d X_t = {\bm F} (X_t) \d t +{\bm \Sigma}(X_t) \d \W + \d  \bm{L}_t,
\ee  
where $\W$ and $\bm{L}_t$ are assumed  to be  respectively a Brownian motion and a  L\'evy process in $\mathbb{R}^d$, that are mutually independent. 
To simplify, we assume here that the covariance matrix $(a_{ij}(\x))= {\bm \Sigma}(\x){\bm \Sigma}(\x)^T$ is a positive definite matrix for each $\x$.

Roughly speaking, a L\'evy process $\bm{L}_t$ on $\mathbb{R}^d$ is a non-Gaussian stochastic process  with independent and stationary increments that experience sudden, large jumps in any direction. 
The probability distribution of the these jumps is characterized by a non-negative Borel measure $\nu$ defined on $\mathbb{R}^d$ and concentrated on  
$\mathbb{R}^d\backslash \{0\}$ that satisfies the property  $\int_{\mathbb{R}^d\backslash\{0\}} \min(1,\y^2)\nu (\d \y) <\infty$.  This measure $\nu$ is called the jump measure of the L\'evy process $\bm{L}_t$. 
Sometimes $X_t$ itself is referred to as a L\'evy process with triplet $({\bm F},{\bm \Sigma},\nu)$. Within this convention, we reserve ourselves the terminology of a L\'evy process to a process with  triplet $(0,0,\nu)$.
We refer to \cite{Applebaum2009} and \cite{peszat2007stochastic} for the mathematical background on L\'evy processes.

Under suitable assumptions on ${\bm F}$, $ {\bm \Sigma}$, and the L\'evy measure $\nu$, the solution $X_t$ to Eq.~\eqref{Eq_Levy_additive}
is a Markov process (e.g.~\cite{protter2005stochastic}) and even a Feller process \cite{kuhn2018solutions}, with associated Kolmogorov operator, which extends the one shown in Eq.~\eqref{Eq_Kop}, taking the following integro-differential form for (e.g.)~$\psi$ in $C^{\infty}_c(\mathbb{R}^d)$ (Courr\`ege theorem \cite{courrege1965forme,kuhn2017levy}):
\begin{widetext}
\bea\label{Eq_Kolmo2_general}
\mathcal{L}_K \psi (\x) &={\bm F} (\x) \cdot \nabla \psi+ \sum_{i,j=1}^d a_{ij}(\x) \partial_{ij} \psi + J \psi(\x), \mbox{with}\\
&J \psi (\x)=\int_{\mathbb{R}^d\backslash\{0\}} \Big[\psi(\x+\y)-\psi(\x)- \y\cdot \nabla \psi (\x) \mathds{1}_{\{\norm{\y}<1\}} \Big] \nu(\d \y),
\eea
\end{widetext}
where $ \mathds{1}_{\{\norm{\y}<1\}}$ denotes the indicator function of the (open) unit ball in $\mathbb{R}^d$; see also \cite[Theorem 3.3.3]{Applebaum2009}. 
 We will call $\mathcal{L}_K $ the Kolmogorov-L\'evy operator to make the distinction with the Kolmogorov operator in the pure diffusive case. 

The first-order term in Eq.~\eqref{Eq_Kolmo2_general} is the drift term caused by the deterministic, nonlinear dynamics. 
The second-order differential operator  represents the diffusion part of the process $X_t$. It is responsible for the continuous component of the process. 

The $J$-term, involving the integral, represents the jump part of the process. It captures the discontinuous jumps that the process experience due to the sudden changes caused by the L\'evy process $\bm{L}_t$.
Its intuitive interpretation breaks down as follows. 
The term, $\psi(\x+\y)-\psi(\x)$, calculates the difference in the test function value before and after the potential jump, capturing the change in the test function due to the jump.

The term $-\y\cdot \nabla \psi (\x) \mathds{1}_{\{\norm{y}<1\}}$ represents a first-order correction for small jumps. It aims to account for the fact that a small jump might not land exactly on the grid point ($\x + \y$),
but somewhere in its vicinity. This term is often referred to as the Girsanov correction. 
Thus, the integral term $J$ accounts for all possible jump sizes ($\y$) within a unit ball, as 
weighted by the L\'evy measure $\nu(\d \y)$.

 The Kolmogorov-L\'evy operator, $\mathcal{L}_K$, such as given by Eq.~\eqref{Eq_Kolmo2_general} generates a  strongly continuous semigroup ($C_0$-semigroup) $T(t)=e^{t \mathcal{L}_K}$ \cite[Theorem 3.5.14]{Applebaum_2019}.
 This semigroup has the following characterization \cite{Applebaum2009}:
\be\label{Eq_Tt}
T(t)\psi=\mathbb{E}\psi(X_t^x),
\ee
i.e.~it provides the expected value of an observable $\psi$ over the stochastic solution paths that connect $x$ in $\mathbb{R}^d$ to $X_t^x$ solving Eq.~\eqref{Eq_Levy_additive} at time $t$. 
The semigroup  gives thus a deterministic macroscopic description of the averaged effects of the combined L\'evy flights and Brownian motions driving the dynamics. 
 
Thanks to semigroup theory, following \cite{Chekroun_al_RP2}, correlation functions and power spectra benefit of decomposition formulas such as given by Eqns.~\eqref{Eq_decomp_corr1} and \eqref{Eq_PSD}, in which the corresponding RP resonances and Kolmogorov modes  are now the spectral elements associated with the isolated part of $\mathcal{L}_K$. This is clarified in Section \ref{Sec_decomp_corr} below. 
 
 The direct numerical approximation of such Kolmogorov-L\'evy operator is however a non-trivial task, requiring in particular a special care to handle the singular integral involved therein \cite{gao2014mean,gao2016fokker}. We take below in Section \ref{Sec_Ulam} another route to estimate the RP resonances and Kolmogorov modes, based instead on the Ulam's method \cite{ulam1960collection}, a data-driven approach that has proven its efficiency for chaotic and pure diffusion dynamics driven by Brownian motion; see \cite{junge2009discretization,Chek_al14_RP,Chekroun_al_RP2}. 

\subsection{Fokker-Planck Equation of L\'evy-driven Dynamics}\label{SubSec_FKPE_Levy}
The Fokker-Planck equation (FPE) provides the governing equation for the transition probability density $p(\x,t) = \mathbb{P}(X_t= \x | X_0= \x_0)$, i.e., the probability that the process $X_t$ has value $\x$ at time $t$ given it had value $\x_0$ at time $0$. We assume this transition probability to be smooth with respect to the Lebesgue volume; see \cite{picard1996existence,priola2009densities,priola2012pathwise} and \cite[Sec.~5.5]{kuhn2017levy} for conditions.

The FPE associated with Eq.~\eqref{Eq_Levy_additive} (in $L^2(\mathbb{R}^d)$) is given by the following non-local integro-differential equation:
\begin{widetext}
\bea\label{Eq_FKPE_nonlocal}
\partial_t p(\x,t) = -\mbox{div} ({\bm F}(\x) p(\x,t))&+\frac{1}{2}\sum_{i,j=1}^d \partial_{ij} \big(a_{ij}(\x) p(\x,t)\big)\\
 &+\int_{\mathbb{R}^d\backslash \{0\}}\Big[p(\x-\y,t)-p(\x,t)+ \y\cdot \nabla p (\x,t) \mathds{1}_{\{\norm{\y}<1\}} \Big] \nu(\d \y).
\eea
\end{widetext}
The integral term in Eq.~\eqref{Eq_FKPE_nonlocal} is associated with the jump processes and is, more specifically, given as the adjoint of the Kolmogorov-L\'evy jump operator $J$ defined in Eq.~\eqref{Eq_Kolmo2_general}.
For conditions ensuring the (full) Kolmogorov-L\'evy operator  $\mathcal{L}_K$ (Eq.~\eqref{Eq_Kolmo2_general}) to be the infinitesimal generator of a strongly continuous semigroup in $L^p$ spaces ($1\leq p < \infty$), we refer to \cite{garroni2002second,chen2010heat,applebaum2021}.

In compact form, Eq.~\eqref{Eq_FKPE_nonlocal} can be written as
\be\label{Eq_FKPE_compact}
\partial_t p=-\nabla \cdot ({\bm F}(\x) p) +\frac{1}{2} \nabla^2 \cdot ({\bm \Sigma}{\bm \Sigma}^Tp) +J^\ast p,
\ee
where $J^\ast p$ denotes the integral term in Eq.~\eqref{Eq_FKPE_nonlocal}. Formally, $\partial_t p=\mathcal{L}_K^\ast p$, where $\mathcal{L}_K^\ast$ denotes the adjoint of the Kolmogorov operator $\mathcal{L}_K$ given by Eq.~\eqref{Eq_Kolmo2_general}.

The expression of $J^\ast$ comes from the duality relation $\langle J \psi,\phi \rangle=\langle \psi,J^\ast \phi \rangle$. Through integration by parts and a change of variables, it can be shown that the adjoint of the jump operator $J$ is exactly the integral term presented in Eq.~\eqref{Eq_FKPE_nonlocal}; see Remark \ref{Rmk_adjoint_sym} below. It is important to note that if the L\'evy jump measure $\nu$ is symmetric ($\nu (\d \y) = \nu (-\d \y)$), then the operator $J$ is self-adjoint, i.e., $J^\ast = J$. This condition is often assumed in certain contexts, but in the general case of L\'evy processes, the use of the adjoint operator $J^\ast$ is strictly necessary.

There exists a well-known equivalent representation of Eq.~\eqref{Eq_FKPE_compact} when $\nu$ is the (symmetric) L\'evy measure:
\be
\nu(\d \y)=c_{d,\alpha} \|\y\|^{-d-\alpha} \d \y,
\ee
associated with an $\alpha$-stable  L\'evy process.  Here $\alpha$ lies in $(0,2)$ and is called the non-Gaussianity index and $c_{d,\alpha}$ is a constant that depends on the dimension and involves the $\Gamma$ function of Euler \cite{Kwasnicki:2017aa}. A $2$-stable ($\alpha= 2$) process is simply a Brownian motion. 

For any such (general) $\alpha$-stable  L\'evy process, one can  recast Eq.~\eqref{Eq_FKPE_compact} by means of the fractional Laplacian $-(-\Delta)^{\alpha/2}$, which is associated with a fat-tailed dispersal kernel (for $\alpha$  in $(0, 2)$) and therefore describes long-distance dependencies. The equivalent formulation is then written as (e.g.~\cite[Eqns.~(57)-(58)]{schertzer2001fractional} and \cite[Eq.~(6)]{priola2012pathwise}): 
\be\label{Eq_FKPE2}
\partial_t p=-\nabla \cdot ({\bm F}(\x) p) +\frac{1}{2} \nabla^2 \cdot ({\bm \Sigma}{\bm \Sigma}^Tp)  -(-\Delta)^{\alpha/2} p,
\ee
where the fractional Laplacian is defined using the following Riesz' representation formula \cite{albeverio2000invariant}:
\bes
-(-\Delta)^{\alpha/2} u (\x)=\mathcal{F}^{-1}(\norm{\bm{k}}^{\alpha}\widehat{u}(\bm{k})),
\ees
where $\widehat{u}$ denotes the Fourier transform of $u$, given by $\widehat{u}(\bm{k})=(2\pi)^{-d/2}\int_{\mathbb{R}^d} e^{\langle \bm{k},\x \rangle} \d \x,$  and $\mathcal{F}^{-1}$ its Fourier inverse. 
Recall that the limiting case $\alpha=2$ corresponds to the Brownian motion which is recovered formally by setting $\alpha=2$ in Eq.~\eqref{Eq_FKPE2}.

\section{Ruelle-Pollicott Resonances and Kolmogorov Modes}\label{Sec_RPs}
In this section, we briefly introduce the theory of Ruelle-Pollicott (RP) resonances, a powerful spectral tool for analyzing the dynamics of complex systems subject to inherent stochasticity \cite{Chekroun_al_RP2}, such as those encountered in climate modeling, turbulent flows, and biological systems \cite{LucariniChekroun2023}. These resonances reveal the fundamental time scales and oscillatory modes of the system, acting as characteristic frequencies and decay rates that govern the relaxation of any correlations, even in the presence of significant noise. Theoretically compelling, RP resonances bridge the gap between observable behavior and the underlying governing equations: they can be extracted from time series data through methods like Markov chain modeling and are mathematically defined as the eigenvalues of the generator of the stochastic process, the Kolmogorov operator \cite{crommelin2006b,Chek_al14_RP,Chekroun_al_RP2,Tantet_al_ENSO}. By characterizing the system's relaxation towards equilibrium, the framework offers crucial insights into the limits of predictability in these stochastic systems. Furthermore, as we will detail in Section \ref{Sec_Green_Kolmo}, the RP resonances and their corresponding eigenfunctions, often referred to as Kolmogorov modes, provide a natural basis for understanding and quantifying the response of complex systems to perturbations.

\subsection{Correlations and power spectra decompositions}\label{Sec_decomp_corr}
As detailed in \cite{Chekroun_al_RP2} for a broad class of stochastic systems driven by Brownian motion, the RP resonances correspond to the isolated eigenvalues of the system's Kolmogorov operator $\mathcal{L}_K$ while the Kolmogorov modes are its eigenfunctions. These modes and resonances hold significant importance. They enable the derivation of rigorous decomposition formulas for both temporal correlation functions and power spectral density functions across a wide range of physical stochastic systems. These decomposition formulas can be made rigorous for general stochastic systems including the hypoelliptic ones  \cite{Chekroun_al_RP2}; see also \cite{dyatlov2015stochastic}.  

We present below the extension of these formulas for SDEs driven by L\'evy processes, of the form given by Eq.~\eqref{Eq_Levy_additive}. 
For the sake of simplicity, we do not (unlike in \cite{Chekroun_al_RP2} for Brownian-SDEs) enter into the precise conditions on $\bm{F}$, $\bm{\Sigma}$, and $\bm{L}_t$ that ensure (i) the SDE to possess a unique ergodic invariant measure $\mu$, and (ii) to generate a $C_0$-semigroup 
 $e^{t \mathcal{L}_K}$ on $L^2_{\mu} (\mathbb{R}^d)$.

The study of existence of invariant probability measures \cite{albeverio2000invariant,kulik2009exponential,behme2015criterion,behme2024invariant}, and ergodic properties of L\'evy processes in terms of the coefficients of the corresponding infinitesimal generator (the Kolmogorov-L\'evy operator) is an ongoing research topic.

We refer to \cite{masuda2004multidimensional,kulik2009exponential,sandric2016ergodicity} for useful criterions regarding   
the exponential ergodicity when the coefficients are locally Lipschitz continuous, and to \cite{xie2020ergodicity} for more general conditions allowing for large jumps.

Once (i) and (ii) are ensured, we can apply \cite[Corollary 1]{Chekroun_al_RP2} and thus obtain for any observables $f$ and $g$ (square-integrable w.r.t.~system's ergodic probability measure): 
\be\label{Eq_decomp_corr1}
C_{f,g}(t)= \sum_{j=1}^N \left(\sum_{k=0}^{m_j-1}a_{jk}^{f,g}  \frac{ t^k}{k!} (\mathcal{L}_K-\lambda_j \textrm{Id})^{k}\right) e^{\lambda_j t} + \mathcal{Q}(t),
\ee
where the reminder  satisfies $\mathcal{Q}(t)\underset{t\rightarrow \infty}\longrightarrow 0$, and the coefficients $ a_{jk}^{f,g}$ are given by \cite[Eq.~(2.11)]{Chekroun_al_RP2}
\bea\label{Eq_coeff_aij}
a_{jk}^{f,g}&=\langle\Phi_j^\ast, g \rangle_\mu \int f(\x) (\mathcal{L}_K-\lambda_j \textrm{Id})^{k}   \Phi_j(\x) \mu (\d \x),
\eea
in which $\Phi_j$ (resp.~$\Phi_j^\ast$) denotes the eigenfunction  (dual eigenfunction) associated with the RP resonances $\lambda_j$, $\langle \cdot,\cdot \rangle_\mu$ denotes the $L^2_\mu$-Hermitian inner product,  while $m_1,\cdots,m_N$ correspond to the (algebraic) multiplicity of these resonances. 
Note that in the above decomposition, we assumed that the observables have been centered, i.e.~that the mean with respect to the ergodic measure $\mu$ has been subtracted.

For these systems, the spectrum of the Kolmogorov operator $\mathcal{L}_K$  is typically contained in the 
left-half complex plane \cite{eckmann2003spectral,Chekroun_al_RP2}, $\{z\in \mathbb{C}\;:\: \Re(z)\leq 0\}$ and its resolvent  $R(z)=(z \mbox{Id}-\mathcal{L}_K)^{-1}$, is a well-defined operator that possesses the following integral representation
\be\label{Eq_resolvent}
R(z) f =\int_{0}^{\infty} e^{-z t} e^{t \mathcal{L}_K} f \d t.
\ee

The RP resonances correspond then to the $N$ poles of the resolvent $R(z)$, and the $m_1,\cdots,m_N$ to the orders of these poles. 
The location of these poles are independent of the observables but their relative importance (through the coefficients  $a_j^{f,g}$ depend on them.  It is noteworthy that the theory of RP resonances has been initially introduced for deterministic chaotic systems 
\cite{ruelle1986locating,pollicott1986meromorphic,keller1999stability,baladi2000positive,gaspard2005chaos} and is an active field of research \cite{faure2011upper,giulietti2013anosov,dyatlov2019mathematical}.

The theory for chaotic systems is however much more challenging as the underlying ergodic measure of physical interest is typically singular with respect to the Lebesgue measure, requiring in particular the use of special function spaces to study the spectral properties of these resonances \cite{butterley2007smooth,faure2011upper,gouezel2006banach}.  
The improved regularity due to noise diffusion simplifies considerably key aspects of the problem  \cite{Hairer_Majda} allowing us in particular to rely here on semigroup theory  \cite{Engel_Nagel} in standard function spaces to derive correlation decomposition formulas such as Eq.~\eqref{Eq_decomp_corr1}; see \cite{Chekroun_al_RP2}.  
RP resonances have been shown also to be instrumental in the diagnosis of efficient stochastic parameterizations for the reduced-order modeling of nonlinear systems with complex dynamics \cite{Kondrashov_al2018_QG,chekroun2021stochastic}. 

A useful  consequence of Eq.~\eqref{Eq_decomp_corr1} is the decomposition of the \emph{power spectrum}, $S_{f, g}(\omega)$, in terms of RP resonances and Kolmogorov modes  following \cite{Chekroun_al_RP2}. The latter is obtained by taking the Fourier transform of the correlation function $C_{f,g}(t)$ (Wiener-Khinchin theorem), which gives e.g.~in the case of poles of order $1$ ($m_1=\cdots=m_N=1$),
\be\label{Eq_PSD}
S_{f, g}(\omega) =\sum_{j=1}^N \frac{a_{j0}^{f,g}   \mathrm{Re}(\lambda_j) }{(\omega - \mathrm{Im}(\lambda_j))^2 + \mathrm{Re}(\lambda_j)^2}+\mathcal{D}(\omega),	\ee
where $\omega$ is a complex frequency and  $\mathcal{D}(\omega)$ denotes some ``background'' function decaying to $0$ as $|\omega|\rightarrow \infty$; see also  \cite{melbourne2007power}. 
Eq.~\eqref{Eq_PSD} shows that the meromorphic extension into the complex plane of the power spectrum $S_{f, g}(\omega) $ when $\omega$ is real,  present poles that are exactly given by the RP resonances. These poles manifest themselves in the (real) power spectrum as peaks that stand out over a continuous background at frequencies given by the imaginary
part of the eigenvalues, when $a_{j0}^{f,g}\neq 0$ and if  the $\lambda_j$ are close enough to the imaginary axis.
 
Thus, the RP resonances and Kolmogorov modes characterize the various ways a system's stochastic dynamics expresses its temporal variability through observables.

 \subsection{Ulam's Estimation of RP Resonances and Kolmogorov Modes}\label{Sec_Ulam}
Analytical formulas for RP resonances and Kolmogorov modes are  scarce (see, e.g., \cite{metafunes2002,gaspard1995,gaspard2002trace,gardiner2009,pavliotisbook2014,Tantet_al_Hopf}). In practice, these resonances are estimated e.g.~from long time series obtained by solving the governing equations (e.g., Eq.~\eqref{Eq_Jin_stoch}); \cite{crommelin2011diffusion,generatorfroyland,Chekroun_al_RP2}.

Leveraging the Ulam method \cite{ulam1960collection}, Markov matrices are crucial for both learning transfer operator properties  and estimating the system's propagator's long-term behavior;  e.g.~\cite{schutte1999direct,junge2009discretization,Froyland2021}. This propagator governs the evolution of probability densities and reveals also how correlations and other system aspects change over time \cite{froyland2003detecting,Chek_al14_RP, generatorfroyland}. This approach applies to both deterministic and stochastic systems \cite{froylandapproximating1998,fishman2002,Chekroun_al_RP2}. 

Ulam's method provides thus a way to estimate the propagator $e^{t\mathcal{L}_K}$  using Markov transition matrices. Eigenvalues of the Kolmogorov-L\'evy  operator are then obtained through logarithm formulas \cite{crommelin2011diffusion,Chekroun_al_RP2}. 
This method consists of subdividing
the state space $\mathcal{X}$, typically shadowing the dynamics' forward attractor, 
 into $N_g$ non-intersecting \emph{cells} or \emph{boxes} $\{B_i\}_{i=1}^{N_g}$ and estimating the dynamics' probability transitions across these boxes. Mathematically, the propagator, $e^{t\mathcal{L}_K}$, for a given transition time $t=\tau$, is approximated by a $N_g\times N_g$ Markov transition matrix $M_{\tau}$,  whose entries are given by \cite{Chekroun_al_RP2}:
\begin{equation}\label{projected transfer operator}
\left[M_{\tau}\right]_{i,j}=\frac{1}{\rho_0(B_i)}\int_{B_i}e^{\tau \mathcal{L}_K}\chi_{B_j}(\x)\d \mu( \x),
\end{equation}
for $i,j=1,\ldots,N_g$, where $\chi_{B_i}$ denotes the indicator function on the box $B_i$, $\mu$ denotes the system's ergodic  invariant probability measure and $\rho_0$, a given initial distribution. The transition matrix is then estimated  by computing a classical maximum likelihood estimator given by \cite{crommelin2011diffusion,schutte1999direct}:
\begin{equation}\label{eq:transition_matrix}
 \left[M_{\tau}\right]_{i,j} = \frac{\#\bigg\{ \Big(\x_{k}\in B_j \Big) \land \Big(\x_{k+\ell} \in B_i \Big) \bigg\}}{\# \Big\{\x_{k} \in B_j\Big\}},
 \end{equation}
 for $i,j = 1,\ldots, N_g$. 
Given a sampling time $\delta t$ at  which the time series $\{\x_k\}_{k=1}^{N_d}$ solving Eq.~\eqref{Eq_Jin_stoch}, is collected, the formula \eqref{eq:transition_matrix} counts the number of observations ($\#\{\cdot\}$)  visiting the box $B_i$ after $\ell= \lfloor \tau/\delta t \rfloor$ iterations, knowing that it already landed in $B_j$.
The resulting operation results into a coarse-graining of the dynamics encoded by $M_\tau$ \cite{crommelin2006b,Chekroun_al_RP2} that incorporates artificial diffusion \cite{generatorfroyland}, of minor impact when $N_d$ and $N_g$ are sufficiently large.

A significant contribution, elaborated in \cite{Chek_al14_RP} and further detailed in \cite{Chekroun_al_RP2}, is the introduction of reduced Ruelle-Pollicott (RP) resonances. This concept provides a rigorous way to define and extract characteristic frequencies and decay rates, even when only a subset of the system's variables is accessible to observation. This is particularly crucial when dealing with high-dimensional, complex systems where measuring the entire state space is often impractical. These reduced resonances are derived from reduced Markov operators acting on functions defined within the observed, lower-dimensional state space. Notably, as demonstrated in \cite[Theorem 3.1]{Chekroun_al_RP2}  and \cite[Theorem A]{Chek_al14_RP} and further explored in \cite{Chekroun_al_RP2,RP_ENSO}, the eigenvalues of these reduced operators can still provide valuable information about the transitions occurring in the full, unobserved system. This is of paramount importance in real-world applications across physics and related disciplines, where experimental or observational limitations often restrict our measurements to only a few key variables or observables.

\section{Linear Response of L\'evy-driven Dynamics}\label{Sec_Green_Kolmo}

\subsection{Green's Function of L\'evy-driven Dynamics}\label{Sec_Green_fcts}

Leveraging the formalism of Section \ref{Sec_FKPE_Levy} for Kolmogorov operators and Fokker-Planck equations for L\'evy-driven stochastic systems of the form of Eq.~\eqref{Eq_Levy_additive}, we can directly extend the FDT relationships discussed in Section \ref{Sec_FDT_Gaussian} to these more general systems. To do so, we follow the approach outlined in Section \ref{Sec_FDT_Gaussian} and consider perturbations to the drift term of the following form:
\bea\label{Eq_pert_SDEgen}
\d X_t=\Big({\bm F} (X_t) + \epsilon g(t) &{\bm G}(X_t)\Big)\d t \\
&+ \bm{\Sigma} (X_t) \d \W+\d \bm{L}_t,
\eea
where $g(t)$ is time-dependent scalar modulation function and ${\bm G}$ is the drift perturbation.

Using Eq.~\eqref{Eq_FKPE_compact}, we obtain that the evolution of probability density functions associated with Eq.~\eqref{Eq_pert_SDEgen} is given by the Fokker-Planck equation
\bea\label{Eq_FKPE_pert}
\partial_t \rho=-\nabla \cdot ({\bm F}(\x) \rho)&- \epsilon g(t) \nabla \cdot ({\bm G}(\x) \rho)\\
&+\frac{1}{2} \nabla^2 \cdot ({\bm \Sigma}{\bm \Sigma}^T\rho) +J^*\rho,
\eea
where $J^* \rho$ denotes the integral term in Eq.~\eqref{Eq_FKPE_nonlocal}.

Let us expand the statistical state, $\rho_\epsilon^t$,  solving the time-dependent Fokker-Planck equation \eqref{Eq_FKPE_pert}:
\be\label{Eq_expansion_density}
\rho_\epsilon^t(\x) = \mu(\x) + \epsilon  \rho_1^t(\x)+ h.o.t.
\ee
where $\mu$, denotes  the reference,  statistical equilibrium given as the system's ergodic probability density for $\epsilon=0$ that we assume here to exist (see \cite{masuda2004multidimensional,kulik2009exponential,sandric2016ergodicity} for conditions). 

To obtain $\rho_1^t$, we plug $\rho_\epsilon^t$ given by Eq.~\eqref{Eq_expansion_density} into Eq.~\eqref{Eq_FKPE_pert} and retain the first-order terms in $\epsilon$. 
By using the semigroup $e^{t \mathcal{L}_K^\ast}$ providing the  solution operator to the Fokker-Planck Eq.~\eqref{Eq_FKPE_pert}, with $\mathcal{L}_K^\ast$ denoting the dual of the Kolmogorov-L\'evy operator (given explicitly as the right-hand side of Eq.~\eqref{Eq_FKPE_nonlocal}), we obtain then
\be
\rho_1^t=\int_0^t e^{(t-s) \mathcal{L}_K^\ast} L_{\bm G} \mu \d s,
\ee
where 
\be\label{Eq_Liouville} 
L_{\bm G} \mu= -  \nabla \cdot\left( {\bm G}  \mu \right).
\ee

The expected value  of 
$\Psi$ at time $t$, for the statistical state $\rho_\epsilon^t$   is  then given by
\begin{equation} \begin{aligned}\label{eq:response function a}
	 \langle \Psi \rangle_{\rho_\epsilon^t}&=\int \Psi(\x) \rho^t_\epsilon(\x) \d \x \\
	&\approx \int \Psi(\x)   \mu(\d \x)  + \epsilon  \int \Psi(\x)  \rho_1^t(\x) \d \x.
\end{aligned}\end{equation}
After integrating by part we arrive at
\bea\label{Eq_integ_by_part_forPsi_rho1}
\int \Psi(\x)  \rho_1^t(\x) \d \x=\int_{-\infty}^{\infty}  \G_{\Psi,G}(t-s) g(s)\d s,
\eea
where $\G_{\Psi,G}$ is the (causal) Green's function given by:
\be\label{eq:Green}
\G_{\Psi,G}(t)=\Theta(t) \hspace{-1ex}\int  \hspace{-1ex} \bigg(e^{t  \mathcal{L}_K}
 \Psi(\x)\big[L_{\bm G}\log(\mu)\big] (\x)\bigg)  \mu(\d \x),
\ee
 in which $\Theta(t)$ is the Heaviside function, $L_G= -\mbox{div} ({\bm G} \cdot )$ and $\mathcal{L}_K$ denotes the Kolmogorov operator.
Structurally, one recovers the same response operator as for the case of stochastic systems driven by Gaussian noise (Section \ref{Sec_FDT_Gaussian}) in which the Kolmogorov operator for pure diffusion (Eq.~\eqref{Eq_Kop}) has been replaced by the  Kolmogorov-L\'evy operator.

Here also, the (genuine) ensemble anomaly at time $t$, $\delta [\Psi](t)=\langle \Psi \rangle_{\rho_\varepsilon^t} - \langle \Psi \rangle_{\mu}$, with respect to the reference state $\mu$, is approximated---at the leading order in $\epsilon$---by the anomalies $\delta^{(1)}[\Psi] (t)$ obtained by the {\it linear response formula}:
\be\label{Eq_LRF}
\boxed{\delta^{(1)}[\Psi] (t) = \epsilon \int_{-\infty}^{\infty}  \G_{\Psi,G} (t-s) g(s)\d s.}
\ee

As discussed in Section \ref{Sec_FDT_Gaussian}, the Green's function $\G_{\Psi,G}(t)$ can also be interpreted as  time-lagged correlations between the observables $\Phi=L_{\bm G}\log(\mu)$  and $\Psi$, involving only the statistics of the unperturbed system. In that sense, it provides a general version of the FDT \cite{Kubo1966,majda2005information} valid for jump-diffusion models.

Using the decomposition of (temporal) correlations in terms of Kolmogorov modes \cite[Corollary 1]{Chekroun_al_RP2} (see Eq.~\eqref{Eq_decomp_corr1}) with $f=L_{\bm G} \log(\mu)$ and $g=\Psi$,  the Green function can then be decomposed as a sum involving the RP resonances and Kolmogorov modes of the unperturbed dynamics. 
In that respect, we recall that 
the action of the semigroup $e^{t \mathcal{L}_K}$ on an observable can be expressed using the eigenfunctions of the Kolmogorov operator:
\beas
e^{t \mathcal{L}_K} \Psi(\x) = \sum_{j=1}^N \sum_{\ell=0}^{m_j-1} e^{\lambda_j t} &\frac{t^\ell}{\ell!} (\mathcal{L}_K - \lambda_j \textrm{Id})^\ell \Phi_j(\x) \langle \Phi_j^\ast, \Psi \rangle_\mu \\
&+ \text{remainder}.
\eeas

Substituting this into the Green's function formula Eq.~\eqref{eq:Green} (for $t \geq 0$ where $\Theta(t) = 1$), we arrive at the following decomposition formula
\begin{widetext}
\bes
\G_{\Psi,G}(t) \approx \Theta(t)\sum_{j=1}^N \sum_{\ell=0}^{m_j-1} \frac{\langle \Phi_j^\ast, \Psi \rangle_\mu}{\ell!} e^{\lambda_j t} t^\ell \int \big[L_{\bm G}\log(\mu)\big](\x) (\mathcal{L}_{K} - \lambda_j \textrm{Id})^{\ell} \Phi_j (\x) \mu(\d \x),
\ees
or in other words
\begin{equation}\label{GreenH}
\G_{\Psi,G}(t) \approx \sum_{j=1}^{N}\sum_{\ell=0}^{m_j-1} \frac{\alpha_{j\ell}(\Psi)}{\ell!} e^{\lambda_jt}t^{\ell}, \quad t\geq 0, 
\end{equation}
with $\G_{\Psi,G}(t)=0$ when $t<0$, and where the coefficients $\alpha_{j\ell}$ are given by:
\begin{equation}\label{Eq_alpha} 
\alpha_{j\ell}(\Psi)=\langle \Phi_j^\ast, \Psi \rangle_{\mu} \int \big[L_{\bm G} \log(\mu)\big](\x)(\mathcal{L}_{K} - \lambda_j \mbox{Id})^{\ell} \Phi_j (\x) \mu(\d\x),
\end{equation} 
\end{widetext}
with $L_G= -\mbox{div} ({\bm G} \cdot )$ and $\mathcal{L}_K$ denoting the Kolmogorov operator associated with the unperturbed system, Eq.~\eqref{Eq_Levy_additive}. 
Thus, the  $\alpha$'s  weight, according to Eq.~\eqref{Eq_alpha}, the contribution of the Kolmogorov modes of the unperturbed dynamics to the response for a given observable $\Psi$ and forcing pattern ${\bm G}$ defining the operator $L_{\bm G}$ given by Eq.~\eqref{Eq_Liouville}.

\subsection{Physical Interpretation of Green's Decomposition Formula}
The Green's function $\G_{\Psi,G}(t)$ embodies the linear response of a chosen observable $\Psi$ of the system to a small, time-dependent perturbation. In our case, the perturbation is introduced through the term $\epsilon g(t) \mathbf{G}(\x)$ in the drift of the stochastic differential equation. Here, $\epsilon$ is a small parameter indicating the perturbation strength, $g(t)$ is the temporal profile of the forcing, and $\mathbf{G}(\x)$ describes the spatial pattern of the forcing. The operator $L_{\bm G} = -\mbox{div} ({\bm G} \cdot )$ quantifies how this spatial pattern influences the probability density of the system.

The Green's function $\G_{\Psi,G}(t)$ essentially tells us how much the average value of the observable $\Psi$ at time $t$ is affected by an impulse perturbation applied at time $t=0$ (represented implicitly by the structure of linear response theory when considering a general $g(t)$ through convolution). In other words, it is the system's "memory" of a past perturbation with respect to the observable $\Psi$ and the forcing pattern $G$. The $\Theta(t)$ function ensures causality: the response only occurs for times $t\geq0$, after the perturbation has been applied (or the impulse occurred at $t=0$).

The decomposition formula reveals that this linear response can be expressed as a sum of fundamental modes, each characterized by a Ruelle-Pollicott resonance $\lambda_j$ and its associated eigenfunctions (Kolmogorov modes). The complex eigenvalues $\lambda_j$ dictate the temporal behavior of each mode. 

Thus, our analysis reveals that this response is not just a monolithic echo; it is  a symphony of different behaviors playing out at various tempos. The decomposition formula unveils these individual components. It tells us that the system's reaction to our perturbation can be understood as a sum of distinct modes, each characterized by a specific Ruelle-Pollicott resonance, denoted by $\lambda_j$. These resonances are like the fundamental frequencies and decay rates inherent to the system's undisturbed, natural dynamics governed by the Kolmogorov operator $\mathcal{L}_K$.

Each resonance $\lambda_j$ holds a crucial piece of the story. Its real part, $\text{Re}(\lambda_j)$, acts like a conductor, dictating how fast a particular mode will fade away over time ($\text{Re}(\lambda_j)\leq 0$ for ergodic systems). The imaginary part, $\text{Im}(\lambda_j)=\omega_j$, sets the rhythm, revealing the natural oscillatory frequency associated with that mode.   Thus, each $\lambda_j$ identifies a characteristic time scale ($1/|\text{Re}(\lambda_j)|$) and frequency of the system's intrinsic dynamics around its equilibrium (described by the ergodic measure $\mu$).
This indicates that near such degenerate frequencies, the response can exhibit polynomial growth multiplied by the exponential, potentially leading to slower decay or more complex temporal behaviors.

These resonances do not exist in isolation. They are intrinsically linked to the system's inherent spatial patterns of fluctuation, which we call Kolmogorov modes, represented by the eigenfunctions $\Phi_j$ (and their duals $\Phi_j^*$). These modes are the natural ways the system "breathes" and fluctuates around its equilibrium state, which is described by the ergodic probability measure $\mu$.

The crucial link between these underlying modes and the actual observed response is provided by the coefficients $\alpha_{j\ell}(\Psi)$. These coefficients act as weights, determining how strongly each Kolmogorov mode contributes to the system's reaction as seen through our chosen observable $\Psi$ and under the influence of our specific perturbation pattern $G$.  The term $\langle\Phi_j^*,\Psi\rangle_\mu$ represents the projection of the observable $\Psi$ onto the dual eigenfunction $\Phi_j^*$ with respect to the ergodic measure $\mu$. This indicates how sensitive the chosen observable is to the spatial pattern of the $j$-th Kolmogorov mode. If this inner product is small, that particular mode will have a weak influence on the observed response $\Psi$.  This  inner product tells us how much the spatial structure of the $j$-th mode "overlaps" with the observable we are looking at. Roughly speaking, if they are very different, this mode might not contribute much to what we observe.

The other part of the coefficient, the integral involving $L_{\bm G}\log(\mu)$ and $(\mathcal{L}_{K} - \lambda_j \mbox{Id})^{\ell} \Phi_j$, describes how the initial push we give to the system (through the forcing pattern $G$, which influences the probability density via $L_{\bm G}\log(\mu)$) excites these underlying Kolmogorov modes. It is like strumming a guitar string ---  depending on where and how one plucks it, one excites different vibrational modes.
Ultimately, this decomposition paints a vivid picture of the system's linear response. By analyzing the resonances with real parts closest to zero, we can identify the most influential modes that will dominate the long-term behavior. Their real parts will tell us about the characteristic response times of the system --- how quickly it reacts and how long the effect lingers. The imaginary parts will reveal if the response is oscillatory and at what frequencies. The coefficients $\alpha_{j\ell}(\Psi)$ then act as filters, showing us that the system's response is not a universal thing but rather depends intimately on the lens through which we look at the response (the observable $\Psi$) and how we are poking the system (the forcing $G$). 
Different observables might be more sensitive to different Kolmogorov modes and thus exhibit different response characteristics to the same forcing. The resonances and Kolmogorov modes act as the building blocks of the system's "impulse response," providing a deep understanding of how perturbations are amplified, damped, and propagate through the system. The coefficients $\alpha_{j\ell}(\Psi)$  then tell us how these fundamental building blocks are weighted to produce the specific response observed for a given forcing and observable.

In summary, this framework provides a profound way to understand the intricate interplay of time scales, frequencies, and spatial patterns that govern how complex stochastic systems react to external influences.

\section{Perturbing the Jump's Law: Non-Local Response Formula}\label{Sec_general_formula}
\subsection{Preliminaries}
To extend the linear response formula by considering perturbations in the jump measure $\nu$, let's denote the unperturbed measure as $\nu_0$ and the perturbation as $\epsilon \delta \nu_t(\d y)$, so the total measure is $\nu_\epsilon = \nu_0 + \epsilon \delta \nu_t$. An example of perturbation can be thought as  $\nu_\epsilon =\nu_0 + \epsilon f(t) \zeta$, where $\zeta$ is a fixed L\'evy measure independent of time $t$; see Remark \ref{Eq_modulation_separation} below.

The Kolmogorov operator $\mathcal{L}_K^\epsilon$ can be written as $\mathcal{L}_K + \epsilon \delta J$ ($\L_K$ given by Eq.~\eqref{Eq_Kolmo2_general}), where $\delta J$ is the first-order change in the jump integral $J$ due to $\delta \nu$:
\be\label{Eq_deltaJ}
\delta J \psi (\x)=\int_{\X} \Gamma_\mu(\x,\y) \delta \nu_t(\d \y), 
\ee
with
\be\label{Eq_Gamma}
\Gamma_\mu(\x,\y)=\Big[\mu(\x+\y)-\mu(\x)- \y\cdot \nabla \mu (\x) \mathds{1}_{\{\norm{\y}<1\}} \Big],
\ee
and $\X=\mathbb{R}^d\backslash\{0\}.$

The perturbed Fokker-Planck equation is $\partial_t \rho_\epsilon^t = (\mathcal{L}_K^\epsilon)^\ast \rho_\epsilon^t$. Using the expansion $\rho_\epsilon^t = \mu + \epsilon \rho_1^t + h.o.t.$ and $\partial_t \mu = (\mathcal{L}_K)^\ast \mu = 0$, we can obtain the equation for $\rho_1^t$ with both perturbations.

The perturbed Kolmogorov operator is $\mathcal{L}_K^\epsilon = \mathcal{L}_K + \epsilon \delta J_t$, where $\delta J_t$ accounts for the perturbation in the jump measure $\delta \nu_t$:
$$ \delta J_t \psi (\x)=\int_{\X} \Gamma_\mu(\x,\y) \delta \nu_t(\d \y). $$
The perturbed Fokker-Planck equation is 
\be\label{Eq_perturbed_FKPE}
\partial_t \rho_\epsilon^t = (\mathcal{L}_K + \epsilon \delta J_t)^\ast \rho_\epsilon^t.
\ee
We use the expansions $\rho_\epsilon^t(\x) = \mu(\x) + \epsilon \rho_1^t(\x) + h.o.t.$ and $\partial_t \mu = (\mathcal{L}_K)^\ast \mu = 0$. Substituting the expansion for $\rho_\epsilon^t$:
$$ \partial_t (\mu + \epsilon \rho_1^t + ...) = ((\mathcal{L}_K)^\ast + \epsilon (\delta J_t)^\ast) (\mu + \epsilon \rho_1^t + ...) $$
We now expand this to the first order in $\epsilon$.

We start with the perturbed Fokker-Planck equation:
$$ \partial_t \rho_\epsilon^t = (\mathcal{L}_K + \epsilon \delta J_t)^\ast \rho_\epsilon^t,$$
and the expansion:
$$ \rho_\epsilon^t = \mu + \epsilon \rho_1^t + O(\epsilon^2),$$
leading to
$$ \partial_t \rho_\epsilon^t = \partial_t \mu + \epsilon \partial_t \rho_1^t + O(\epsilon^2).$$
Since $\mu$ is the equilibrium density, $\partial_t \mu = 0$. Thus,
$$ \partial_t \rho_\epsilon^t = \epsilon \partial_t \rho_1^t + O(\epsilon^2).$$
Now, we substitute the expansions into the right-hand side (RHS) of the perturbed Fokker-Planck equation (Eq.~\eqref{Eq_perturbed_FKPE}):
$$ \text{RHS} = (\mathcal{L}_K + \epsilon \delta J_t)^\ast (\mu + \epsilon \rho_1^t + O(\epsilon^2)),$$
which, using the linearity of the adjoint operator, gives:
\beas
\text{RHS} = (\mathcal{L}_K)^\ast (\mu + &\epsilon \rho_1^t + O(\epsilon^2)) \\
&+ \epsilon (\delta J_t)^\ast (\mu + \epsilon \rho_1^t + O(\epsilon^2)).
\eeas
Expanding further:
\beas
\text{RHS} = (\mathcal{L}_K)^\ast \mu + &\epsilon (\mathcal{L}_K)^\ast \rho_1^t + O(\epsilon^2) \\
&+ \epsilon (\delta J_t)^\ast \mu + \epsilon^2 (\delta J_t)^\ast \rho_1^t + O(\epsilon^3).
\eeas
We know that $\mu$ is the equilibrium distribution for $\mathcal{L}_K$, which means $(\mathcal{L}_K)^\ast \mu = 0$. So the equation becomes:
\be
 \text{RHS} = \epsilon (\mathcal{L}_K)^\ast \rho_1^t + \epsilon (\delta J_t)^\ast \mu + O(\epsilon^2).
 \ee
Now, equating the LHS and RHS:
$$ \epsilon \partial_t \rho_1^t + O(\epsilon^2) = \epsilon (\mathcal{L}_K)^\ast \rho_1^t + \epsilon (\delta J_t)^\ast \mu + O(\epsilon^2) $$
To obtain the equation for $\rho_1^t$ to the first order in $\epsilon$, we divide by $\epsilon$ and ignore higher-order terms:

 \be
 \partial_t \rho_1^t = (\mathcal{L}_K)^\ast \rho_1^t + (\delta J_t)^\ast \mu.
\ee
The solution for $\rho_1^t$ is then given by:
\bea\label{Eq_rho1}
\rho_1^t = \int_0^t e^{(t-s) (\mathcal{L}_K)^\ast} &g(s) L_{\bm G} \mu \d s \\
&+ \int_0^t e^{(t-s) (\mathcal{L}_K)^\ast}  (\delta J_s)^\ast \mu \d s. 
\eea
To derive the response formula, we need thus to evaluate the action of the adjoint  $(\delta J_s)^\ast$ on $\mu$. 

\subsection{The Adjoint Formula}
 To find the formula for the adjoint, $(\delta J_s)^\ast$, we start with the definition of the adjoint operator through the duality relation:
    $$ \int (\delta J_s \psi)(\x) \phi(\x) d\x = \int \psi(\x) ((\delta J_s)^\ast \phi)(\x) d\x $$
The left-hand side (LHS) is:
    $$ \text{LHS} = \int \left( \int_{\X} \Gamma_\mu(\x,\y) \delta \nu_s(\d \y) \right) \phi(\x) d\x $$
     We can interchange the order of integration:
    $$ \text{LHS} = \int_{\X} \delta \nu_s(\d \y) \int  \Gamma_\mu(\x,\y) \phi(\x) d\x $$
Now, we examine each term in the inner integral (due to $\Gamma_\mu$ given by Eq.~\eqref{Eq_Gamma}) separately.
 The two first terms are easy to handle. 
For the term, $\int \psi(\x+\y) \phi(\x) \d\x$, we use the change of variable $\z = \x + \y$, so $\x = \z - \y$ and $\d\x = \d\z$. The integral becomes $\int \psi(\z) \phi(\z - \y) \d\z$. Changing the integration variable back to $\x$, we have $\int \psi(\x) \phi(\x - \y) \d\x$. The second term  gives  trivially   $ - \int \psi(\x) \phi(\x) d\x$. It is the third term, $-\int(\y\cdot \nabla \psi (\x)) \mathds{1}_{\{\norm{\y}<1\}} \phi(\x) d\x$, that requires more care.

To handle this term, we use the integration by parts for the $i$-th component:
{\small 
    \beas
    - \int \partial_i \psi (\x) y_i \mathds{1}_{\{\norm{\y}<1\}} \phi(\x) d\x & = \int \psi(\x) \partial_i (\phi(\x) y_i \mathds{1}_{\{\norm{\y}<1\}}) d\x \\
    &= \int \psi(\x) \partial_i \phi(\x) y_i \mathds{1}_{\{\norm{\y}<1\}} d\x.
    \eeas
    }
            Summing over all components $i$, we get $$\int \psi(\x) (\y \cdot \nabla \phi(\x)) \mathds{1}_{\{\norm{\y}<1\}} d\x.$$

Substituting these back into the expression for LHS:
\begin{widetext}
\bea
 \text{LHS} &= \int_{\X} \delta \nu_s(\d \y) \int \Big[\psi(\x) \phi(\x - \y) - \psi(\x) \phi(\x) + \psi(\x) (\y \cdot \nabla \phi(\x)) \mathds{1}_{\{\norm{\y}<1\}} \Big] d\x\\
 \text{LHS} &= \int \psi(\x) \left( \int_{\X} \Big[\phi(\x - \y) - \phi(\x) + \y \cdot \nabla \phi(\x) \mathds{1}_{\{\norm{\y}<1\}} \Big] \delta \nu_s(\d \y) \right) d\x.
 \eea
\end{widetext}
   
    Comparing this with the RHS:
    $$ \text{RHS} = \int \psi(\x) ((\delta J_s)^\ast \phi)(\x) d\x $$
 We can identify the adjoint operator $(\delta J_s)^\ast$ acting on $\phi$:
 \be\label{Eq_adjoint}
 \boxed{\Big[(\delta J_s)^\ast \phi\Big](\x)=  \int M_\phi(\x,\y)\delta \nu_s(\d \y),}
 \ee
 with 
 \be\label{Eq_Mphi}
 M_\phi(\x,\y)= \Big[\phi(\x - \y) - \phi(\x) + \y \cdot \nabla \phi(\x) \mathds{1}_{\{\norm{\y}<1\}} \Big].
 \ee

\br\label{Rmk_adjoint_sym}
In this remark, we show that $((\delta J_s)^\ast \phi)(\x) = (\delta J_s \phi)(\x)$,  when the measure $\delta \nu_s$ is symmetric. Let's change the integration variable $\y \rightarrow -\y$ in the expression for $((\delta J_s)^\ast \phi)(\x)$. Since $\delta \nu_s$ is symmetric, $\delta \nu_s(-\d \y) = \delta \nu_s(\d \y)$.
\begin{widetext}
\begin{align*} ((\delta J_s)^\ast \phi)(\x) &= \int_{\mathcal{X}} \Big[\phi(\x - \y) - \phi(\x) + \y \cdot \nabla \phi(\x) \mathds{1}_{\{\norm{\y}<1\}} \Big] \delta \nu_s(\d \y) \\ &= \int_{\mathcal{X}} \Big[\phi(\x - (-\z)) - \phi(\x) + (-\z) \cdot \nabla \phi(\x) \mathds{1}_{\{\norm{-\z}<1\}} \Big] \delta \nu_s(\d (-\z)) \quad (\text{where } \z = -\y) \\ &= \int_{\mathcal{X}} \Big[\phi(\x + \z) - \phi(\x) - \z \cdot \nabla \phi(\x) \mathds{1}_{\{\norm{\z}<1\}} \Big] \delta \nu_s(\d \z). 
\end{align*}
\end{widetext}
Replacing $\z$ with $\y$ again,  we get:
\be
((\delta J_s)^\ast \phi)(\x) = \int \Gamma_{\mu}(\x,\y)\delta \nu_s(\d \y), 
\ee
with $\Gamma_{\mu}$ given by Eq.~\eqref{Eq_Gamma}.  This is exactly the expression for $(\delta J_s \phi)(\x)$; see Eq.~\eqref{Eq_deltaJ}. Therefore, $((\delta J_s)^\ast \phi)(\x) = (\delta J_s \phi)(\x)$, which implies $((\delta J_s)^\ast = \delta J_s$. This directly proves that the operator $\delta J_s$ is self-adjoint when the measure $\delta \nu_s$ is symmetric.
\er

\subsection{The Extended Response Formula}
We are now in position to derive our general response formula.
Using Eq.~\eqref{eq:response function a} with $\rho_1^t$ now given by Eq.~\eqref{Eq_rho1}, we obtain, after integrating by part (Eq.~\eqref{Eq_integ_by_part_forPsi_rho1}), that the anomaly in the expected value of $\Psi$ is given by
\beas
\delta^{(1)}[\Psi] (t) &= \epsilon_1 \int_{-\infty}^t \G_{\Psi,G} (t-s) g_1(s)\d s \\
&+ \epsilon_2 \underbrace{\int \Psi(\x) \left( \int_0^t e^{(t-s) (\mathcal{L}_K)^\ast} (\delta J_s)^\ast \mu  \d s \right) \d \x}_{I_2},
\eeas
where the first term accounts for perturbation in the drift part (Eq.~\eqref{Eq_LRF}), and the term $I_2$ accounts for the perturbation in the jump's law.

To determine $I_2$, we first switch the order of integration:
$$ I_2=\epsilon \int_0^t \d s \int \Psi(\x) (e^{(t-s) (\mathcal{L}_K)^\ast} (\delta J_s)^\ast \mu \d \x.$$
Using the duality relation $\langle f,A^\ast g \rangle=\langle Af,g \rangle$:
$$I_2= \epsilon \int_0^t \d s \int (e^{(t-s) \mathcal{L}_K} \Psi)(\x) ((\delta J_s)^\ast \mu)(\x) \d \x.$$
Now we need the form of $(\delta J_s)^\ast \mu$. We know that for the adjoint of the jump operator $J$ acting on a density, we have:
{\small 
\bes
J^\ast \rho(\x) = \int_{\X} \Big[\rho(\x-\y)-\rho(\x)+ \y\cdot \nabla \rho (\x) \mathds{1}_{\{\norm{\y}<1\}} \Big] \nu(\d \y).
\ees
}
Applying this to $\delta J_s$ and $\mu$, we obtain 
\be
 (\delta J_s)^\ast \mu (\x) = \int_{\X} M_\mu(\x,\y)\delta \nu_s(\d \y),
\ee
with $M_\mu(\x,\y)$ given by Eq.~\eqref{Eq_Mphi}, with $\phi=\mu.$

The integral $I_2$ becomes then:
{\small
\beas
I_2 &= \epsilon \int_0^t \d s \int (e^{(t-s) \mathcal{L}_K} \Psi)(\x) \left( \int_{\X} M_\mu(\x,\y) \delta \nu_s(\d \y) \right) \d \x \\
&= \epsilon \int_0^t \d s \int_{\X} \delta \nu_s(\d \y) \int (e^{(t-s) \mathcal{L}_K} \Psi)(\x) M_\mu(\x,\y) \d \x.
\eeas
}
Defining the Green's function for the perturbation in the Jump's law, $\nu$, as:
\be
\boxed{\G_{\Psi,\nu}(t, \y) = \Theta(t) \int (e^{t \mathcal{L}_K} \Psi)(\x) M_\mu(\x,\y) \d \x,} 
\ee
with $M_\mu(\x,\y) = \mu(\x-\y)-\mu(\x)+ \y\cdot \nabla \mu(\x) \mathds{1}_{\{\norm{\y}<1\}}$.

The extended linear response formula is then:
\fcolorbox{black}{faintpink}{%
    \begin{minipage}{0.9\linewidth}
    \setlength{\abovedisplayskip}{0pt}
    \setlength{\belowdisplayskip}{0pt}
    \begin{align*}
    \delta^{(1)}[\Psi] &(t) = \epsilon \int_{-\infty}^t \mathcal{G}_{\Psi,G} (t-s) g(s)\d s \tag{gLRF}\label{Eq_gLRF}\\
    &+ \epsilon \int_{-\infty}^t \d s \int_{\X} \mathcal{G}_{\Psi,\nu}(t-s, \y) \delta \nu_s(\d \y).
    \end{align*}
    \end{minipage}%
}

This formula for the first-order perturbation in an observable $\Psi$, $\delta^{(1)}[\Psi](t)$, offers a powerful physical lens into how systems with both continuous and jump-like dynamics respond to external influences. It shows us that the total response is a graceful superposition of two distinct physical effects: one arising from a {\it gradual, local nudge} and another from a {\it sudden, non-local shocks} (due to the jumps).

The first term, containing the Green's function $\mathcal{G}_{\Psi,G}(t)$, represents the {\it local response}. This is the familiar part of linear response theory, mirroring how systems react to continuous forces or a gentle, smooth push. The perturbation here is a change in the system's deterministic drift, which in turn impacts the overall invariant measure and thus affects the observable $\Psi$. While this perturbation can be globally applied across the system's state space, its effect at any given point $\bm{x}$ depends only on the properties of the system at that point, not on what happens at distant points $\y \neq \x$. The Green's function $\mathcal{G}_{\Psi,G}(t)$ acts as a susceptibility kernel, elegantly quantifying how this local, continuous perturbation propagates through time to affect the observable $\Psi$. This term captures the expected, diffusive-like behavior.

The second term, with its Green's function $\mathcal{G}_{\Psi,\nu}(t, \y)$, reveals the unique character of jump processes by describing the {\it non-local response}. This is the system's reaction to a {\it shock}---a sudden, instantaneous change in the jump law that can displace it to a distant state. This shock is characterized by the function $M_\mu(\x, \y)$, which is sensitive to both the size of the jump $\y$ and the underlying, unperturbed equilibrium measure $\mu$. The Green's function $\mathcal{G}_{\Psi,\nu}(t, \y)$ then measures the correlation between this sharp, non-local event and the subsequent, continuous evolution of the observable. It is a key physical insight that while the perturbation itself is a sudden shock, the overall response it generates is a smooth, continuous function. This continuity arises because the response averages over the probabilities of all possible jump sizes, effectively translating the abruptness of a single jump into a well-behaved, analytical framework for the entire system's reaction.

In essence, the generalized linear response formula (Eq.~\eqref{Eq_gLRF}) provides a complete picture by separating these two fundamental types of system responses. It beautifully demonstrates that a system with jumps reacts differently to a continuous force than it does to a sudden, wide-ranging shock, and its overall response is the simple sum of these two distinct effects.

As with the drift-based response, we can gain a deeper understanding of the jump-based Green's function, $\G_{\Psi,\nu}(t, \y)$, by applying the modal decomposition of the semigroup $e^{t\L_K}$ This reveals that the response to a jump perturbation is also a symphony of the system's fundamental Ruelle-Pollicott resonances $\lambda_j$ and their associated Kolmogorov modes $\Phi_j$.

By substituting the modal expansion of the semigroup into the Green's function formula, we arrive at the decomposition:
\be
\G_{\Psi,\nu}(t, \y) \approx \sum_{j=1}^{N}\sum_{\ell=0}^{m_j-1} \frac{\beta_{j\ell}(\Psi, \y)}{\ell!} e^{\lambda_jt}t^{\ell}, 
\ee
where the new coefficients $\beta_{j\ell}$ are given by:
\bes
\beta_{j\ell}(\Psi, \y) = \langle \Phi_j^\ast, \Psi \rangle_\mu \underbrace{\int (\mathcal{L}_K - \lambda_j \mbox{Id})^{\ell} \Phi_j (\x) M_\mu(\x,\y) \d\x}_{J_{j\ell}(\y)}.
\ees
The integral, $J_{j\ell}(\y)$, is the central piece that links the jump perturbation to the system's intrinsic dynamics. It provides a measure of the coupling strength (an $L^2$ product) between the perturbation between two key functions. The first one,  $(\mathcal{L}_K - \lambda_j \mbox{Id})^{\ell} \Phi_j (\x)$ isolates a specific spatial pattern associated with the Ruelle-Pollicott resonance $\lambda_j$.  For simple eigenvalues (when $\ell=0$), this is simply the eigenfunction $\Phi_j(\x)$ itself. For degenerate eigenvalues, the higher-order terms in the series represent generalized eigenfunctions that capture the more complex spatial structures of the resonance.

The second function, $M_\mu(\x,\y) $, is a fundamental object that quantifies the non-local change in the system's equilibrium measure $\mu$ at a location $\x$ due to a jump of size $\y$. It is the mathematical signature of the perturbation's physical impact.

The integral, as an $L^2$-product, measures the geometric overlap between the spatial pattern of the Kolmogorov mode and the spatial impact of the jump. If this integral vanishes, it means that for a jump of size $\y$, the geometric impact of the perturbation is orthogonal to the shape of the mode. This signifies that the jump of size $\y$ does not excite that specific mode. This is the structural reason why some jumps may excite certain modes rather than others. The coefficients $\beta_{j\ell}$ reveal that the non-local nature of the jump perturbation, dependent on both starting position $\x$ and jump size $\y$, creates a unique coupling mechanism. A jump can instantaneously connect two spatially distant parts of the system, and its ability to excite a particular mode depends on the collective properties of the mode's spatial pattern across all possible jump locations. The $L^2$-product captures this complex, non-local interaction, providing a rigorous and unambiguous link between the physical forcing and the (unperturbed) system's Kolmogorov modes.

\br\label{Rmk_integbyparts}
For the jump perturbation case, we used the concept of duality (which is analogous to integration by parts in function spaces) to move the adjoint operator from the time evolution part to the part involving the perturbation:
The anomaly due to the jump perturbation was:
$$ \epsilon \int \Psi(\x) \left( \int_0^t e^{(t-s) (\mathcal{L}_K)^\ast} \delta J \mu \d s \right) \d \x $$
We then changed the order of integration and used the duality relation:
$$ \int \Psi(\x) (e^{(t-s) (\mathcal{L}_K)^\ast} \phi(\x)) \d \x = \int (e^{(t-s) \mathcal{L}_K} \Psi)(\x) \phi(\x) \d \x $$
where $\phi(\x) = (\delta J \mu)(\x)$. This step is essentially the integration by parts that allowed us to express the result in terms of the forward evolution under $\mathcal{L}_K$.
Therefore, the integration by parts occurred implicitly through the use of the adjoint operator and its duality property. It is not a direct integration by parts involving spatial derivatives as in the drift case which leads to $L_{\bm G}\log(\mu)$.
\er
\br\label{Eq_modulation_separation}
We consider the special case where the perturbation to the L\'evy measure is separable in time and space: 
$$ \delta \nu_t(\d\y) = \epsilon' f(t) \zeta(\d \y),$$ 
where $\epsilon' f(t)$ is the time modulation and $\zeta(\d\y)$ is a fixed L\'evy measure (independent of time $t$ and location $\x$). 
The second, non-local term of Eq.~\eqref{Eq_gLRF} is: 
$$ \delta^{(1)}[\Psi]_{\nu} (t) = \epsilon' \int_{-\infty}^t \d s \int_{\mathcal{X}} \mathcal{G}_{\Psi,\nu}(t-s, \y) \delta \nu_s(\d \y). $$ 
Substitute the form of the perturbation $\delta \nu_s(\d\y) = f(s) \zeta(\d\y)$ into the expression: 
$$ \delta^{(1)}[\Psi]_{\nu} (t) = \epsilon' \int_{-\infty}^t f(s) \d s \left[ \int_{\mathcal{X}} \mathcal{G}_{\Psi,\nu}(t-s, \y) \zeta(\d \y) \right].$$ 
The term in the square brackets is a convolution in the jump-size space $\y$. Since this inner integral only depends on the time difference $\tau=t-s$, we define a new, simpler Green's function, $\mathcal{R}_{\Psi, \zeta}(\tau)$, that absorbs the spatial non-locality: 
\be
\mathcal{R}_{\Psi, \zeta}(\tau) = \int_{\mathcal{X}} \mathcal{G}_{\Psi,\nu}(\tau, \y) \zeta(\d \y).
\ee

Now, let's substitute the definition of the jump Green's function, $\mathcal{G}_{\Psi,\nu}(\tau, \y)$: 
$$ \mathcal{R}_{\Psi, \zeta}(\tau) = \Theta(\tau) \int_{\mathcal{X}} \d \x (e^{\tau \mathcal{L}_K} \Psi)(\x) \left[ \int_{\X} M_\mu(\x,\y) \zeta(\d \y) \right].$$

The inner integral over $\y$ is precisely the action of the adjoint of the jump operator ($\mathcal{L}_{\zeta}^\ast$) associated with the perturbing measure $\zeta$, acting on the unperturbed invariant measure $\mu$, according to: 
$$ \mathcal{L}_{\zeta}^\ast [\mu](\x) = \int_{\X} M_\mu(\x,\y) \zeta(\d \y).$$ 
This results in the simplified form for the combined term: 
\be\label{Eq_Green_R}
\boxed{\mathcal{R}_{\Psi, \zeta}(\tau) = \Theta(\tau) \int_{\mathcal{X}} (e^{\tau \mathcal{L}_K} \Psi)(\x) \left( \mathcal{L}_{\zeta}^\ast [\mu](\x) \right) \d \x}.
\ee

The second term of Eq.~\eqref{Eq_gLRF} is now a simple convolution in time, exactly matching the functional form of the first (drift) term: 
$$ \delta^{(1)}[\Psi]_{\nu} (t) = \epsilon' \int_{-\infty}^t f(s) \mathcal{R}_{\Psi, \zeta}(t-s) \d s.$$ 
The full generalized LRF becomes:
\beas
\delta^{(1)}[\Psi] (t) = \epsilon \int_{-\infty}^t &\mathcal{G}_{\Psi,G} (t-s) g(s)\d s \\
& + \epsilon' \int_{-\infty}^t \mathcal{R}_{\Psi, \zeta}(t-s) f(s)\d s.
\eeas
This simplification shows that a perturbation to the jumps' law ($\delta \nu_t$) that is separable in time and space shares the key structural attribute of a perturbation of the drift term. The overall response is indeed made of two parts.
A contribution that is a convolution of the drift Green's function $\mathcal{G}_{\Psi,G}$ with the time-modulation $g(s)$. The core forcing is derived from the perturbation of the deterministic vector field on the invariant measure (the term $L_{\bm G}\log(\mu)$ in Eq.~\eqref{eq:Green}). 
 A contribution stemming from the jump law perturbation,  which results into a convolution of a new Green's function $\mathcal{R}_{\Psi, \zeta}$ with the time-modulation $f(s)$. The core forcing is here derived from the adjoint of the perturbing jump operator ($\mathcal{L}_{\zeta}^\ast$) acting on the invariant measure. 

This highlights that both types of perturbations ultimately introduce a generalized forcing term that acts upon the invariant measure, and the system responds to this forcing through the same unperturbed dynamics ($e^{\tau \mathcal{L}_K}$) acting on the observable $\Psi$; cf Eq.~\eqref{eq:Green} with Eq.~\eqref{Eq_Green_R}. This is a powerful demonstration of the unified structure of response theory, even in the presence of non-local L\'evy dynamics.
\er

\subsection{Diffusive Limit from Jump's Law Perturbations}
To illustrate the theory, we consider a perturbation of the form,
$$
\delta \nu(\d\y) = \lambda K(\x, \x+\y)  \d\y,
$$
where $\x$ represents the system's current location or state, and $\y$ is the vector describing the jump's size and direction.
The intensity of the jumps is given by the jump kernel $K(\x, \x') = \lambda K(\x, \x')$, in which $\lambda$ is the jump rate. In this scenario, the perturbation only affects the jump process. Therefore, the first term of the generalized linear response formula, which describes the continuous response to a change in the deterministic flow, vanishes.  
The entire linear response is thus captured by the second, non-local term.

We choose $K$ to be the Green's function of the operator $I-\eta \Delta$, with $\eta>0$. 
This choice ensures $K$ is translationally invariant, such that $K(\x, \x+\y) = K_{\eta}(\y)$, with a Fourier transform given by $\widehat{K}_{\eta}(\mathbf{k}) = \frac{1}{1 + \eta k^2}$, where $k=|\mathbf{k}|$.

Substituting this perturbation into the response formula, and performing a change of variables ($\tau = t-s$), we get the simplified expression for the first-order perturbation:
\bes
\delta^{(1)}[\Psi] (t) = \epsilon \lambda  \int_{0}^\infty \d\tau \int_{\mathcal{X}} \mathcal{G}_{\Psi,\nu}(\tau, \y) K_{\eta}(\y) \d\y.
\ees
This simplified formula provides a clear physical picture. The parameter $\eta$ acts as a correlation lengthscale, filtering the jump sizes. It governs the  characteristic radius of influence of a jump: the perturbation is most influential for jumps with "small" sizes ($|\mathbf{y}| \lesssim \sqrt{\eta}$), with its effect diminishing rapidly for jumps larger than this length scale.

The overall response is a time-integrated sum of a convolution between the non-local Green's function  $\mathcal{G}_{\Psi,\nu}$ and the translationally invariant jump kernel $K_{\eta}$.  This convolution selectively amplifies or suppresses the system's reaction to jumps based on their size, effectively filtering the system's response based on the spatial scale of the perturbation.


A fascinating aspect of this formula emerges when we consider the double limit where the correlation lengthscale $\eta \to 0$ and the perturbation rate $\lambda= 1/\eta$. Physically, this corresponds to the perturbation inducing an infinite number of infinitesimally small jumps. 
As a result, the discrete jump process begins to resemble a continuous, diffusive process, providing an important check for the consistency of our theory with existing linear response frameworks for standard stochastic diffusion \cite{Hairer_Majda,Santos2022}.

To formalize this, we must evaluate the limit of the integral:
$$ \lim_{\eta\to 0} \delta^{(1)}[\Psi](t) = \lim_{\eta\to 0} \epsilon \frac{1}{\eta} \int_0^\infty \d\tau \int_{\X} \mathcal{G}_{\Psi,\nu}(\tau, \y) K_{\eta}(\y) \d\y. $$

 The expression becomes:
 \be\label{Eq_limit_tofind}
 \lim_{\eta\to 0} \delta^{(1)}[\Psi](t) = \epsilon \lim_{\eta\to 0} \frac{1}{\eta} \int_0^\infty \hspace{-1ex}\d\tau \int_{\X} \d\y K_{\eta}(\y) N(\y),
 \ee
with $N(\y)=\int_{\mathbb{R}^d} \d\x (e^{\tau \mathcal{L}_K} \Psi)(\x) M_{\mu}(\x,\y) \d\y.$

We can exchange the order of integration and focus on the integral over $\mathbf{y}$:
$$ I = \lim_{\eta\to 0} \frac{1}{\eta} \int_{\X} K_{\eta}(\y) M_{\mu}(\x,\y) \d\y. $$
The Taylor expansion is a local approximation, which would not be valid for an integral over all of $\mathbb{R}^d$ in general. However, as we showed earlier, in the limit $\eta \to 0$, the Kernel $K_\eta(\mathbf{y})$ approaches the Dirac delta function, $\delta(\mathbf{y})$. This means that the jump kernel becomes highly concentrated around $\mathbf{y} = 0$, and the integral is therefore dominated by the region where the Taylor expansion is most accurate. This allows us to safely truncate the series and neglect the higher-order terms, as their contributions to the total integral become vanishingly small.

The Taylor expansion of $\mu(\x-\y)$ around $\y = \mathbf{0}$ (assuming $\mu$ smooth \cite[Sec.~5.5]{kuhn2017levy}), gives:
$$ \mu(\x-\y) = \mu(\x) - \y\cdot\nabla\mu(\x) + \frac{1}{2} \sum_{i,j} y_i y_j \frac{\partial^2\mu}{\partial x_i \partial x_j} + O(|\y|^3). $$
Substituting this expansion into the definition of $M_{\mu}$, its expression  becomes:
\beas
 M_{\mu}(\x,\y)=\bigg[ -&\y\cdot\nabla\mu(\x) + \frac{1}{2} \sum_{i,j} y_i y_j \frac{\partial^2\mu}{\partial x_i \partial x_j}  \bigg] \\
 &+ \y\cdot \nabla \mu(\x) \mathds{1}_{\{\norm{\y}<1\}} + O(|\y|^3). 
 \eeas
 As we take the limit $\eta \to 0$, the kernel $K_\eta(\y)$ becomes a Dirac delta function, concentrating the integral at $\mathbf{y}=\mathbf{0}$. In this vanishingly small neighborhood, the indicator function $\mathds{1}_{\{\norm{\y}<1\}}$ is essentially 1. The first-order terms, $-\mathbf{y}\cdot\nabla\mu(\x)$ and $\mathbf{y}\cdot\nabla\mu(\x)$, exactly cancel each other out. The expression inside the integral simplifies to just the second-order term from the Taylor expansion:
$$ I = \lim_{\eta\to 0} \frac{1}{\eta} \int_{\X} K_{\eta}(\y) \left[ \frac{1}{2} \sum_{i,j} y_i y_j \frac{\partial^2\mu}{\partial x_i \partial x_j} + O(|\y|^3) \right] \d\y. $$
The second moment of a probability distribution is related to the second derivative of its characteristic function (the Fourier transform of the distribution) at the origin:
\begin{equation}
\int_{\mathbb{R}^d} y_i y_j K_\eta(\mathbf{y}) \d\mathbf{y} = -\left. \frac{\partial^2}{\partial k_i \partial k_j} \widehat{K}_\eta(\mathbf{k}) \right|_{\mathbf{k}=0}.
\end{equation}
The Fourier transform of our kernel is $\widehat{K}_{\eta}(\mathbf{k}) = \frac{1}{1 + \eta \mathbf{k}^2}$, where $\mathbf{k}^2 = k_1^2 + k_2^2 + \dots + k_d^2$. We can compute the first derivative with respect to $k_i$:
$$
\frac{\partial \widehat{K}_\eta}{\partial k_i} = - \frac{1}{(1+\eta \mathbf{k}^2)^2} (2\eta k_i),
$$
leading to the second derivative:
$$
\frac{\partial^2 \widehat{K}_\eta}{\partial k_j \partial k_i} = - \frac{2\eta \delta_{ij}}{(1+\eta \mathbf{k}^2)^2} + \frac{4\eta k_i \cdot 2\eta k_j}{(1+\eta \mathbf{k}^2)^3}.
$$
Evaluating this expression at $\mathbf{k}=0$, we find:
$$
\left. \frac{\partial^2 \widehat{K}_\eta}{\partial k_j \partial k_i} \right|_{\mathbf{k}=0} = -2\eta \delta_{ij}.
$$
Therefore, the integral of the second moment is:
$$
\int_{\mathbb{R}^d} y_i y_j K_\eta(\mathbf{y}) \d\mathbf{y} = -(-2\eta \delta_{ij}) = 2\eta \delta_{ij}.
$$
Substituting this back into our expression for the integral $I$:
\beas
I& = \frac{1}{2}\sum_{i,j} \frac{\partial^2\mu}{\partial x_i \partial x_j} \left( \lim_{\eta\to 0} \frac{1}{\eta} \int y_i y_j K_\eta(\y)\d\y \right)\\
&= \frac{1}{2}\sum_{i,j} \frac{\partial^2\mu}{\partial x_i \partial x_j} (2\delta_{ij}) = \sum_{i} \frac{\partial^2\mu}{\partial x_i^2} = \Delta\mu(\x).
\eeas 
The entire inner integral over $\mathbf{y}$ in Eq.~\eqref{Eq_limit_tofind}, therefore reduces to the action of the Laplacian on the equilibrium measure.

Substituting this result back into the expression for the first-order perturbation, we obtain the full characterization of the response in this limit:
\bes
\lim_{\eta\to0} \delta^{(1)}[\Psi] (t) = \epsilon \int_{0}^\infty \d\tau \int_{\mathbb{R}^d} (e^{\tau \mathcal{L}_K} \Psi)(\x) (\Delta\mu)(\x) \d\x.
\ees
This is the key insight. The perturbation to the jump process, which was initially described by a non-local response and a convolution over jump sizes, has transformed into a response that is now purely local in the jump size $\y$. In this limit, the system's reaction to the perturbation is equivalent to its reaction to a continuous, diffusive-like process. The Central Limit Theorem provides the theoretical basis for this, as the sum of a large number of independent, small-amplitude random variables tends to a Gaussian distribution. Thus, we have shown how a jump-based perturbation can, in a specific limit, produce a new diffusive term that impacts the overall system response.

It is instructive to compare the result of our jump-based formalism with the linear response of a system whose diffusion term is directly perturbed \cite{Santos2022}. A common form for such a perturbation is given by a modification to the SDE:
$$
\d \x(t) = F(\x)\d t + \Sigma(\x) \d \W+ \epsilon g(t)\Gamma(\x)\d \W,
$$
for which $\Sigma (\x)$ is a invertible and symmetric $d\times d$ diffusion matrix, 
and $\Gamma(\x)$ is a $d\times d$ matrix defining the correction to the background noise, and for our purposes, we can assume $g(t) = 1$. 
The linear response for this system is known to be \cite{Santos2022}:
$$
\delta^{(1)}[\Psi](t) = \epsilon\int_{-\infty}^{t} e^{(t-s)L_0^\ast} \Psi(x)\mathcal{L}_1 \mu(x)\d\x \d s,
$$
where $L_0$ is the Kolmogorov operator of the unperturbed system, and the perturbation operator $\mathcal{L}_1$ is defined as:
$$
\mathcal{L}_1 \rho= \frac{1}{2}\nabla^2 : (\Sigma\Gamma^{\text{T}}+\Gamma\Sigma^{\text{T}}) \rho.
$$
 For our result to be consistent with this diffusive perturbation formula, we must be able to choose a matrix $\Gamma$ such that the perturbation operator $\mathcal{L}_1$ becomes $\Delta$, namely, find $\Gamma$ such that:
 $$
\frac{1}{2} (\Sigma\Gamma^{\text{T}}+\Gamma\Sigma^{\text{T}}) = \mathbf{I}.
$$ 
 Due to our assumptions on $\Sigma$, this equation can be solved with  $\Gamma(\x)=[\Sigma(\x)]^{-1}.$
Thus, in the diffusive limit, our jump's law perturbation---with a kernel $K$ given as the Green's function described above---produces a new diffusive term of the form $\epsilon [\Sigma(\x)]^{-1}\d \W$.
\section{Shear-induced chaos in an ENSO model}\label{ENSO}

ENSO is  a major large-scale climate pattern, exhibiting complex interannual to decadal variability driven by coupled atmosphere-ocean processes \cite{Neelin_al98,McPhaden2020,Timmermann2018_NatureENSO}. El Ni\~no events, characterized by warming of the tropical Pacific Ocean and weakening trade winds, occur every few years. These events disrupt global climate patterns, altering precipitation \cite{Trenberth2020} and cloud coverage worldwide \cite{liu2023opposing} with impacts on ecosystems, fisheries, and human societies.

Accurately modeling and predicting ENSO, as well as assessing its response to climate change, remains a crucial challenge in climate science  \cite{Timmermann2018_NatureENSO,Cai2021}. Simple dynamical models have proven remarkably effective in capturing key features of this complex phenomenon, despite its inherent complexity and incomplete understanding \cite{Jin2020}. Our analysis builds on this tradition, offering a novel perspective on one of these models by leveraging recent advances in stochastic modeling, specifically state-dependent jump processes \cite{Chekroun_al22SciAdv}. More precisely, our focus is on the Jin's ENSO recharge oscillator model which has proven to be an instrumental  building block in recent breakthroughs achieved in ENSO forecasts compared to global climate models and the most skilful artificial intelligence forecasts  \cite{zhao2024explainable,vialard2025nino}.

\subsection{The Jin's recharge oscillator model of ENSO}\label{Sec_Jin_model}
Thus, we consider the Jin's recharge oscillator model \cite{jin1997equatorial} which is 
a coupled model for the SST anomaly, $T$, averaged over the central to eastern equatorial Pacific and the thermocline depth anomaly in the western Pacific $h$. The model reads as
\begin{align}\label{Eq_Jin}
    \begin{split}    
         \frac{\mathrm{d}h}{\mathrm{d}t} &= -r h - \alpha b T + p_1(t) \\
         \frac{\mathrm{d}T}{\mathrm{d}t} &= RT + \gamma h - e \left(h + bT \right)^3 +p_2(t).
    \end{split}
\end{align}

When the perturbation terms are zero ($p_i =0$), this model is based on the concept of the ocean's "recharge" process, whereby heat content is stored and released in the equatorial Pacific basin.  
The model divides the equatorial Pacific into two pools: a western pool and an eastern pool. The western pool is characterized by a deeper thermocline ($h$), while the eastern pool has a shallower thermocline. The model emphasizes the interplay between ocean and atmosphere. Changes in sea surface temperature (SST) in one pool influence wind patterns, which, in turn, affect the thermocline depth in the other pool. The core of the model is the concept of "recharge," where warm water accumulates in the western pool. Eventually, this warm water is "discharged" eastward, leading to a warming of the eastern Pacific and the development of an El Ni\~no event \cite{Timmermann2018_NatureENSO}.

The Jin model's nonlinearity reflects the nonlinear vertical distribution of the temperature in the tropical upper ocean. Such a nonlinear dependence on subsurface temperature in the (western) thermocline depth is obtained by adding a term proportional to  $-h^3$ (the eastern thermocline depth)  into the SST equation following ideas commonly employed in ENSO modeling \cite{BH89,Zebiak_al87,Neelin_al98} which through ocean-atmosphere coupling through winds  leads to the nonlinearity in the SST eastern equation of Eq.~\eqref{Eq_Jin}; see \cite[Eqns.~(2.1) and (3.1)]{jin1997equatorial} for details. This nonlinearity is crucial for capturing the oscillatory nature of ENSO.

 In its standard parameter regime (i.e., observationally fitted parameters), the model shows damping with a negative linear growth rate, where oscillations are excited by stochastic forcing. As the ocean-atmosphere coupling parameter $\mu$ is further increased (Table \ref{Table_model_param}), the model supports a limit cycle known as the ENSO fundamental oscillation \cite{jin1997equatorial} that emerges through a Hopf bifurcation. This periodic oscillation  is shown in Figure \ref{Fig_Jin_isochrons} along with a few of its isochrons  characterizing the nonlinear relaxation towards it; see Section \ref{Sec_iso_shear}.  
Our study takes place for stochastic perturbations, perturbing this limit cycle. The results presented in Sections \ref{Sec_iso_shear} and \ref{Sec_RPs_Jin} below, do not change qualitatively from the more realistic case of damped oscillations excited by stochastic forcing as dealt with in \cite{Tantet_al_Hopf} in the case of pure diffusion (no-jump), and the Hopf normal form.

 When the perturbation terms $p_i$ are stochastic, they are often sought for accounting for  
uncertainties, variability, and potential missing physics that might be difficult to capture deterministically in such simplified, conceptual models \cite{csg11}.
Intuitively, the presence of stochastic terms can help capture the intermittent nature of ENSO events, by attributing 
the  irregular occurrence of strong El Ni\~no or La Ni\~na episodes to the interplay between stochastic fluctuations and nonlinear ENSO  features \cite{roulston2000response,lengaigne2004triggering,fedorov2015impact}, encoded here by the ENSO limit cycle in the Jin's recharge oscillator model.

The plausibility of a  stochastic forcing as a mechanism for ENSO irregularity has been argued in many studies; see \cite{vialard2025nino} for a recent review. 
Physical origins of such a forcing include large-scale synoptic atmospheric transients such as the Madden Julian Oscillation  \cite{batstone2005characteristics} or wind bursts \cite{roulston2000response,fedorov2002response,lengaigne2004triggering,fedorov2015impact}. Dynamically, the idea is to explicitly separate the slow and fast modes in the atmosphere and add the latter to simple deterministic models as a stochastic forcing term.
Other sources of stochasticity include processes associated with atmospheric/moist convective disturbances whose timescales can vary from hours to weeks.  
Typically, the additional fast-mode random forcing disrupts the slow scales and convert the original periodic or damped oscillation supported by the deterministic model into an irregular one with a diversity of El Ni\~no events. Such dynamical behaviors have been previously documented through a combination of data-driven models  built from observational data \cite{penland1995optimal,kondrashov2005hierarchy,CKG11,chen2016diversity} as well as through a hierarchy of ENSO models
including general circulation models \cite{lengaigne2004triggering,fedorov2015impact}, intermediate-complexity models \cite{blanke1997estimating,chen2018observations,eckert1997predictability,thual2016simple,roulston2000response,zavala2003response}, or other conceptual models \cite{chen2022multiscale,chen2023rigorous,chekroun2024effective}.

In this study, we aim at analyzing the type of ENSO irregularity that is produced by applying a new type of stochastic parameterizations designed in   \citep{Chekroun_al22SciAdv} to produce enriched temporal variability out of periodic oscillations.

\subsection{Shear-induced chaos caused by jump-diffusion processes}\label{Sec_iso_shear}

\begin{figure}
    \centering
   \includegraphics[height=0.3\textwidth, width=.3\textwidth]{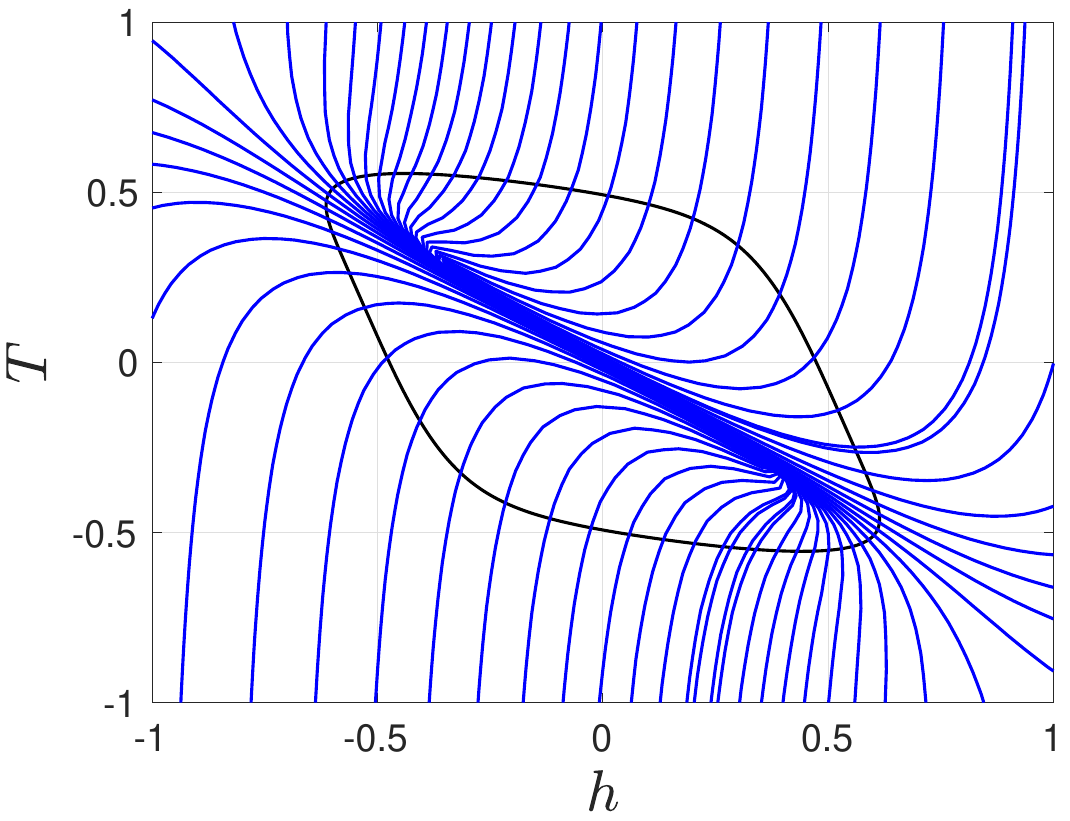}
    \caption{Jin model's limit cycle (black curve) and a few of its isochrons (blue curves). Recall that an isochron gives the  locus of all points sharing the same asymptotic phase when converging towards the limit cycle \cite{guckenheimer1975isochrons}.  The model parameters are those used in \cite{jin1997equatorial} and listed in Table \ref{Table_model_param} with $\delta=0.6$. Throughout this study,  the parameter $\delta$ is used as our  bifurcation parameter since it relates to the ocean-atmosphere coupling parameter $\mu$ used by Jin in \cite{jin1997equatorial} ($\mu=\frac{2}{3}(1+\delta)$ in this study, see Table \ref{Table_model_param}).}
    \label{Fig_Jin_isochrons}
\end{figure}

Reference \cite{Chekroun_al22SciAdv} introduced a general framework for generating solutions with enhanced temporal variability by applying stochastic perturbations to systems exhibiting a fundamental oscillation (limit cycle). The limit cycle's isochrons \cite{guckenheimer1975isochrons,Ashwin2016}, representing level sets of the oscillation's phase function, 
are pivotal in determining the system's response to spike \cite{winfree1980geometry,izhikevich2007dynamical} or  stochastic perturbations \citep{Chekroun_al22SciAdv}. The more their twist is pronounced, the more the system is susceptible to produce shear-induced chaos \cite{lin2008shear,Young2016},
with a stochastic attractor characterized by stretching and folding in the state-space, indicative of (stochastic) Smale's horseshoes. Reference \citep{Chekroun_al22SciAdv} introduces a practical method to systematically enhance stochastically the isochrons' twist,  promoting thus the emergence of such chaotic behavior.
This leads to solutions with increased temporal variability, complexity, and broader frequency spectra, improving model realism for various applications. 
         Figure \ref{Fig_Jin_isochrons} shows a few isochrons computed for the Jin model's limit cycle following the algorithm given in \cite[Chapter 10]{izhikevich2007dynamical}. We refer to \cite{guillamon2009computational,detrixhe2016fast} for other algorithms to compute isochrons in dimension greater than two (iso-surfaces). 
  
\begin{figure}[h!]
    \centering
   \includegraphics[height=0.07\textwidth, width=.4\textwidth]{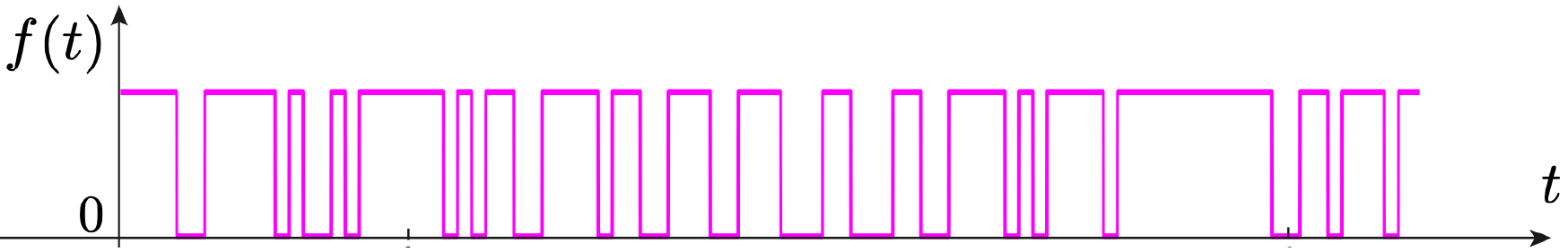}
    \caption{A realization of the comb noise $f(t)$ used in Eq.~\eqref{Eq_F}.}
    \label{fig:jump}
\end{figure}

Guided by the stochastic parameterization approach of  \cite{Chekroun_al22SciAdv}, we propose thus to apply to the Jin model (Eq.~\eqref{Eq_Jin}), the following state-dependent stochastic disturbances:
\begin{equation}\label{Eq_F} 
    \begin{split}
        p_1(t) &= \sigma \dot W^1_t - D  T \Big(h^2+T^2+a h\Big) f(t),\\
        p_2(t) &= \sigma \dot W^2_t + D  h \Big(h^2+T^2 + \beta T\Big) f(t),
    \end{split}
\end{equation}
where $f(t)$ is a jump process that corresponds to a random comb signal, where the activation events (represented by the pulses) occur at random times. More precisely, given a firing rate $f_r$  in $(0,1)$, and duration $\Delta t > 0$, we define thus $f(t)$ as the following real-valued jump process:
\begin{equation}\label{Eq_f}
f(t) = \mathds{1}_{\{\xi_n\leq f_r\}}, \quad n \Delta t \leq t < (n+1)\Delta t, 
\end{equation}
where $\xi_n$ is a uniformly distributed random variable taking values in $[0,1]$ and $\mathds{1}_{\{\xi \leq f_r\}}=1$ if and only if $0 \leq \xi\leq f_r$.   A realization of $f(t)$ is shown in Fig.~\ref{fig:jump}.  This type of jump process is also known as a dichotomous Markov noise \cite{bena2006dichotomous}, or a two-state Markov jump process in certain fields \cite{stechmann2011stochastic}. It is encountered in many applications \cite{horsthemke1984noise}.
Note that the Wiener processes $W_t^{j}$ and the jump process $f(t)$ are taken to be mutually independent.

These stochastic terms are aimed at capturing nonlinear and intermittent processes that are not represented in the simplified Jin's recharge oscillator model. Similar stochastic terms involving a Heaviside function have been considered in the recent literature on ENSO recharge oscillators \cite{vialard2025nino}, and other stochastic ENSO modeling \cite{thual2016simple}.
Such intermittent processes might include complex ocean-atmosphere interactions, the role of oceanic current and eddies, or  atmospheric teleconnections. For instance, the term multiplied by the jump process $f(t)$ in the $p_2$-term, affecting the temperature equation in Eq.~\eqref{Eq_Jin}, can be thought as caused by nonlinear and intermittent thermocline upwelling and advective feedback processes. 
In the original Jin model \cite{jin1997equatorial}, the term $-\alpha b T$ in the $h$-equation expresses a simple proportional relation between the  wind stress and SST anomalies. The multiplicative jump term in $p_1(t)$ of Eq.~\eqref{Eq_F} aims thus at accounting for extra nonlinear and feedback mechanisms between the wind stress and SST anomalies which are present in more elaborated,  spatially-extended models of ENSO; see \cite{Zebiak_al87,Jin_al93_part1,Jin_al93_part2,Dijkstra05,cao2019mathematical} and references therein.

\begin{figure*}[htbp]
    \centering
   \includegraphics[height=0.65\textwidth, width=.9\textwidth]{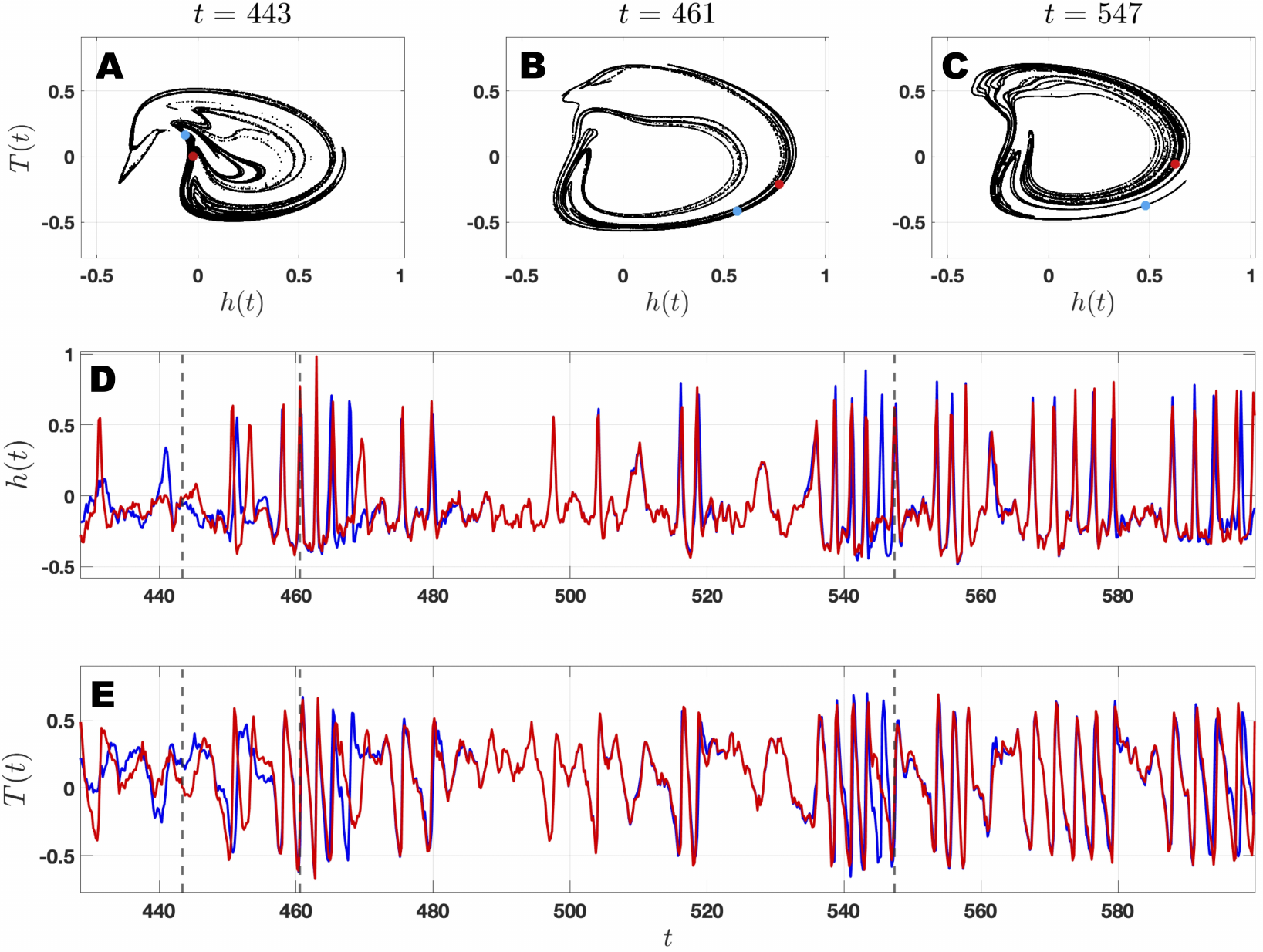}
    \caption{{\bf Stochastic strange attractor.} Here, the snapshots attractors shown in panels A, B, and C are computed from the stochastic Jin model (Eq.~\eqref{Eq_Jin_stoch}). Their time instants at which they are computed out of $10^6$ data points,  are marked by the vertical dashed lines shown in panels D and F.  These latter panels show two solutions emanating from two distinct initial conditions (blue and red curves), but driven by the same noise path, in both the $h$- and $T$-variable.   
 One observes here on-off synchronization phenomenon illustrating a well-known signature of stochastic chaos \cite{csg11}. The corresponding locations of these time series on the snapshot attractors shown in panels A, B, and C, are marked by blue and red dots. 
    The model's parameters values used in these computations are those given in Table \ref{Table_model_param} with $\delta=0.5$ while the noise parameters are given in Table \ref{Table_noise_param} (Case B).}
    \label{fig:chaoticPBA}
    \vspace{-.2cm}
\end{figure*}

 We mention that the effects of state-dependence noise on ENSO dynamics is an ongoing research topic \cite{jin2007ensemble,sardeshmukh2015understanding}, and in that sense our noise proposal in Eq.~\eqref{Eq_F} involving jump process and different state-dependences than in previous studies,   is aimed at providing new insights on this topic.

We write finally Eq.~\eqref{Eq_Jin} subject to stochastic disturbance \eqref{Eq_F}    into the following compact form:
\bea\label{Eq_Jin_stoch}
\dot{X_t}={\bm F} (X_t) + \sigma \dot{\W} +D {\bm B}(X_t) f(t),
\eea
where $X_t=(h,T)^{tr}$, $\dot{\W}=(\dot{W}_t^1,\dot{W}_t^2)^{tr}$, and the  
components of the drift term ${\bm F}=(F_1,F_2)^{tr}$ are given by
\bea\label{Eq_Fdrift}
 F_1(h,T)&=-r h - \alpha b T\\
F_2(h,T)&=RT + \gamma h - e \left(h + bT \right)^3,
\eea
while the multiplicative factor of the jump process is given by
\be\label{Eq_B}
{\bm B}(X_t)=\begin{pmatrix}
-T(h^2+T^2+ah)\\
 h(h^2+T^2+ \beta T) 
\end{pmatrix}.
\ee

As detailed in  \cite{Chekroun_al22SciAdv} in a general context, generating shear-induced chaos from Eq.~\eqref{Eq_Jin_stoch} using  perturbations of the form $D {\bm B}(\x) f(t)$ requires a careful selection of the firing rate. This rate must be sufficiently low to enable nonlinear relaxation towards the limit cycle while remaining high enough to maintain the impact of random kicks. Both aspects are essential for stretch-and-fold dynamics to take place. The perturbation magnitude $D$ also plays a critical role. In agreement with theoretical predictions of \cite{Chekroun_al22SciAdv}, appropriate combinations of random kicks (controlled by $f_r$ and $D$) and fast stochastic fluctuations (governed by $\sigma$) drive strong interactions between the stochastic forcing $D {\bm B}(\x) f(t)$ and the Jin model's nonlinear dynamics near its limit cycle. This interplay leads eventually to shear-induced chaos, as evidenced by the stretching and folding patterns shown in Figure \ref{fig:chaoticPBA}A-C. The specific noise parameters for this chaotic regime are listed in Table \ref{Table_noise_param} (Case B).

\begin{figure}[htbp]
    \centering
      \includegraphics[height=0.22\textwidth, width=.22\textwidth]{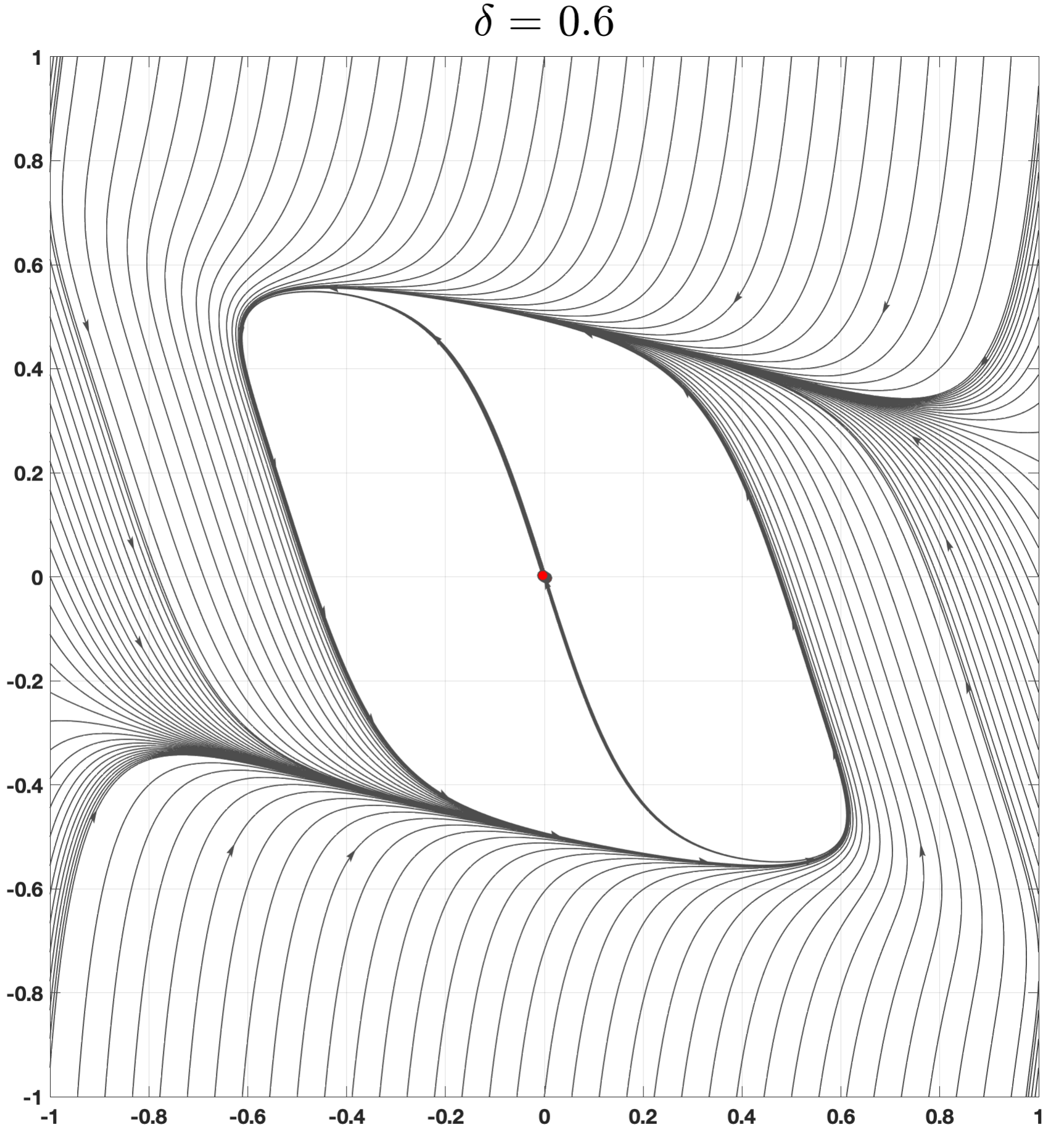}
         \hspace{-1ex}
         \includegraphics[height=0.22\textwidth, width=.22\textwidth]{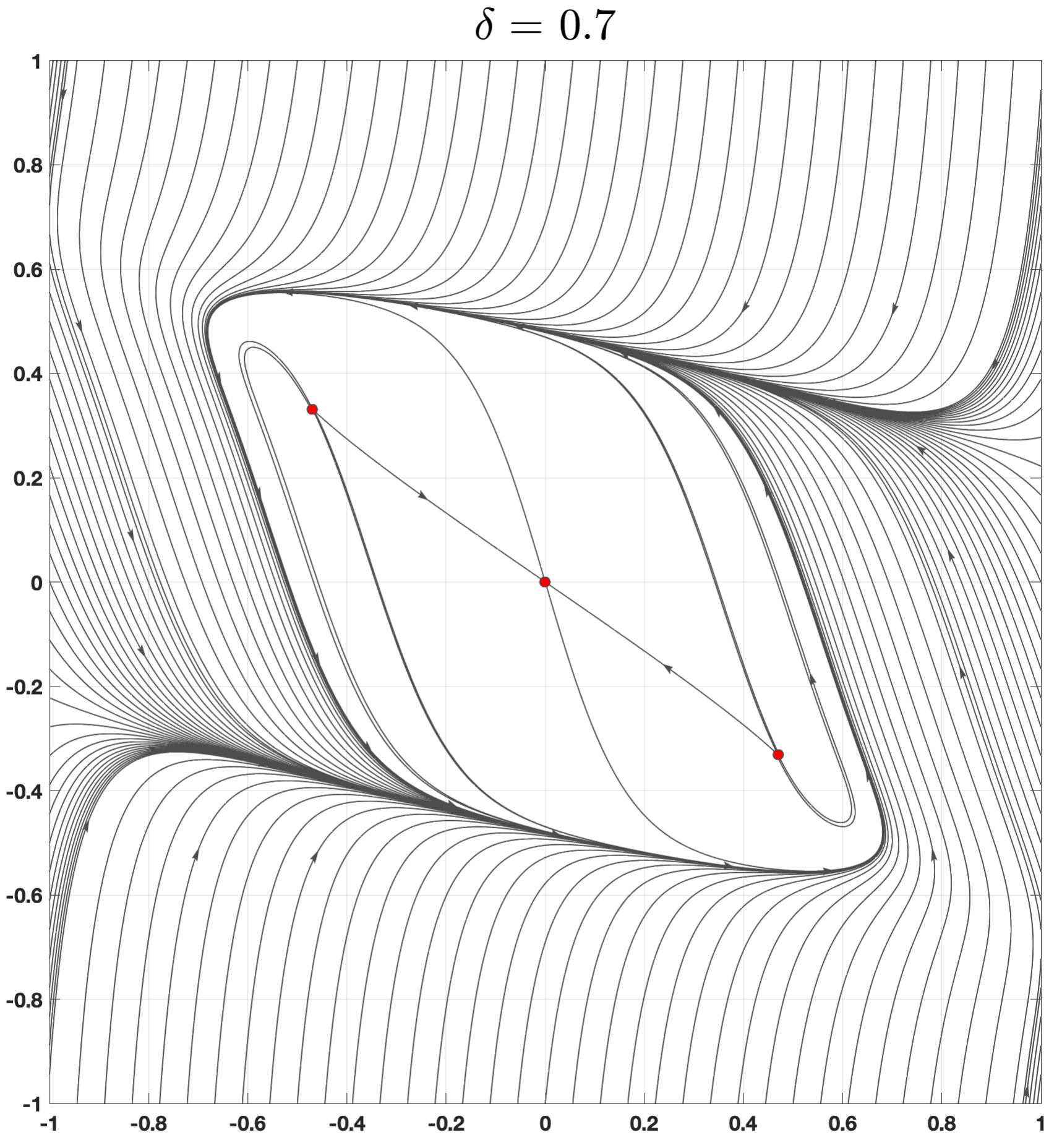}
    \caption{{\bf Jin model's vector field dependence on $\delta$}. A few streamlines of the deterministic Jin model  shown for different values of $\delta$. The unstable steady states are shown by red dots. A bifurcation of such steady states occurs between $\delta=0.6$ and $\delta=0.7$, while the stable limit cycle remains.}
    \label{Fig_Jin_bif}
    \vspace{-.1cm}
\end{figure}

\begin{table}[tbh!]
\caption{Model parameters. For the units we refer to \cite{jin1997equatorial}.}
\label{Table_model_param}
\centering
\begin{tabular}{ccccccccccccc}
\toprule\noalign{\smallskip}
$c$ & $\gamma$ & $b$ & $R$ & $r$ & $\alpha$ & $e$      \\
\noalign{\smallskip}\hline\noalign{\smallskip}
1 & $0.75$ & $\frac{5}{3}(1+\delta)$ & $\gamma b -c$  & $0.25$ & 0.125 & $0.2$   \\
\noalign{\smallskip} \bottomrule
\end{tabular}
\end{table}
Note that in our notations, the bifurcation control parameter $\mu$ used by Jin in \cite{jin1997equatorial} is recovered as $\mu=\frac{2}{3}(1+\delta)$. The parameter $\delta$ is thus our bifurcation parameter here. 
For $\delta=0.5$ used in most of our numerical experiments below the system exhibits a stable limit cycle that bifurcated from an unstable equilibrium. As noted by Jin in \cite{jin1997equatorial} the  system exhibits the emergence of two new unstable steady states as $\delta$ approaches $\delta=0.7$; see Fig.~\ref{Fig_Jin_bif}.

\begin{table}[tbh!]
\caption{Noise parameters}
\label{Table_noise_param}
\centering
\begin{tabular}{cccccccc}
\toprule\noalign{\smallskip}
 {} & $D$ & $\sigma$  & $f_r$  &  $\Delta t$ & $a$ & $\beta$ \\
\noalign{\smallskip}\hline\noalign{\smallskip}
{\bf Case A:} & $0$ & $0.1$ & N/A & N/A & N/A &N/A  \\
\noalign{\smallskip}\hline\noalign{\smallskip}
{\bf Case B:} & $(3.5)^2$ & $0.1$ & 0.7 & $10^{-2}$ &  0.7 & 0.1\\
\noalign{\smallskip} \bottomrule
\end{tabular}
\end{table}

Our goal is to provide new insights on this type of stochastic chaos by computing the underlying Kolmogorov modes, linear response and Green functions from statistical physics. To do so, we generalize below these concepts to L\'evy diffusion processes from which shear-induced chaos becomes a special case (Section \ref{Sec_jump-driven_dynamics}). 
As shown in Section \ref{Sec_linear_response_Green}, the theory is particularly useful for predicting via linear response theory how changes 
in ocean-atmosphere coupling (i.e.~in the parameter $\delta$) affect the system's behavior.


\subsection{Kolmogorov operator of the stochastic Jin's model}\label{Sec_jump-driven_dynamics}
We go back now to the case of the stochastic Jin's model given by Eq.~\eqref{Eq_Jin_stoch} and write down the corresponding Kolmogorov-L\'evy operator.  The difference with what precedes lies in the state-dependent form of the jump perturbations. Still, the formalism recalled above allows us to deal with this situation with ease as explained below. 

To do so, we first write down the Kolmogorov operator associated with Eq.~\eqref{Eq_Jin_stoch} in the case $D=0$ (no jump). It is given by (see Eq.~\eqref{Eq_Kop}) 
\be\label{Eq_Kolmo}
K_\sigma= {\bm F_1}(h,T) \partial_h + {\bm F_2}(h,T) \partial_T +\frac{\sigma^2}{2} \Delta, 
\ee
where $\Delta=\partial_h^2+ \partial_T^2$. 

Since in Eq.~\eqref{Eq_Jin_stoch}, the (scalar) jump perturbations are multiplied by the vector $D{\bm B}(\x)$, the Kolmogorov operator Eq.~\eqref{Eq_Kolmo2_general} takes here the form
\begin{widetext}
\be\label{Eq_Kolmo2}
\mathcal{L}_K \psi (\x) =K_\sigma \psi(\x)+ \int_{s\in \mathbb{R}\backslash \{0\}}\bigg(\psi(\x+s D{\bm B}(\x))-\psi(\x)-sD\mathds{1}_{\{|s|<1\}} \nabla \psi(\x)\cdot {\bm B}(\x)\bigg) \nu(\d s),
\ee
\end{widetext}
where the jump measure  $\nu$ is the measure associated with the real-valued jump process $f(t)$.
The latter is the Dirac delta $\lambda \delta_1$, where $\lambda$ is the intensity of the Poisson process associated with the jump process. Recall that this  intensity is the limit of the probability of a single event occurring in a small interval divided by the length of that interval as the interval becomes infinitesimally small. 

Consider a small time interval $(t,t+s)$. The probability of an event occurring in this interval is given by the probability that the random variable $\xi_n$ falls within the range $[0, f_r]$. Since $\xi_n$ is uniformly distributed, this probability is simply $f_r$. The expected number of events in the interval $(t, t+s)$ is the product of the probability of an event and the length of the interval. Therefore, the expected number of events is $ s \times f_r$ and $\lambda=f_r$.This means that the expected number of activation events in any given time interval is proportional to the length of that interval and the firing rate. 

As a consequence, Eq.~\eqref{Eq_Kolmo2} simplifies to
\be\label{Eq_Kolmo_Jin}
\mathcal{L}_K \psi (\x) =K_\sigma \psi(\x)+f_r \bigg(\psi(\x+{\bm B}(\x))-\psi(\x)\bigg).
\ee
The next section (Section \ref{Sec_RPs})  introduces the notions of Ruelle-Pollicott (RP) resonances and Kolmogorov modes for general Kolmogorov operators $\mathcal{L}_K $ such as  given by Eq.~\eqref{Eq_Kolmo2_general}. 
The particular case of the Kolmogorov operator given by Eq.~\eqref{Eq_Kolmo_Jin} is dealt with in Section \ref{Sec_RPs_Jin}.

\subsection{Stochastic Jin model's RP resonances and Kolmogorov modes}\label{Sec_RPs_Jin}    
 In this Section, we apply the Ulam's approach recalled in Section \ref{Sec_Ulam} to simulations of
the stochastic Jin model (Eq.~\eqref{Eq_Jin_stoch}) obtained via an Euler-Maruyama scheme with a time-step $\delta t= 10^{-2}$.
Two cases of stochastic disturbances are considered as listed in Table \ref{Table_noise_param}.  
The underlying time series $\x_{k}$ are obtained by collecting $N_d=10^7$ sequential data points (after transient removal) by sampling  every $10$ time-steps numerical solutions to Eq.~\eqref{Eq_Jin_stoch} made of $10^8$ data points.
Markov transition matrices $M_\tau$ with $\tau=3$ and built up out of $N_g=5746$ cells shadowing the resulting cluster of data points in the $(h,T)$-plane, are then estimated according to the Ulam's procedure recalled above. The dominant spectral elements are computed using standard routines. The corresponding eigenvalues are shown below after taking the logarithm.

\begin{figure}[htbp]
    \centering
       \includegraphics[height=0.25\textwidth, width=.5\textwidth]{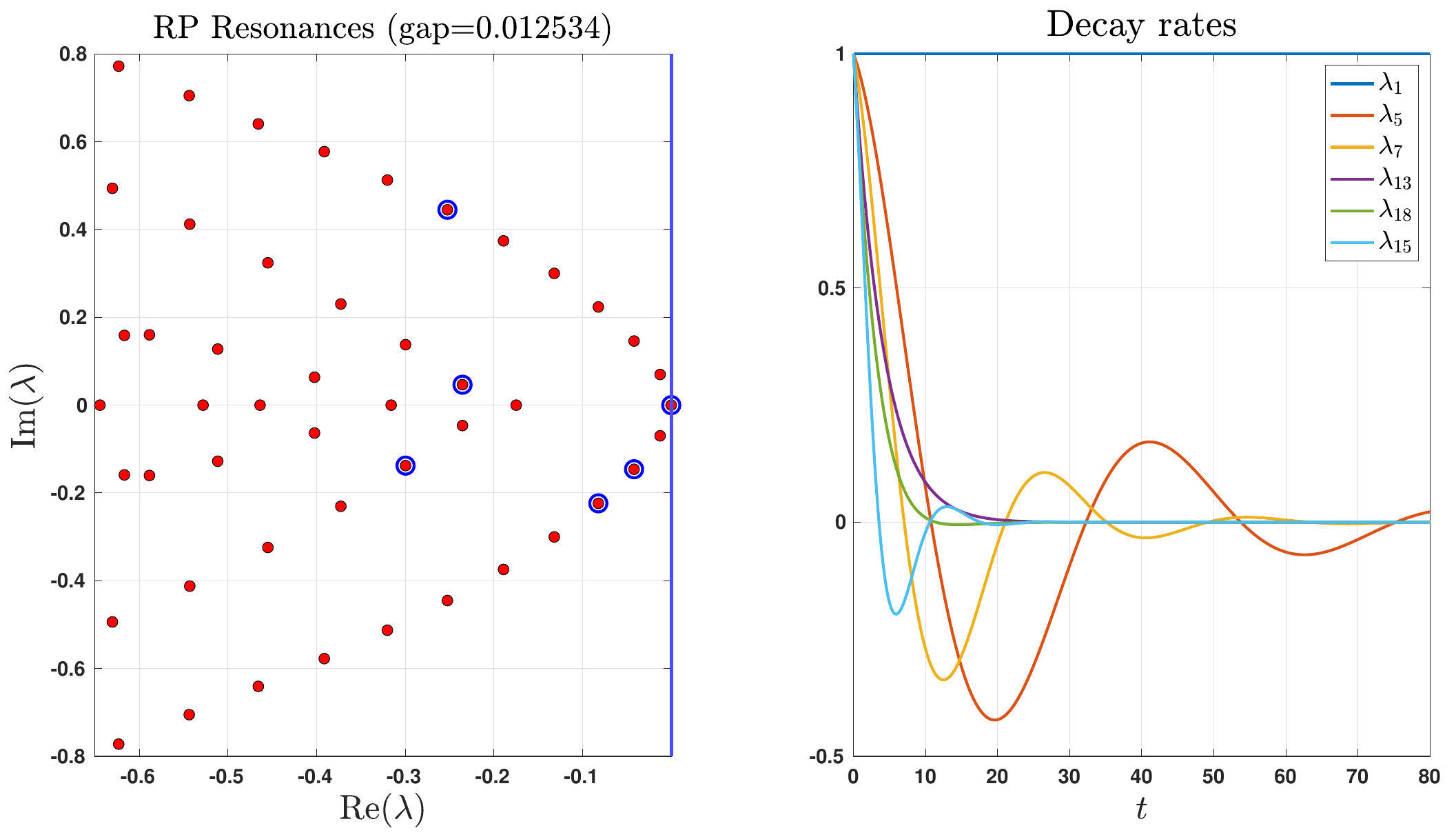}
    \caption{{\bf RP resonances and decay rates.} Here, shown for the Kolmogorov operator $K_\sigma$(Eq.~\eqref{Eq_Kolmo}) in a purely diffusive case, $D=0$; Case A in Table  \ref{Table_noise_param}. The RP resonances share  features  (parabola and triangular shapes) exhibited by an Hopf normal form in presence of small additive white noise  \cite{Tantet_al_Hopf}. The light-blue line in the right panel corresponds to the non-decaying eigenvalue $\lambda_1=0$. The vertical blue line in the left panel indicates the imaginary axis.}
    \label{Fig_RP_res_diffusion}
\end{figure}

\begin{figure}[htbp]
    \centering
        \includegraphics[height=0.3\textwidth, width=.48\textwidth]{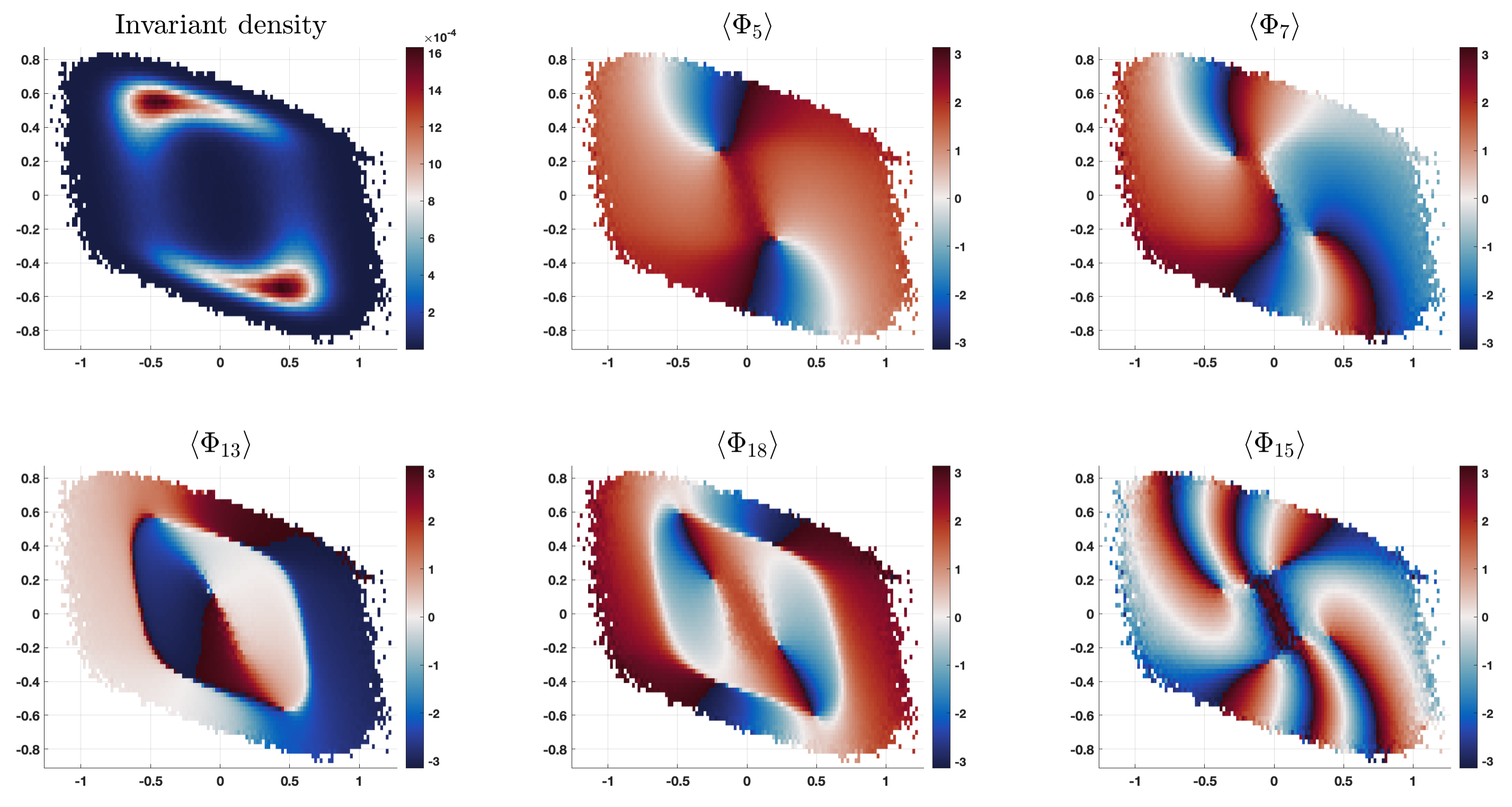}
    \caption{{\bf Kolmogorov Modes.} A few Kolmogorov modes associated with the RP resonances circled in blue in Fig.~\ref{Fig_RP_res_diffusion}.}
    \label{Fig_Kolmo_modes_diffusion}
\end{figure}

\begin{figure}[htbp]
    \centering
        \includegraphics[height=0.25\textwidth, width=.48\textwidth]{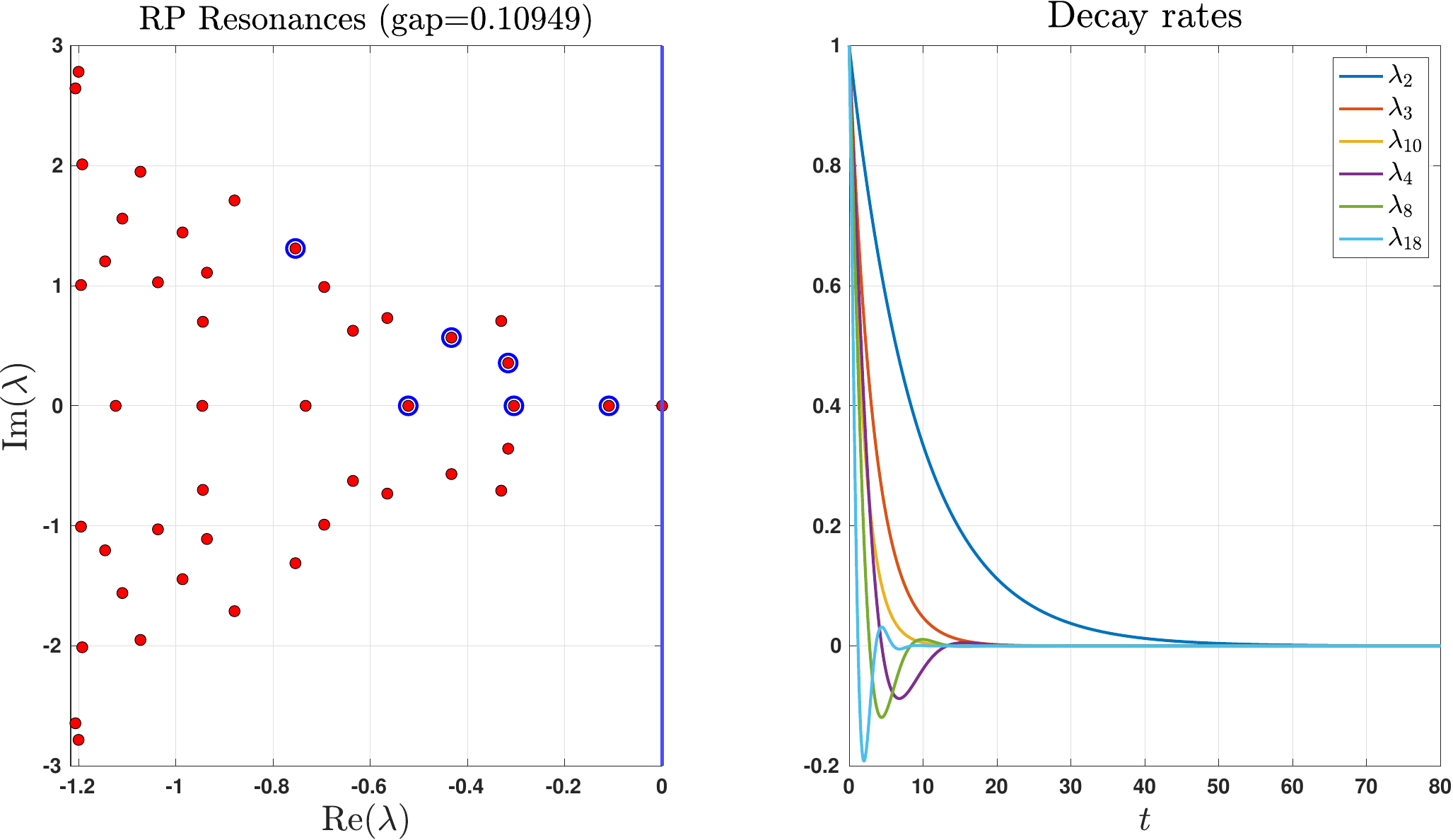}
    \caption{{\bf RP resonances and decay rates.} The model and noise parameters correspond to the shear-induced chaos shown in  Fig.~\ref{fig:chaoticPBA} (Case B in Table  \ref{Table_noise_param}). The spectral gap $\mathfrak{g}=\text{Re}(\lambda_2)$ (distance to the imaginary axis of the rightmost non-zero RP resonance) is noticeably increased here compared to Case A (Fig.~\ref{Fig_Kolmo_modes_diffusion}), indicative of a stronger mixing in the state space and faster decay of correlations.}
    \label{Fig_RP_resonances_twist}
\end{figure}

\begin{figure*}[htbp]
    \centering
        \includegraphics[height=0.4\textwidth, width=.9\textwidth]{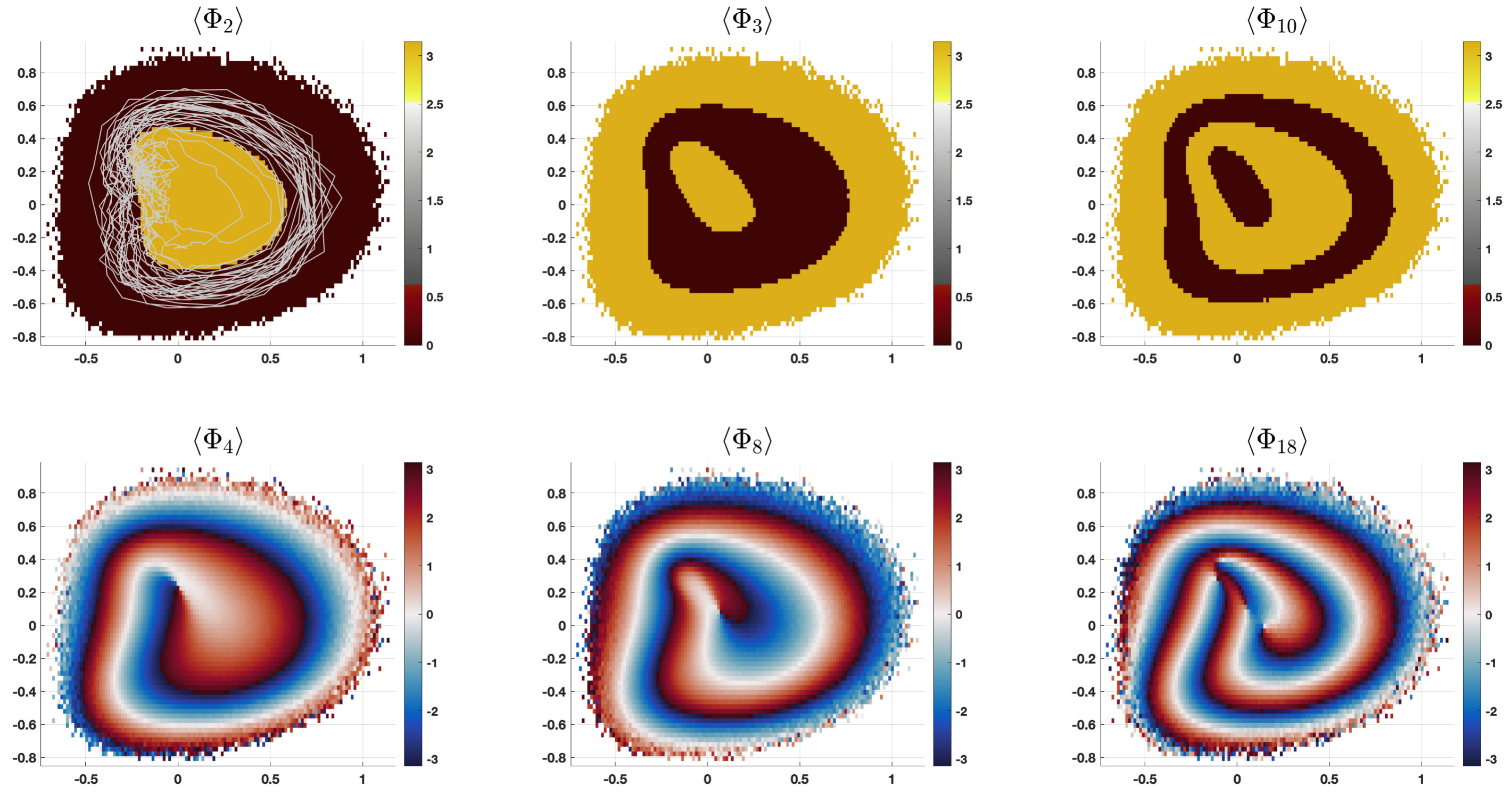}
    \caption{{\bf Kolmogorov Modes.} A few Kolmogorov modes associated with the RP resonances circled in blue in Fig.~\ref{Fig_RP_resonances_twist}, that is for the case of the shear-induced chaos of  Fig.~\ref{fig:chaoticPBA}. The top row shows the phase of real modes while the bottom row shows that of complex modes.   A segment of the time series shown in Fig.~\ref{fig:chaoticPBA} is here superimposed on $\Phi_2$ as the thin grayish curve. It mainly resides within the brown region corresponding to the most visited region by the dynamics. The complex modes' phase exhibits spiraling patterns reflecting the underlying stretch-and-fold dynamics displayed by the stochastic attractor shown.}
    \label{Fig_Kolmo_modes_twist}
\end{figure*}

The results are shown in Figures \ref{Fig_RP_res_diffusion}-\ref{Fig_Kolmo_modes_diffusion} for Case A, and  in Figures \ref{Fig_RP_resonances_twist}-\ref{Fig_Kolmo_modes_twist}, for Case B. 
The phase (in $(0, 2\pi)$) of the Kolmogorov modes are shown for the RP resonances circled in blue. 
Note that in Case A ($D=0$) one recovers the parabolic shape formed by the dominant resonances, enveloping a triangular array of resonances as in the case of the Hopf normal form subject to a small additive white noise \cite{Tantet_al_Hopf}. As explained in \cite{Tantet_al_Hopf}, the triangular array of resonances is associated with the unstable steady state while the parabola is associated with the random stable limit cycle.  
The phase of two modes composing this triangular array are shown as $\langle \Phi_{13}\rangle$ and $\langle\Phi_{18}\rangle$ in  Fig.~\ref{Fig_Kolmo_modes_diffusion}.  Their structures differ notably from those associated with the parabola of resonances. The latter modes whose phase is shown  as  $\langle \Phi_5\rangle$, $\langle\Phi_7\rangle$, and  $\langle\Phi_{15}\rangle$ in Figure \ref{Fig_Kolmo_modes_twist} are characterized by level sets organized by the unperturbed system's isochrons (see Fig.~\ref{Fig_Jin_isochrons})---here orthogonal to these, as already pointed in \cite{Tantet_al_Hopf} for the case of the Hopf normal form. Note that the isochrons are known to provide exactly the level sets of certain Koopman eigenfunctions for pure deterministic oscillations \cite{mauroy2012use}.

The modes and resonances associated with Case B, i.e.~when the state-dependent jump-noise is activated, show different attributes. The spectral gap ($\mathfrak{g}=\text{Re}(\lambda_2)$) is noticeably increased in Case B compared to Case A, indicative of a stronger mixing in the state space and faster decay of correlations; cf Figure \ref{Fig_RP_res_diffusion} and \ref{Fig_RP_resonances_twist}. We observe also that the parabola and triangular array formed by the resonances in Case A are now destroyed, giving rise to another discrete geometry of the spectrum. 
A remarkable feature though is the presence of real resonances that extend until the deep part of the spectrum as one moves leftward in the complex plane. The corresponding real modes display positive and negative areas that are better revealed when plotting their phase as shown in the top row of Figure \ref{Fig_Kolmo_modes_twist}, with phase equals to $\pi$ corresponding to negative values (yellow zones) and phase equals to $0$ corresponding to positive values (brown zones).    

Mode $\Phi_2$, associated with the slowest decaying eigenvalue $\lambda_2$ in Figure \ref{Fig_RP_resonances_twist}, reveals two distinct regions. The brown region corresponds to the most frequently visited area of the state space by the dynamics of Eq.~\eqref{Eq_Jin_stoch}. Moving leftward in the complex plane, the real modes delineate concentric regions of positive and negative values, corresponding to varying sojourn time statistics. While a detailed analysis of these statistics is beyond this paper's scope, the capability of real Kolmogorov modes to identify regions with potentially rare events (characterized by short sojourn times) warrants attention.

The phase of the (purely) complex Kolmogorov modes reveal striking features. Their phase is noticeably exhibiting spiraling patterns reflecting  the underlying stretch-and-fold dynamics displayed by the stochastic attractor (Figure \ref{fig:chaoticPBA}A-C), with the order of these spiraling patterns (its ``winding number'') that increases as one moves  leftward in the complex plane; see bottom row of Figure \ref{Fig_Kolmo_modes_twist}. 
Overall, they also account for the rotation imposed by the specific form of the state-dependent component in Eq.~\eqref{Eq_F}, i.e.~the ${\bm B}$-term (Eq.~\eqref{Eq_B}).

Recall that $T(t)=e^{t\mathcal{L}_K}$ provides the deterministic macroscopic description of the averaged effects of the combined L\'evy flights and Brownian motions (see Eq.~\eqref{Eq_Tt}).  By using the decomposition theorem of $C_0$-semigroups in our context \cite[Theorem V.3.1]{Engel_Nagel} and since the RP resonances are here simple, we obtain, by summing up over the $N$ dominant RP resonances, that
\be\label{Eq_decomp_semigroup}
T(t)=\sum_{j=1}^N  e^{\lambda_j t}\langle \Phi_j^\ast|_{\mu}   +R_N(t),
\ee
with $R_N(t)$ decaying exponentially to $0$ at a controlled rate as $N\rightarrow\infty$; see \cite[Theorem 1]{Chekroun_al_RP2}. Here $\langle \Phi_j^\ast|_{\mu}$ denotes the weighted "half"-inner  product  with respect to the system's (ergodic) invariant measure $\mu$, by the  $j^{th}$-Kolmogorov mode of the Fokker-Planck operator given by the RHS of Eq.~\eqref{Eq_FKPE_nonlocal}, and associated with the RP resonance $\lambda_j$.    

Thus for any smooth observable $\psi$, since $T(t)\psi$ is also equals to $\int_{\mathbb{R}^d} \psi(\x) p(t,\x) \d \x$ with $p$ solving the  Fokker-Planck equation  Eq.~\eqref{Eq_FKPE_compact}, we arrive at the formula
\be
\int_{\mathbb{R}^d} \psi(\x) p(t,\x) \d \x=\sum_{j=1}^N  \langle \Phi_j^\ast|\psi\rangle_{\mu} e^{\lambda_j t} +\epsilon_N(t),
\ee
 with $\epsilon_N(t)$ decaying exponentially to $0$ at a controlled rate as $N\rightarrow\infty$. In particular, this formula provides the rate of convergence of any moment statistics (for instance by taking $\psi(\x)=\|\x\|^p$) and allows for identifying the Kolmogorov mode(s) controlling this rate of decay.  This semi-analytical/empirical observation hints towards an actual exponential ergodicity for the actual stochastic Jin's model Eq.~\eqref{Eq_Jin_stoch} whose rigorous proof goes beyond the scope of this paper; see \cite{masuda2004multidimensional,kulik2009exponential,sandric2016ergodicity} for theoretical frameworks.

Recall also that the Kolmogorov modes provide the characteristic patterns in the state space, that encode the system's variability due to the decomposition formula of power spectra Eq.~\eqref{Eq_PSD}.
Adopting the language of climate change, they provide the modes of the system's natural variability, prior a perturbation is applied and are thus of utmost importance.

Note that although the Ulam's approach enables us to reveal with accuracy key features of the Kolmogorov modes here, it performs poorly to resolve the modes' details at the periphery of the available data points. 
Other data-driven approaches using e.g.~the extended dynamic mode decomposition could be used to address this issue \cite{williams2015data,klus2018data}, although such alternative method could encounter other difficulties boiling down to the appropriate choice of basis functions, tied to the intricate spiraling structures exhibited by the (purely complex) Kolmogorov modes lying in the deeper part of the spectrum.

The next section highlights another key property of Kolmogorov modes:
 their ability to encode the system's response to general perturbations of the drift term  (${\bm F}$)  through Green functions.

\subsection{Linear response of shear-induced chaos caused by jump-diffusion}\label{Sec_linear_response_Green}
In this section, we apply the  linear response framework of Section \ref{Sec_Green_Kolmo} to the 
jump-diffusion Jin model given by  
Eqns.~\eqref{Eq_Jin_stoch}-\eqref{Eq_B}.
We are interested in predicting 
 the response of this jump-diffusion model to small perturbations of the drift term ${\bm F}$ in this model.

We consider the physically relevant situation where the ocean-atmosphere coupling parameter $\delta$ is allowed to be varied as $\delta\rightarrow\delta +\epsilon g(t)$.  This parameter variation impacts the two components of the drift ${\bm F}$ in Eq.~\eqref{Eq_Fdrift} as follows: 
\begin{subequations} 
\begin{align}\label{deltadrift}
    F_1 \to F_1 &- \epsilon g(t) \alpha \frac{5}{3} T \\ 
    F_2 \to F_2  &+ \epsilon g(t) \frac{5}{3}T  \left( \gamma - 3 e \left(h + b T \right)^2 \right)\nonumber \\&+ \epsilon^2 g(t)^2\frac{25}{3} \left(h+bT \right)T^2\nonumber \\& + \epsilon^3g(t)^3 \left( \frac{5}{3}T\right)^3.\label{deltadriftb}
\end{align} 
\end{subequations}

At first sight, due to the presence of higher-order terms in $\epsilon$,  this choice of parameter variation does not seem to fit within  the linear response framework of 
Section \ref{Sec_Green_Kolmo}, in which ${\bm F}(\x)$ is perturbed into ${\bm F}(\x)+\epsilon g(t){\bm G}(\x)$. However, LRT aims at assessing  the impacts on statistics at the leading order in $\epsilon$ and therefore  accounting for  contributions of the quadratic and cubic terms in $\epsilon$ from Eq.~\eqref{deltadriftb}, are  irrelevant at this level of approximation.  
We can thus theoretically frame our experiments as corresponding, 
up to linear terms in $\epsilon$, to the following perturbation:
\beas
  F_1 \to F_1 &+ \epsilon g(t)G_1,  \mbox{ with }  G_1 =  -  \alpha \frac{5}{3} T\\ 
    F_2 \to F_2  &+ \epsilon g(t) G_2,  \mbox{ with }  G_2 =  \frac{5}{3}T  \left( \gamma - 3 e \left(h + b T \right)^2 \right).
\eeas

 We first adopt an empirical, direct approach following  \cite{gritsun2017} to estimate the system's response to parameter variations.  In that respect, we sample $M$ points from a long unperturbed system run  ($\epsilon=0$), distributed according to the invariant measure $\mu $. For each ensemble member, we apply a perturbation $\delta \to \delta + \epsilon  g(t)$ and estimate the response $\Delta \langle \Psi \rangle_{g, \delta + \epsilon} = \langle  \Psi \rangle_{g, \delta + \epsilon} - \langle \Psi \rangle_0$, where $ \langle  \cdot \rangle_{g, \delta + \epsilon}$ denotes the ensemble average over the perturbed system relative to the forcing $g(t)$ and amplitude $\epsilon$.

Then, following  \cite{gritsun2017}, we extract the linear component of the response using a centered difference approximation:
\begin{equation}
\label{eq: estimation linear response}
	\delta_\epsilon^{(1)}[\Psi] (t) \approx \frac{\Delta \langle \Psi \rangle_{g, \delta + \epsilon} - \Delta \langle \Psi \rangle_{g, \delta - \epsilon} }{2}.
\end{equation}
For small but finite $\epsilon$, this approximation effectively eliminates the quadratic response term and provides a robust estimate of the linear component.

For a given forcing $g(t)$, we implement two sets of response experiments with amplitudes $\pm \epsilon$. Extracting a linear response signal from numerical experiments involves careful considerations and trade-offs. The perturbation amplitude $\epsilon$ should be large enough to maximize the signal-to-noise ratio but small enough to avoid nonlinear effects.  A large ensemble size $M$ is also crucial to reduce fluctuations arising from the system's  chaotic dynamics and its inherent stochasticity tied to jump process. Generally, smaller perturbation amplitudes require larger ensemble sizes to confidently extract a linear response signal.

We apply this empirical approach to the case of a step perturbation $g(t)= \Theta(t)$ \cite{Lucarini2017}, corresponding to a sudden change in the ocean-atmosphere coupling parameter. We evaluate the system's response for different perturbation magnitudes, $\epsilon = 0.02  k \times  \delta$, where $k=1,\cdots,5$.

\begin{figure}[htbp]
    \centering
    \includegraphics[height=0.4\textwidth, width=.5\textwidth]{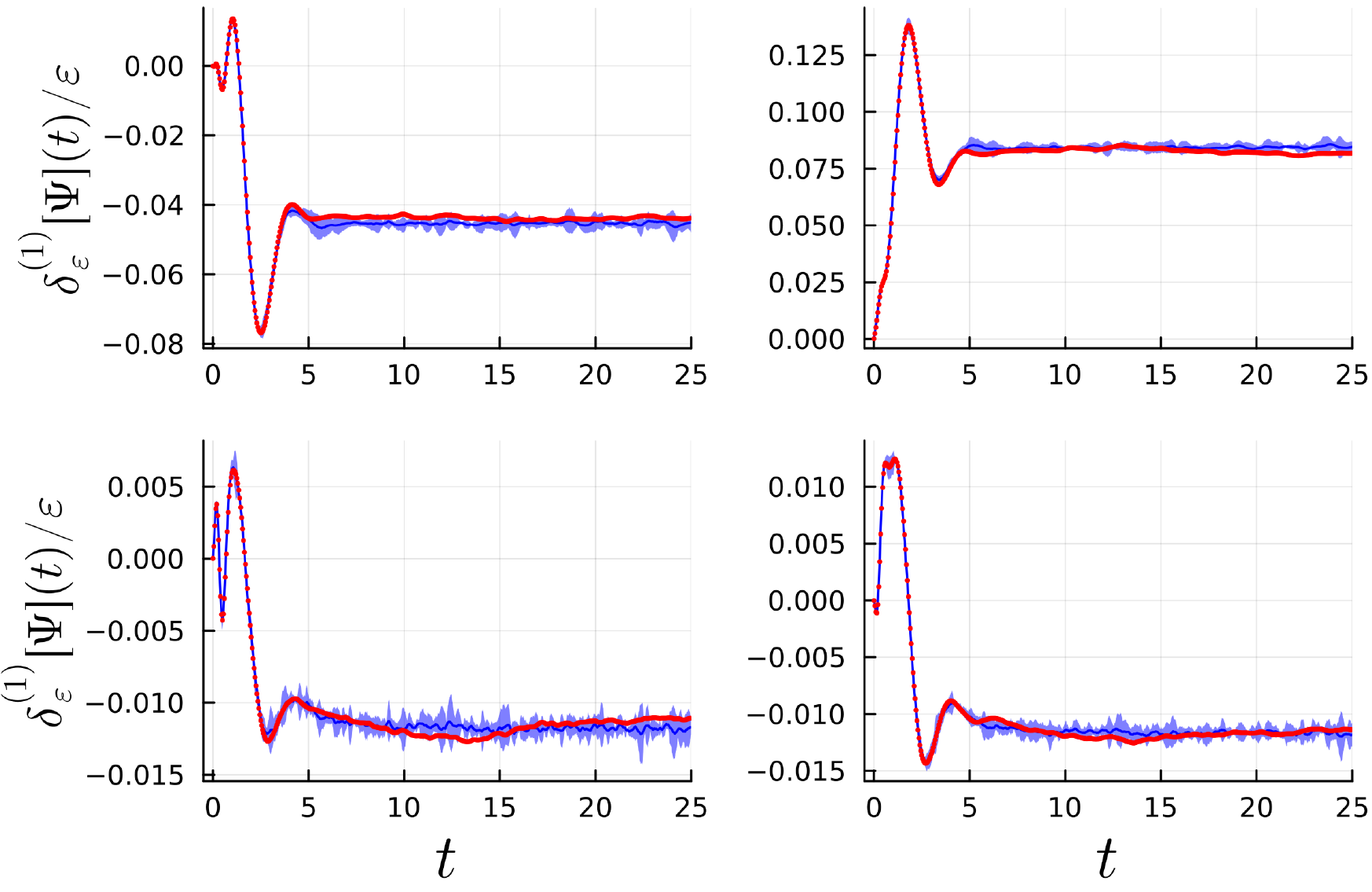}
    \caption{Linear response to a step perturbation for different observables of the system. From top to bottom, left to right $\Psi= h, T, h^2 , T^2$. The blue curves (and shaded area) show the centered difference approximations of the response as given by Eq.~\eqref{eq: estimation linear response}. The red dots show the system's response through the Green's function formalism of Section \ref{Sec_Green_fcts}; see Text.}   
    \label{fig:Linear Response}
\end{figure}

Figure  \ref{fig:Linear Response}  shows that the linear response can be reliably extracted for perturbations up to $10\%$ of the original $\delta$ value. This indicates a robust range of validity for the linear response approximation. The blue curve represents the mean rescaled response $\delta_\epsilon^1[\Psi](t)/\epsilon$  across different $\epsilon$ values, while the shaded area denotes the fluctuations estimated, at each point in time, as $\pm  (\max_{\epsilon} \delta_\epsilon^1[\Psi]/\varepsilon- \min_{\epsilon}\delta_\epsilon^1[\Psi]/\varepsilon)$. These results confirm that the system (Eq.~\eqref{Eq_Jin_stoch}), despite its discontinuous jump process component (${\bm B}(\x) f(t)$), exhibits smooth statistical dependence on the perturbation parameter $\epsilon$ and, consequently, on the ocean-atmosphere coupling parameter $\delta$.

To further validate LRT, we compare these empirical results with the Green function formalism of Section \ref{Sec_Green_fcts}. In that respect, we compute the Green's function of the system, from carefully designed response experiments and use it to predict the response to a step perturbation. According to LRT, the response to any time-dependent forcing $g(t)$ can be reconstructed from the Green's function $ \G_{\Psi,G}(t)$ as given by Eq.~\eqref{Eq_LRF}.

To estimate the Green's function, we apply a short-duration perturbation $g(t) = 1/\delta t$ over a single time step $\delta t$. By considering perturbations of different amplitudes, we reliably estimate $ \G_{\Psi,G}(t)$ (not shown). The response to a step perturbation  $g(t) = \Theta(t)$ is then calculated using:
\begin{equation}
    \delta^{(1)}[\Psi] (t) = \int_{0}^{t} \G_{\Psi,G}(s) \mathrm{d}s,
\end{equation}
due to Eq.~\eqref{Eq_LRF}. 

Figure  \ref{fig:Linear Response} demonstrates excellent agreement between the brute-force response (blue curve) and the estimate obtained from the Green's function (red dots), further supporting the validity of linear response theory.

The state-dependent jump process ${\bm B}(\x) f(t)$ and its interaction with the system's nonlinearity can introduce persistent fluctuations, making it difficult to isolate the smooth response properties. To reliably estimate the linear response, we employed a large ensemble size ($M\approx 10^8$ for step perturbations and $M \approx 5 \times 10^9$ for the Green's function). Despite the effects caused by the jump process, the underlying linear response signal remains detectable.

We focused on the response of highly fluctuating variables like powers of $h$ and $T$. For spatially-extended, high-dimensional systems, estimating the Green's function for spatially averaged observables may require fewer ensemble members, as spatial averaging can mitigate jump-induced fluctuations. This is particularly relevant in cases like convective parametrization, where local conditions determine transitions between convective and non-convective states, often lacking significant spatial coherence. Recent studies on using LRT for climate response predictions have shown that increased model complexity and spatial resolution can reduce the required ensemble size \cite{Lucarini2017,Lembo2020}.

\section{Response in an Energy Balance Climate Model Driven by an $\alpha$-stable Process}\label{Ghil-Sellers}
Here, we propose a further climate-centered application of the linear response theory developed in Section \ref{Sec_Green_Kolmo} by testing whether it is possible to use linear response theory to perform accurate climate change projections in a simple yet physically relevant climate model. 

The Ghil-Sellers (GS) model \cite{Ghil1976} is a one-dimensional energy balance model (EBM) \cite{North1981} developed to understand the latitudinal distribution of temperature and its feedbacks. It is a foundational model in climate science, providing key mathematical insights into the Earth system's multistability. This includes the famous ice-albedo feedback, which explains the co-existence of warm and snowball states \cite{HoffmanSchrag}; see discussions in \cite{Bodai2015,Ghil2020}. Recently, it has been used as a test case for studying the statistical mechanical underpinnings of optimal fingerprinting for climate change detection and attribution \cite{Hasselmann1993b,Hegerl1996,Allen2003}.

The GS model is a spatially-extended system, written as a reaction-diffusion equation for the zonally-averaged surface temperature $T(x,t)$. Here, $x=2\phi/\pi$ is a normalized geographical latitude that lies in $[-1,1]$, and $t$ denotes time. The model is described by the following partial differential equation (PDE):
\begin{equation}\label{eq:pde}
\begin{split}
\partial_tT &= \mathcal{D}[x,T,\partial_xT, \partial_{xx}T]\\
&= \frac{1}{c(x)}\left(\frac{2}{\pi}\right)^2\frac{1}{\cos\left(\frac{\pi x}{2}\right)}\partial_x\left(\cos\left(\frac{\pi x}{2}\right)k(x,T)\partial_xT\right) \\
&+ \frac{1}{c(x)}\mu_s Q(x)\left(1 - \alpha(x,T)\right) \\
&- \frac{1}{c(x)}\sigma T^4\left(1 - m\tanh(c_3T^6)\right),
\end{split}
\end{equation}
where one considers with boundary and initial conditions given by $\partial_xT(-1,t) = \partial_xT(1,t) = 0$ and $T(x,0) = T_0(x)$, respectively. The nonlinear operator $\mathcal{D}$ can be explained as follows. The function $c(x)$ is the effective heat capacity of the atmosphere, land, and ocean per unit surface area at $x$. The first term describes the meridional heat transport as a diffusive law, with $k(x,T)$ incorporating the effects of sensible and latent heat transport. The second term represents the solar energy input, where $\mu_s$ is the reduced solar constant ($\mu_s=1$ for present-day climate), $Q$ is the irradiance and $\alpha$ is the albedo, whose parametrization in terms of $T$ allows for the potential onset of the tipping behaviour.

The third term describes the infrared radiation emission to space, and is described by a Boltzmann's law, modified by a factor describing the greenhouse effect, whose intensity is controlled by $m$. It will serve as our main control parameter. The model is described in \cite{Bodai2015}, where all the parameterization are carefully described.

Adding a stochastic forcing to such models is a common practice aimed at providing a qualitative representation of the effect of unresolved scales of motion \cite{hasselmann1976,LucariniChekroun2023}. In previous work \cite{Lucarini_Chekroun_PRL24}, we proved it is possible to construct accurate Green's functions for a stochastically forced version of the GS model using a Gaussian white noise field, $\eta(x,t)$, which satisfied $\langle\eta(x,t)\rangle =0$ and $\langle\eta(x,t)\eta(x',t')\rangle = C(x)\delta(x-x')\delta(t-t')$, with $C(x)>0$.

In the spirit of the current contribution, we move beyond Gaussian forcing. Here, we consider a spatio-temporal stochastic field, $\eta(x,t)$, that at each location is distributed according to an $\alpha$-stable process. This choice of forcing allows us to model intermittent and extreme events that are not captured by traditional Gaussian noise. We choose $\eta(x,t)\sim\mathcal{S}(\alpha,\beta,\gamma(x),\mu_s=\beta\gamma(x)\tan(\pi\alpha/2))$ which ensures the forcing has zero mean, preventing unphysical drifts in the temperature field. Additionally, the noise is assumed to be spatially and temporally uncorrelated, $\eta(x,t)\indep\eta(x',t')$ if $(x,t)\neq(x',t')$. The meaning of the $\alpha$-stable parameter is discussed in further detail below.

Following the methodology in \cite{Lucarini_Chekroun_PRL24}, we implement the model with a latitudinal spatial discretization of $5^{\circ}$, which results in a phase space of $M=37$ temperature variables. We represent spatial derivatives using standard centered differences and discretize time with a step of $\Delta t= 1$ day, or $86400$ s.

\begin{figure*}
    \centering
a)   \includegraphics[width=.3\textwidth]{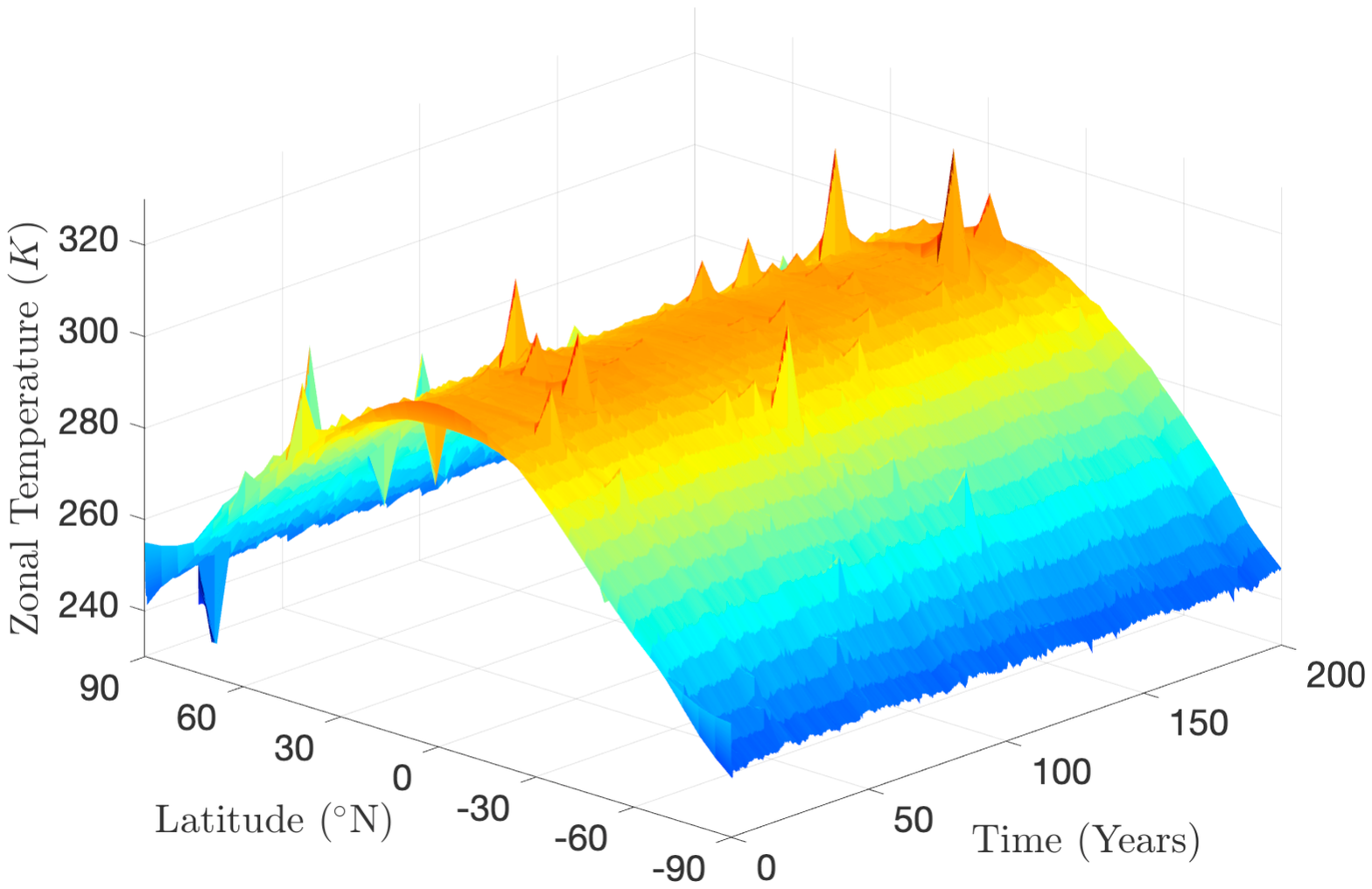}
  b)  \includegraphics[width=.3\textwidth]{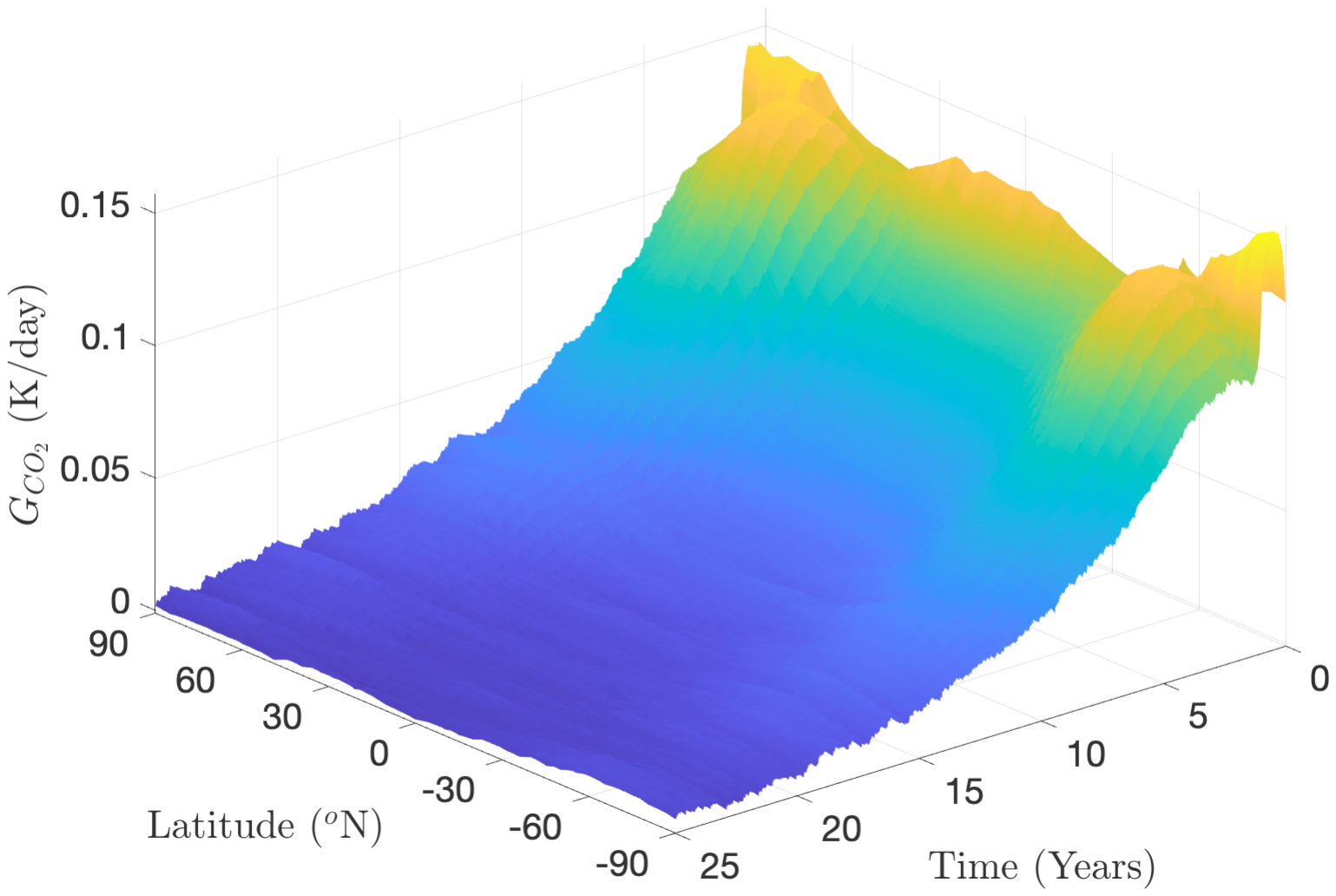}
 c)  \includegraphics[width=.3\textwidth]{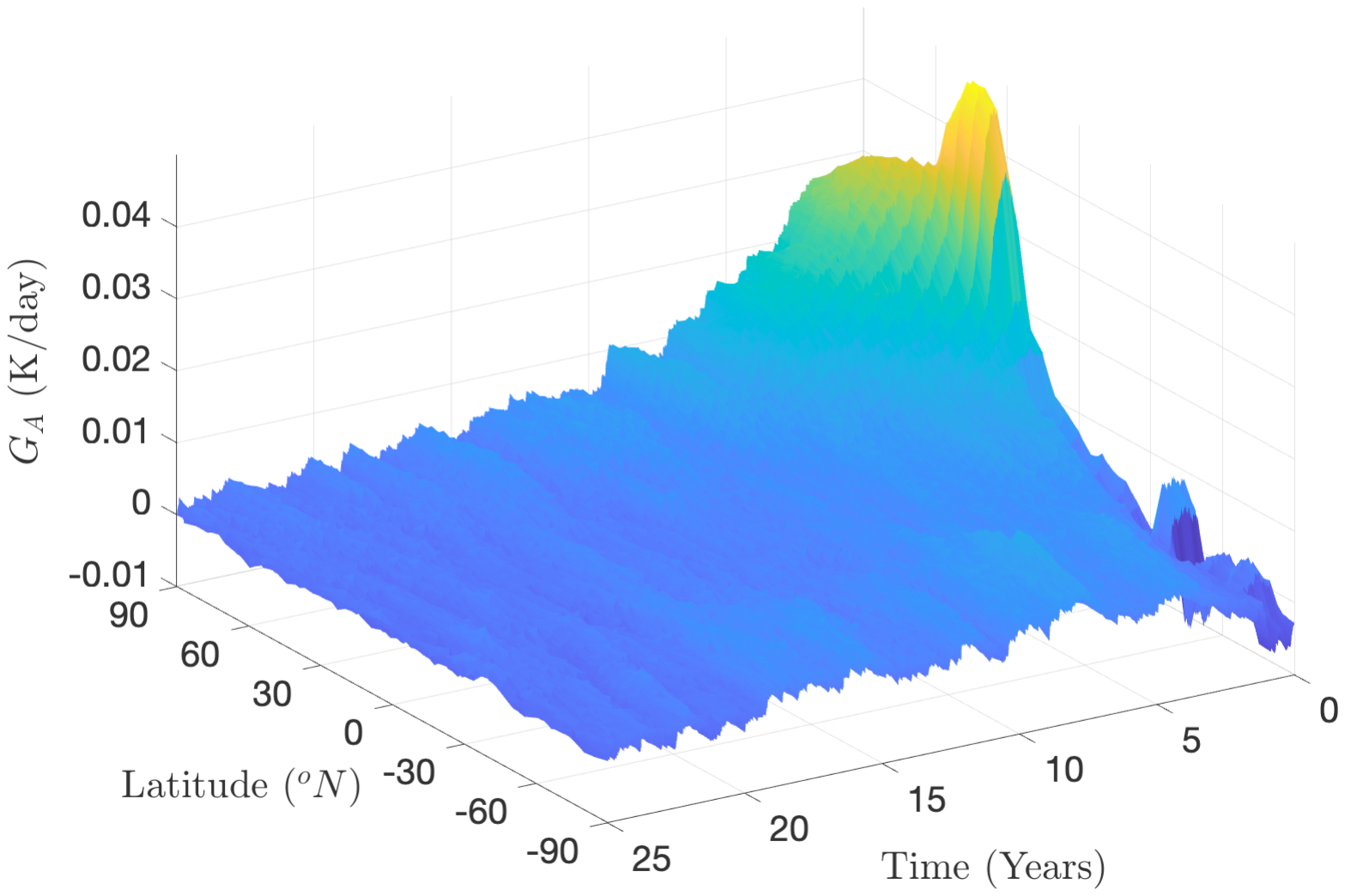}\\
e)   \includegraphics[width=.3\textwidth]{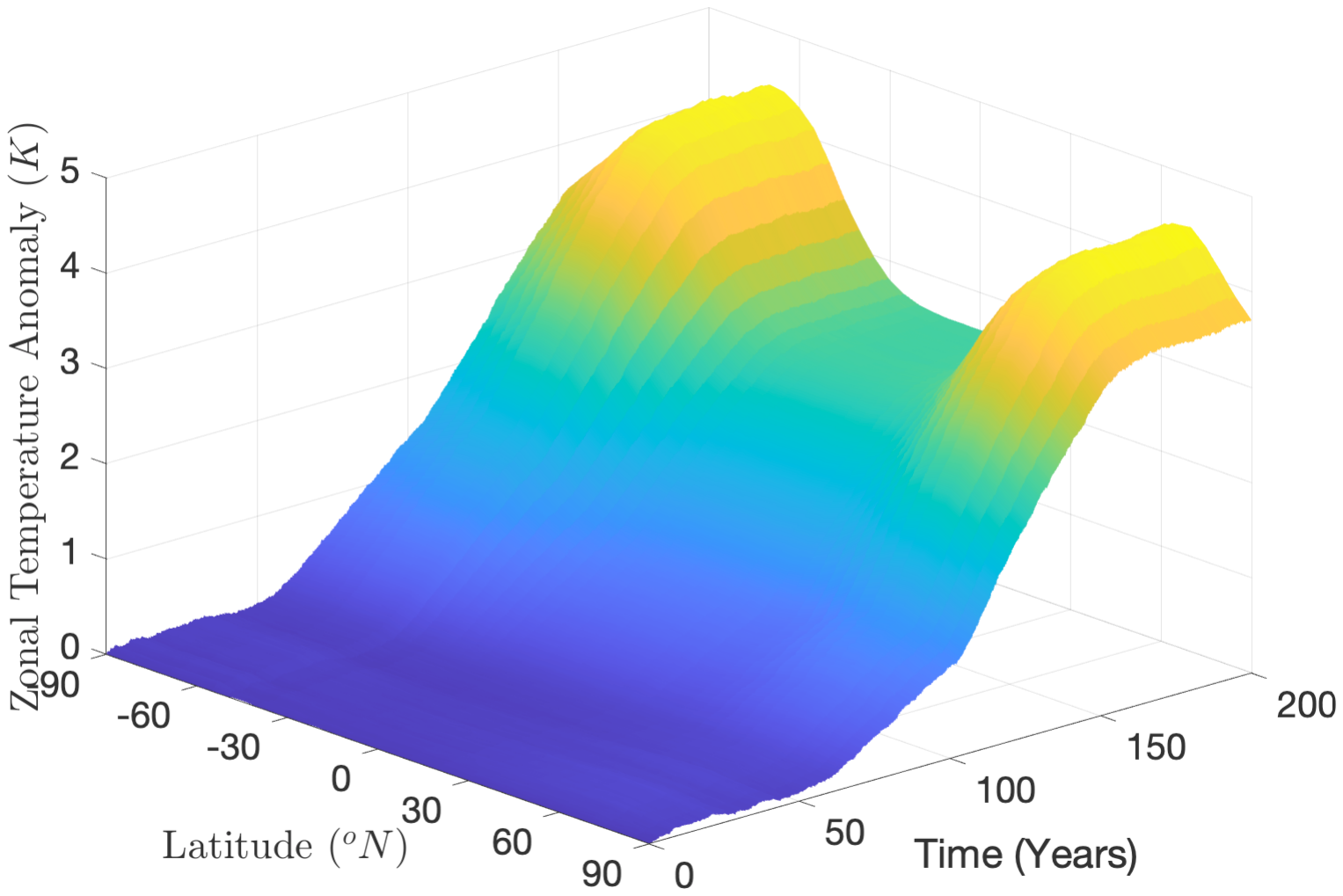}
 f)  \includegraphics[width=.3\textwidth]{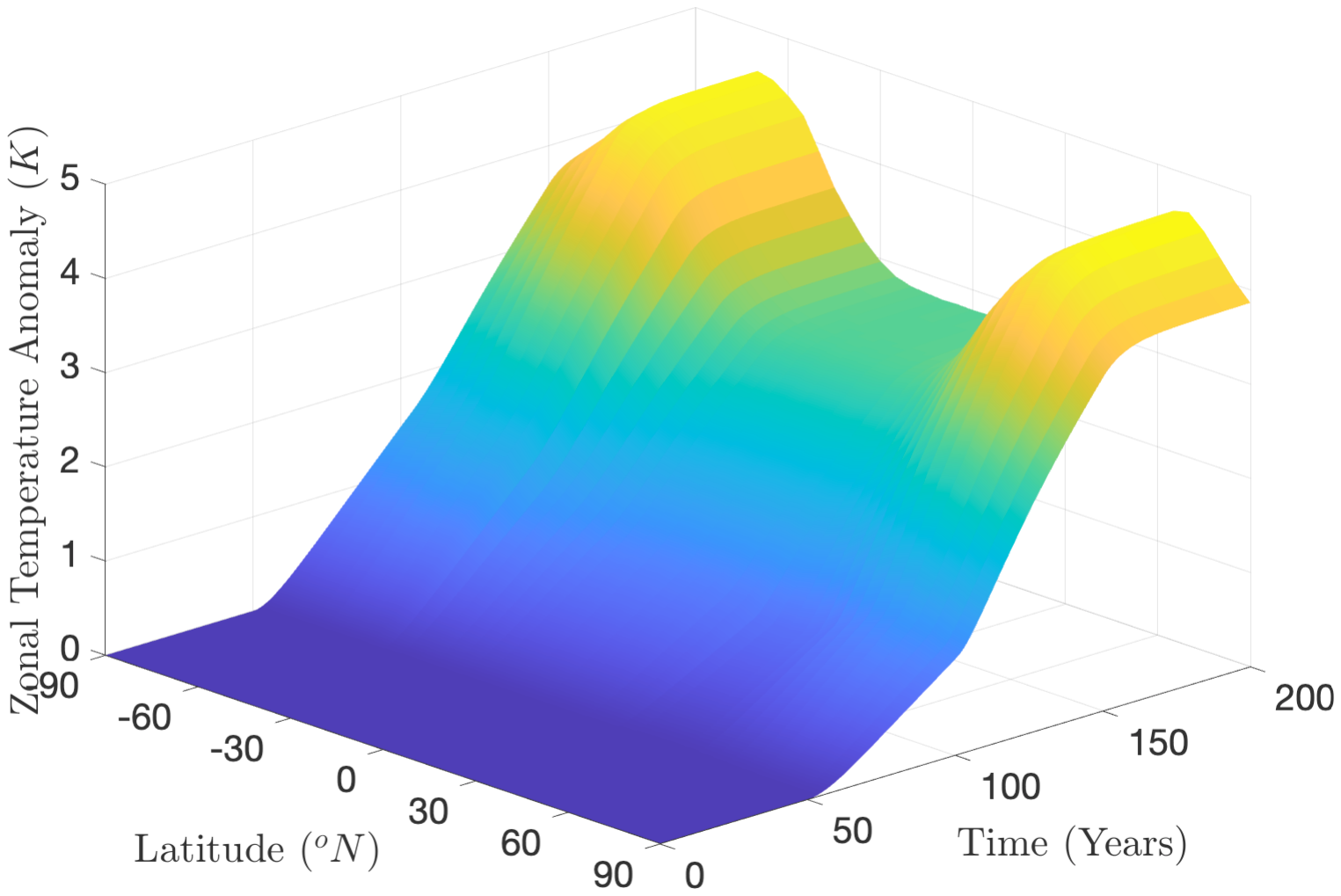}
 g)  \includegraphics[width=.3\textwidth]{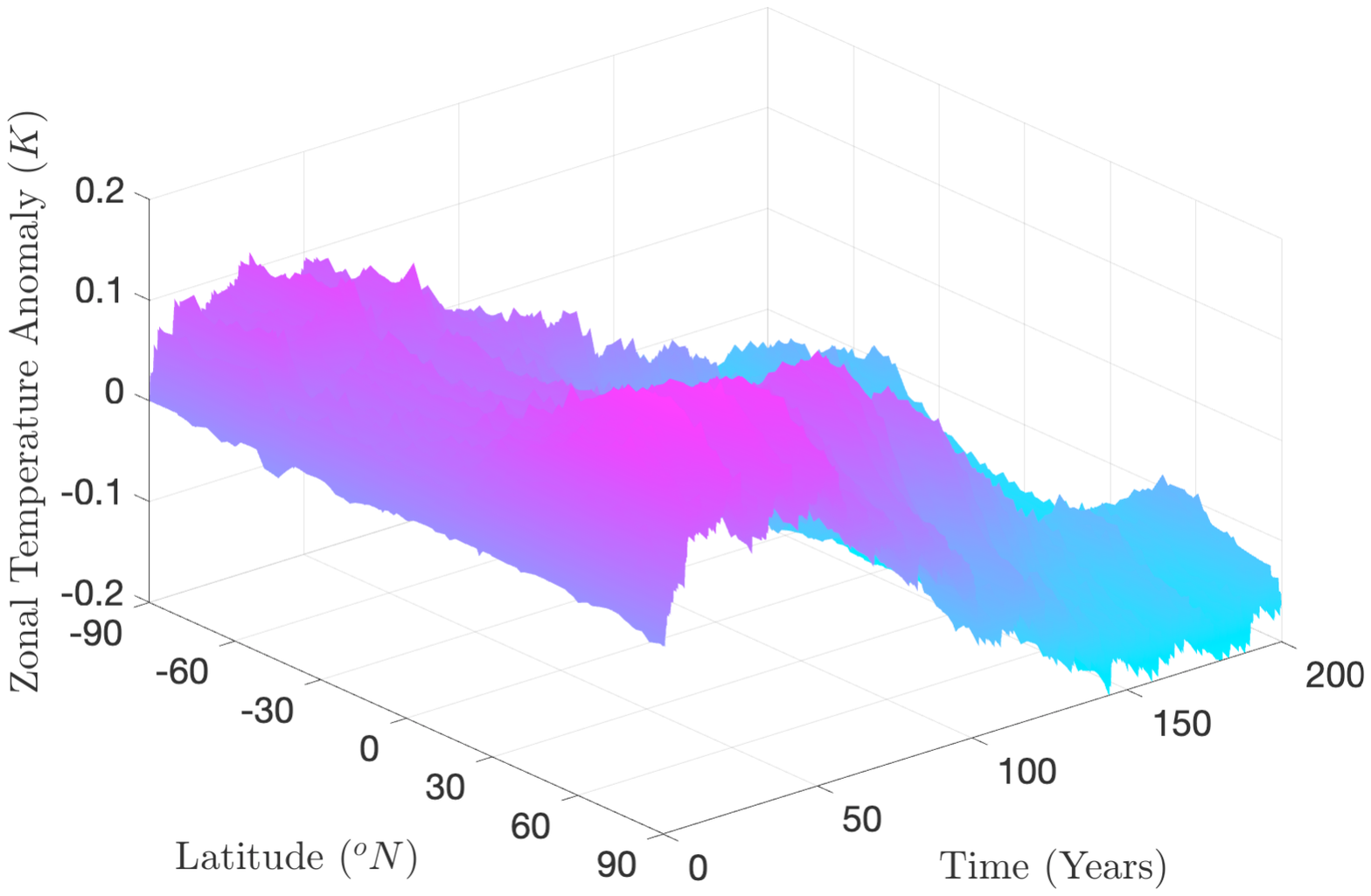}
 \caption{{\bf Ghil-Sellers model forced with space-time stochastic forcing locally distributed as an asymmetric $\alpha$-stable noise: Response vs simulations}. Panel a): Example of the evolving temperature field in steady state conditions. Note the substantial size of the peaks due to jumps, emphasizing a non-negligible amount of background noise. Panel b): Green function, $G_{CO_2}(T_j)$, associated with  $CO_2$ change. Panel c):  Green function, $G_{A}(T_j)$, associated with radiative effect of aerosols.  Panel e): Predicted response of the zonal temperature field using our LRT formalism. Panel f):  Ensemble average of zonal temperature anomaly field from brute-force simulations. Panel g): Bias between brute-force simulations of climate change and the prediction obtained by LRT.}\label{GhilSellers}
\end{figure*}

The resulting discretized equation is:
\begin{equation}
\mathbf{T}(t+\Delta t)=\mathbf{F}(\mathbf{T})\Delta t + (\Delta t)^{1/\alpha} \mathbf{L(t)}.
\end{equation}
Here, $\mathbf{T}=(T_1,\ldots,T_M)$, $\mathbf{F}$ is the spatially discretized version of the operator $\mathcal{D}$, and $\mathbf{L}=(L_1(t),\ldots,L_M(t))$, where $L_j(t)\sim\mathcal{S}(\alpha,\beta,\gamma',\mu_s=\beta\gamma'\tan(\pi\alpha/2))$ for all $j=1,\ldots,M$, and $L_j(t)\indep L_k(t')$ if $(j,t)\neq(k,t')$. For our simulations, we choose $\alpha=1.75$ and $\beta=0.5$. This choice of parameters gives a relatively modest presence of jump processes (whose relevance increases as $\alpha$ is decreased) and a prevalence of positive large fluctuations. The latter is important for preventing a tipping from the warm reference state to the snowball state, as discussed in \cite{Lucarini2022}. We set $\gamma'=1/30000$. With these parameters, we construct the reference steady state of the system; an example of an evolving space-time field is shown in Figure \ref{GhilSellers}a). 

We then follow \cite{Lucarini_Chekroun_PRL24} step-by-step in terms of the forcings and protocols for computing response operators. The only---and extremely relevant---difference in this study is in the properties of the underlying steady state. We consider two distinct climate change experiments: a) an increase in the greenhouse constant $m$, which succinctly describes a buildup of CO$_2$ concentration; and b) a localized reduction of incoming radiation in the region [$25^\circ N, 45^\circ N$], which represents in a very simplified way the radiative effect of aerosols injected into the atmosphere of the northern hemisphere's low-to-mid latitudes.

The specific perturbations for the climate change experiment, starting from steady-state conditions, are as follows:
\begin{itemize}
\item A linear increase of $m$ (build of CO$_2$ concentration) over 100 years, from an initial value of $m=m_0=0.5$ to $m=m_0+\delta m$ with $\delta m=0.01$. The value of $m$ is then kept constant afterwards.
\item In the region [$25^\circ N, 45^\circ N$], the solar constant $\mu_s$ is multiplied by a factor $\mu_s'$. This factor decreases linearly over 50 years from the initial value $\mu_{s,0}'=1$ to $\mu_s'=1-\delta\mu_s'$ with $\delta\mu_s'=0.012$. The perturbation then decays with an exponential law with a characteristic time of 20 years.
\end{itemize}

These experiments are repeated 10,000 times and the results are averaged, yielding the zonal temperature anomaly field shown in Fig.~\ref{GhilSellers}d.

Separately, we compute the Green's functions $G_{CO_2}(T_j)$ and $G_{A}(T_j)$ for $j=1,\ldots,M$, using the protocol reported in \cite[Appendix D]{Lucarini_Chekroun_PRL24} and also used in \cite{Lucarini2017,Lembo2020}. In this case, we also use 10,000 ensemble members. The resulting Green's functions are shown in Figs.~\ref{GhilSellers}b and \ref{GhilSellers}c, respectively.

By applying linear response theory via Eq.~\eqref{Eq_LRF} at each location, using the time modulation described above for the perturbations applied to $m$ and $\mu_s$, and after summing the two contributions, we obtain the predicted response of the temperature field shown in Fig.~\ref{GhilSellers}e. The bias between the direct numerical simulation of climate change and the prediction obtained by linear response theory is very modest across the considered time frame, as shown in Fig.~\ref{GhilSellers}e.

We conclude that linear response theory applies convincingly even when the Ghil-Sellers model is modified to include a spatial stochastic forcing field that is locally distributed as an $\alpha$-stable process, in agreement with the theory presented in this contribution.

A notable drawback of including jump processes, as opposed to considering standard Gaussian perturbations as we did in \cite{Lucarini_Chekroun_PRL24}, is that a much larger number of ensemble members is needed (10,000 instead of 100). This is required for both (a) estimating the climate response from numerical simulations and (b) estimating the Green's functions. This is not surprising given the nature of the stochastic process we are considering.

Finally, we note that the Green's functions in Figs.~\ref{GhilSellers}b-c are similar---but not identical---to the corresponding ones shown in \cite[Figs 1e-f]{Lucarini_Chekroun_PRL24}. This is to be expected, as the correlation properties of the reference steady states are different for Gaussian versus $\alpha$-stable background stochastic forcings. We also note that the unit of measure for the Green's function should read K/day instead of K/year, as was erroneously reported in the previous paper.

\section{Discussion}\label{conclu}
Understanding the class of dynamical systems for which linear response theory can be applied and deriving applicable linear response formulas is a significant area of research across mathematics, physics, and various fields of science that deal with complex systems. Response theory has broad applications, including predicting future changes in response to external forcings, assessing sensitivity to parameter changes for model tuning, and addressing optimization problems like anticipating adversarial attacks, i.e.~understanding which perturbations can cause the largest response in a system, see \cite{Antown2018,Antown2022}. 

As recalled in the Introduction, for a large class of diffusion processes, response operators can be expressed as time-lagged correlations between system's observables, generalizing the classical FDT. By employing the Kolmogorov operator formalism, we have clarified the link between forced and free fluctuations and decomposed response operators into terms associated with specific modes of variability (with their own decay of correlation) of the unperturbed system. This approach has the added benefit of clarifying the conditions under which critical transitions emerge through the dominant spectral gap \cite{Chek_al14_RP} and defining the critical mode associated with the occurrence of the  divergent behavior \cite{Lucarini2020,Chekroun_al_RP2,LucariniChekroun2023,Zagli2024}.

Using the Kolmogorov operator formalism (Section \ref{Sec_Kolmo_Levy}), we have successfully extended response theory to mixed jump-diffusion models (Sections \ref{Sec_Green_Kolmo} and \ref{Sec_general_formula}), which involve both Gaussian and discontinuous stochastic forcing (jumps). While the inclusion of jumps introduces nonlocal terms into the Fokker-Planck equation (Section \ref{SubSec_FKPE_Levy}), the linearity of the equation remains intact, allowing us to apply perturbation techniques. Moreover, the decomposition of response operators using Kolmogorov modes provides a clear framework for analysis and decomposition of the response operator in terms of modes of variability, still applies (Sections \ref{Sec_RPs},  \ref{Sec_Green_Kolmo}, and \ref{Sec_general_formula}). These results are in agreement with recent extensions of response theory derived for finite-state Markov processes \cite{Lucarini2025PTRSA}.

We have applied our theoretical framework to analyze the response to parameter variations, of the Jin's ENSO recharge model \cite{jin1997equatorial} subject to state-dependent jumps and additive white noise (Section \ref{Sec_linear_response_Green}). These jump-diffusion perturbations are aimed at capturing intermittent processes accounting for extra nonlinear and feedback mechanisms between the wind stress and SST anomalies which are present in more elaborated,  spatially-extended models of ENSO \cite{Zebiak_al87,Jin_al93_part1,Jin_al93_part2,Dijkstra05,cao2019mathematical}. 
They  proceed from the general framework of \cite{Chekroun_al22SciAdv}, and have been shown to produce generically shear-induced chaos \cite{Young2016} through the subtle interaction of jump-diffusion processes and nonlinear dynamics.
Within this framework, we successfully constructed response operators and verified their accuracy.

We also computed the RP resonances and Kolmogorov modes for the shear-induced chaotic dynamics displayed by our jump-diffusion perturbed version of the Jin's ENSO recharge model (Section \ref{Sec_RPs_Jin}). There, we have shown that 
such mixed stochastic perturbations responsible for the emergence of chaos, induce a larger spectral gap than in the case where the nonlinear dynamics is only subject to white noise disturbances (cf Figures \ref{Fig_RP_res_diffusion} and \ref{Fig_RP_resonances_twist}). This increase in the spectral gap is synonymous of an enhancing of the phase-space mixing characterized by faster decay of correlations. Remarkably, in the jump-diffusion case, the Kolmogorov modes obtained from (Ulam) approximations of the Markov semigroup (Eq.~\eqref{Eq_Tt}) display stretching and folding features (Fig.~\ref{Fig_Kolmo_modes_twist}) characteristics of the underlying pullback attractors (Fig.~\ref{Fig_Jin_bif}). Roughly speaking, the expectation operator involved in Eq.~\eqref{Eq_Tt} does not erase these important dynamical characteristics of the dynamics. Rather, these  structures are still somehow encoded within  the Kolmogorov modes, emphasizing the dynamical relevance of these modes. 
Our response operators could also be highly valuable for evaluating various modeling components (nonlinear and stochastic) to further enhance the impressive long-range ENSO forecast skills of  the extended nonlinear recharge oscillators,  recently introduced in  \cite{zhao2024explainable}; see also \cite{vialard2025nino}.

Due to the generic character of jump-diffusion perturbations borrowed from \cite{Chekroun_al22SciAdv}, and the generality of our response operators (Sections \ref{Sec_Green_Kolmo} and \ref{Sec_general_formula}), it is expected that such operators are still accurate for a broad class of nonlinear systems subject to such stochastic disturbances. Following   \cite{Chekroun_al22SciAdv}, those include nonlinear systems supporting a high-dimensional limit cycle, to which  the jump-diffusion perturbations of \cite{Chekroun_al22SciAdv} are guaranteed to produce shear-induced chaos as long as a center manifold can be computed; see also \cite{lu2013strange}.

High-dimensional limit cycles are found in various partial differential equations (PDEs) and time-delay models encountered in the study of climate dynamics. They play an important role in ENSO dynamics  \cite{TSCJ94,tziperman1995irregularity,JinEtal96,Galanti_al00,CGN17,cao2019mathematical,Tantet_al_ENSO}, the description of cloud-rain oscillations \cite{koren2011aerosol,koren2017exploring,chekroun2020efficient}, or other basic geophysical flows \cite{simonnet2003low,Dijkstra05,GCS08}. We refer to \cite{MW14,dijkstra2015,chekroun2020efficient,chekroun2022transitions,chekroun2024effective} for center manifold calculations and generalizations \cite{CLM20_closure,chekroun2023optimal} in such high-dimensional settings (both for PDEs and time-delay models), and to \cite{CLW15_vol1,CLW15_vol2,chekroun2023transitions,chekroun2025unravel}, for stochastic PDEs.

As a further test of the relevance of our results, we have constructed response operators and used them to successful compute climate projections using a modified version of the Ghil-Sellers  model \cite{Sellers,Ghil1976,Bodai2015}. Indeed, here the one-dimensional reaction-diffusion equation describing the latitudinal budget and meridional transfer of energy across the domain of the climate system is forced using a spatio-temporal stochastic field distributed according to a L\'evy noise law. This indicates the relevance of response theory for treating the problem of climate change also in the case one considers complex background stochastic forcings, going well beyond the usual Gaussian \textit{ansatz}. 

While originally primarily studied in finance, interest in mixed jump-diffusion processes has expanded to various fields of science and technology, including e.g.~biology and epidemiology. Singular perturbations also arise in models with complex decision-making structures, such as those found in climate models. For instance, the parametrization of subscale convection in the ocean and atmosphere often involves "if-then" statements to assess the stability of geophysical flows. Our results suggest that, despite these strong nonlinearities in the model formulations, linear response theory can still be applied. This strengthens the argument for using this approach to perform climate change projections with models of varying complexity and to assess the proximity to tipping points \cite{Ghil2020,LucariniChekroun2023}.

Finally, our results build up on the recent findings presented in \cite{Lucarini_Chekroun_PRL24} and provide foundational support for the use of optimal fingerprinting methods for climate change detection and attribution also in the case one considered complex stochastic forcings as treated here. This strengthens one of the key aspects of the science behind climate change. Optimal fingerprinting is in fact a statistical methodology that  is not in principle restricted to climate applications. Hence, the findings presented in this paper extend to a large class of complex system the possibility of establishing a pathway for linking causally observed change signal with acting forcings.

\begin{acknowledgements}
This work has been supported by the Office of Naval Research (ONR) Multidisciplinary University Research Initiative (MURI) grant N00014-20-1-2023, and by the National Science Foundation grant DMS-2407484.
This work has been also supported by the European Research Council (ERC) under the European Union's Horizon 2020 research and innovation program (grant agreement No. 810370).  
VL acknowledges the partial support provided by the Horizon Europe Project ClimTIP (Grant No. 100018693), by the ARIA SCOP-PR01-P003 - Advancing Tipping Point Early Warning AdvanTip project, by the European Space Agency project PREDICT (contract 4000146344/24/I-LR), and by the EPSRC project LINK (Grant No. EP/Y026675/1). NZ has been supported by the Wallenberg Initiative on Networks and Quantum Information (WINQ).
\end{acknowledgements}

\bibliographystyle{ieeetr}
\bibliography{Super_Bib_v2MC_VL}
\end{document}